# Enhanced White-Light Emission from Self-Trapped Excitons in Antimony and Bismuth Halides through Structural Design

Philip Klement[1], Lukas Gümbel[1], Meng Yang[2], Jan-Heinrich Littmann[1], Tatsuhiko Ohto[3], Hirokazu Tada[4], Sangam Chatterjee[1], Johanna Heine[5,*]

[1] Institute of Experimental Physics I and Center for Materials Research (ZfM), Justus Liebig University Giessen, Heinrich-Buff-Ring 16, 35392 Giessen, Germany.

[2] Department of Chemistry and mar.quest|Marburg Center for Quantum Materials and Sustainable Technologies, University of Marburg, Hans-Meerwein-Straße, 35043 Marburg, Germany.

[3] Graduate School of Engineering, Nagoya University, Aichi 464-8603, Japan.

[4] Graduate School of Engineering Science, Osaka University, Japan.

[5] Institute of Chemistry, School V, University of Oldenburg, Germany.

**ABSTRACT:** Lead halide perovskites have catalyzed the rise of main-group metal halide materials as promising candidates for next-generation optoelectronics, including solar cells, light-emitting diodes, lasers, sensors, and photocatalysts. Among these, efficient light-emission arises from self-trapped excitons, wherein excited states induce transient lattice distortions that localize excitons. However, the complex interplay of factors, such as lattice distortions, lattice softness, and electron-phonon coupling dynamics, obscures the direct structure-property relationships complicating the targeted material design. In this study, we advance the understanding of self-trapped exciton (STE)-based emission in hybrid antimony and bismuth halides, emphasizing the interplay of structural and electronic factors that enhance white-light emission. We systematically vary composition, anion dimensionality, connectivity, and the organic cation and find that the presence of Bi/Sb and Cl in edge-sharing anion motifs promotes white-light emission and optimal electron-phonon coupling. Chlorides outperform bromides, and organic cations, such as CMA and BZA, only subtly influence optical behavior by altering lattice dynamics and rigidity, resulting in tunable emission characteristics without compromising STEs. This work deepens the understanding of the emission mechanisms in hybrid halide perovskites and establishes guiding principles for tailoring optoelectronic properties, paving the way for advanced materials with enhanced white-light emission for next-generation optoelectronic applications.

## INTRODUCTION

Metal-halide materials, particularly lead-halide perovskites, have garnered significant interest due to their outstanding optoelectronic properties and potential applications in next-generation solar cells.[1] Derived from this class of materials, low-dimensional metal-halide perovskites featuring self-trapped excitons (STE) promise to revolutionize solid-state and display lighting and radiation detection due to their intrinsic broadband white-light emission spectrum, high quantum efficiencies, and absence of self-absorption.

Efficient light-emission was initially explored in nanoparticles of 3D inorganic lead halide perovskites, such as $CsPbBr_3$ which feature an anionic substructure of corner-sharing octahedra extending in all spatial directions.[2] Bright photoluminescence (PL) is also reported for 2D organic-inorganic perovskites like $(C_4H_9NH_3)_2PbBr_4$, where layered inorganic anions are separated by organic cations.[3,4] This specific arrangement forms a quantum-well-like structure that provides the confinement for efficient emission processes.[5] Typically, spectrally narrow free-exciton emission is observed in these <100> perovskites, where corner-sharing $EX_6$ octahedra (M = metal, typically $Sn^{2+}$ or $Pb^{2+}$; X = halogen, Cl, Br, or I) compose the flat inorganic layer. The emission wavelength can be tuned by altering the slab thickness of the inorganic layer,[6] the constituting metal,[7] the halogen,[8] or by modifying the distortion within layers.[9]

In 2014, Karunadasa *et al.* reported (N-MEDA)[$PbBr_4$] (N-MEDA = N-methylethane-1,2-diammonium)[10] and (EDBE)[$PbX_4$] (EDBE = 2,2′ -(ethylenedioxy)-bis(ethylammonium); X = Cl, Br)[11] as the first 2D perovskites exhibiting broad white-light emission with a photoluminescence quantum yield (PLQY) of 9%. Since this initial discovery, additional metal-halide materials showing similar emission properties have been reported,[12] including compounds featuring chain-like (1D)[13] and molecular (quasi-“0D”)[14–16] anions.[17,18] Potential applications include the use as phosphors or electroluminescent materials in light-emitting diodes,[19–21] scintillators,[22] solar concentrators,[23] remote thermometry[24,25] and sensing.[26]

The emission mechanism is attributed to STEs that form due to the soft lattice and strong electron-phonon coupling in these materials.[27] In this process, an excited electron-hole pair causes an elastic lattice-distortion and traps by its own deformation potential from which it recombines radiatively. The exciton-trapping by this transient defect leads to broad emission bands and large Stokes shifts. STE formation in metal halide materials is believed to depend on optimal electron-phonon coupling, which should neither be too weak nor too strong, with a suggested Huang-Rhys factor of 10–40.[28]

Previously, STE (also sometimes referred to as small polarons or in molecular materials as excimers) have been studied in solid noble gases,[29] binary metal halides such as NaCl,[30] organic semiconductors,[31] and, under specific circumstances, in solid state semiconductors such as ZnSe:Te.[32]

The emergence of STE emission has been extensively investigated in 2D perovskites.[33,34] The distortion of the inorganic layer appears to play a crucial role, although direct comparisons between different types of layers found in <100>, <110> and <111> perovskites remain challenging. The constituting halogens or halogen mixtures are also significant, providing another tool for fine-tuning the emission color.[35] On the other hand, 0D perovskites such as $Cs_4SnBr_6$ and $(Ph_4P)_2SbCl_5$,[36,37] are promising white-light emitters with high or even ideal quantum efficiencies. These compounds are conceptually more simpl to analyze than their 2D counterparts, and guidelines for their design have been established: Efficient STE emission is observed in compounds containing $5s^2$ ions such as $Sn^{2+}$ and $Sb^{3+}$ with chlorido or bromido ligands, low coordination numbers, and sufficient separation of the anions through the use of bulky organic cations.[38–41]

However, 1D and 0D perovskites remain significantly underexplored, and their performance lags lead-halide perovskites. A major challenge is to understand the relationship between structural factors and white-light emission. Comprehensive structure-property relationships for STE emission could significantly affect the emission efficiency and characteristics, impacting potential applications in optoelectronics.

In this work, we present the synthesis and detailed analysis of the structural and optical properties of a series of 1D and 0D antimony and bismuth halide materials incorporating the cyclohexylmethylammonium (CMA) cation. The synthesized compounds include $[CMA]_4[Sb_2Cl_{10}]$ (**1**), $[CMA]_4[Bi_2Cl_{10}]$ (**2**), $[CMA]_3[SbBr_5]Br$ (**6**), and $[CMA]_3[BiBr_5]Br$ (**7**). We also have prepared both new and literature-known compounds using the benzylammonium (BZA) cation to facilitate direct comparisons among closely related materials. These include $[BZA]_6[Sb_2Cl_{10}]Cl_2$ (**3**), $[BZA]_4[Bi_2Cl_{10}]$ (**4**),[42] $[BZA]_3[BiCl_5]Cl$ (**5**),[43] $[BZA]_2[SbBr_5]$ (**8**)[44], $[BZA]_3[SbBr_6]$ (**9**), and $[BZA]_3[BiBr_6]$ (**10**). [45].

We explore the relationship between the crystal structure and optical properties across all synthesized compounds, finding STE emission to dominate the luminescence properties. We attribute the spectrally broad photoluminescence from STEs observed in in 8 out of 10 compounds to strong electron-phonon coupling in these materials. The Huang-Rhys factors range from 4.9 to 22 and the phonon energies from 38 to 120 meV. Analyze the STE emission with respect to variations in metal, halide, organic cation, and anion connectivity infers that STE emission is primarily associated with edge-connected, chloride-containing anions. Some direct comparisons are possible within the chloride edge-sharing motif: At a fixed cation (CMA), Sb is bright while Bi is dark (**1** vs **2**), at fixed metal (Bi), switching CMA→BZA restores emissivity (**2** vs **4**). Thus, the apparent metal effect is filtered by cation-controlled lattice dynamics, which can move the system into the moderate S, higher $E_{Ph}$ window favorable for radiative STE emission. Across the broader series, cations remain secondary, but within a fixed anion motif they can be decisive.

## METHODS

**General.** All reagents were used as received from commercial sources. CHN analysis was carried out on an *Elementar* CHN-analyzer. Powder patterns were recorded on a *STADI MP* (*STOE* Darmstadt) powder diffractometer, with CuKα1 radiation with λ = 1.54056 Å at room temperature in transmission mode (see Figures S20-S28). IR spectra were measured on a *Bruker Tensor 37* FT-IR spectrometer equipped with an ATR-Platinum measuring unit (see Figures S29-S34). Thermal analysis was carried out by simultaneous DTA/TG on a *NETZSCH STA 409 C/CD* in the temperature range of 25 to 1000 °C with a heating rate of 10 °C $min^{-1}$ in a constant flow of 80 ml $min^{-1}$ Ar (see Figures S35-S43). Optical absorption spectra were recorded on a *Varian Cary 5000* UV/Vis/NIR spectrometer in the range of 300–800 nm in diffuse reflectance mode employing a Praying Mantis accessory (*Harrick*).

**General synthesis procedure.** Under aerobic conditions, the respective metal oxides were dissolved in hydrohalic acid solution, followed by the addition of amine. The resulting clear reaction solution was heated to reflux for 30 minutes, filtered and left undisturbed for crystallization, which occurred within one to three days. Crystals were isolated by filtration, washed with small amounts of cold glacial acetic acid and pentane and dried under vacuum. Reagent amounts and yields for individual reactions are given in tables S1 and S2 in the SI.

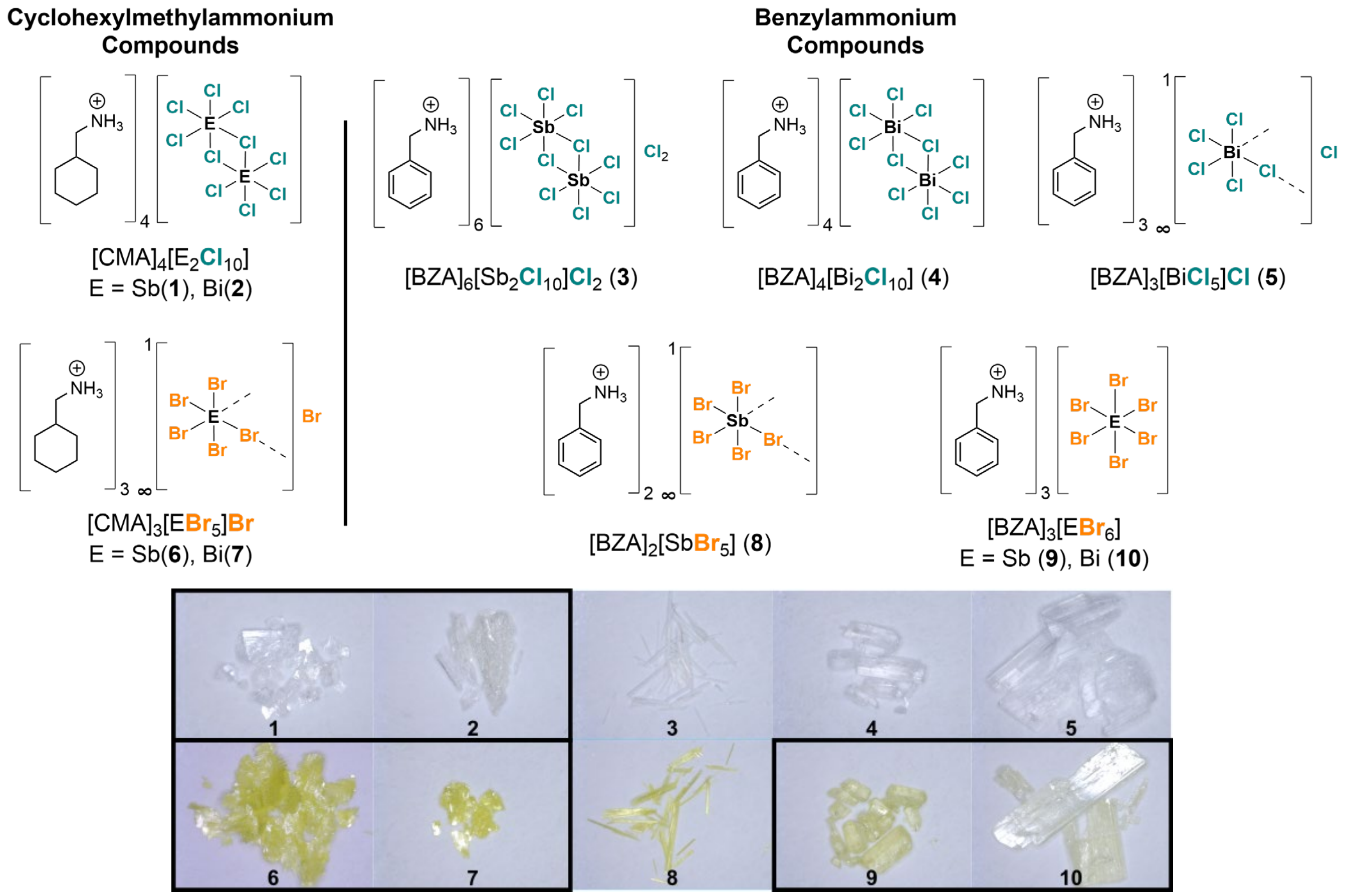

**Figure 1:** Overview of the cyclohexylammonium (CMA) and benzylammonium (BZA) metal halide compounds with schematics shown on top and crystal samples shown below, with isostructural pairs marked with a black frame. The chain-like nature of the $[EX_5]^{2-}$ anion is denoted by polymeric notation.

Synthesis and analysis data for compound **5** ia provided in a previous publication. [43] Compounds **8**, **9** and **10** were prepared according to modified literature procedures to allow for the use of metal oxides and amine as starting materials.[44] [45] The synthesis and single crystal structure at room temperature of **4** has been reported before.[42] We measured the single crystal structure at 100 K to supplement the literature data. The phase purity of all compounds was confirmed via CHN and PXRD analysis (see SI).

Single crystals for measurement were directly obtained from the mother liquor prior to filtration. Thermal analysis and IR data indicate that **1** and **2** lose water upon drying (see SI).

**X-ray Crystallography.** Single crystal X-ray determination was performed at 100 K on a on a Bruker Quest D8 diffractometer with microfocus MoKα radiation and a Photon 100 (CMOS) detector or a STOE IPDS-2 diffractometer equipped with an imaging plate detector system using $MoK_{\alpha}$ radiation with graphite monochromatization. The structures were solved using direct methods, refined by full-matrix least-squares techniques and expanded using Fourier techniques, using the ShelX software package within the OLEX2 suite.[46–49] All non-hydrogen atoms were refined anisotropically unless otherwise indicated. Hydrogen atoms were assigned to idealized geometric positions and included in structure factors calculations. Pictures of the crystal structures were created using DIAMOND.[50] Additional details on individual refinements are reported in the SI, Tables S3-S8 and Figures S1-S6. Crystallographic data has been deposited at the CCDC[51] as 2479960 (**1**), 2479964 (**2**), 2479965 (**3**), 2479962 (**4**), 2479961 (**6**) and 2479963 (**7**).

**Optical characterization.** Optical measurements were carried out at a low temperature of 4 K with the samples in vacuum. For μ-reflectance measurements, we utilized light emitted from a combined deuterium-tungsten lamp. The light was focused onto the sample using a $CaF_2$ lens resulting in an approximately 250 μm diameter spot size. The reflected light was collected by the same lens and directed into the spectrometer. We subtracted the background reflectance intensity ($R_{bg}$) from the sample reflectance intensity ($R_{sample}$) and normalized it using the reflectance intensity from a reference Si substrate ($R_{ref}$) to obtain reflectance spectra. The normalized reflectance was calculated as $R = \frac{R_{\mathrm{sample}}-R_{\mathrm{bg}}}{R_{\mathrm{ref}}-R_{\mathrm{bg}}}$ and the corresponding absorption as $A = 1 - R$. For μ-photoluminescence (PL) measurements, samples were excited using a 325 nm (3.82 eV) laser. The beam was focused into a 1.2 μm spot using a 40× microscope objective lens with a numerical aperture of 0.6, and the excitation power density was 20 W $cm^{-2}$. For photoluminescence excitation spectroscopy (PLE), samples were excited using a 5 kHz laser amplifier system with an optical parametric amplifier. The wavelength range was 310–600 nm with a spectral width of 5 nm, and the excitation power was 1.9 μW. For PLE analysis, the PL intensity was integrated from 620–730 nm to yield one spectrum per wavelength step. The sample was at room temperature under ambient conditions. Further details on the optical setup can be found in the SI.

**Density functional theory calculations.** The first-principles calculations were performed with the projected augmented wave method implemented in the VASP code.[52,53] We used Perdew-Burke-Ernzerhof (PBE) functional.[54] The plane wave energy cutoff was set to 500 eV. A gamma-centered 2×2×2 k-point grid was employed for the experimentally obtained structure of compounds **1**, **2**, **6**, and **7**.

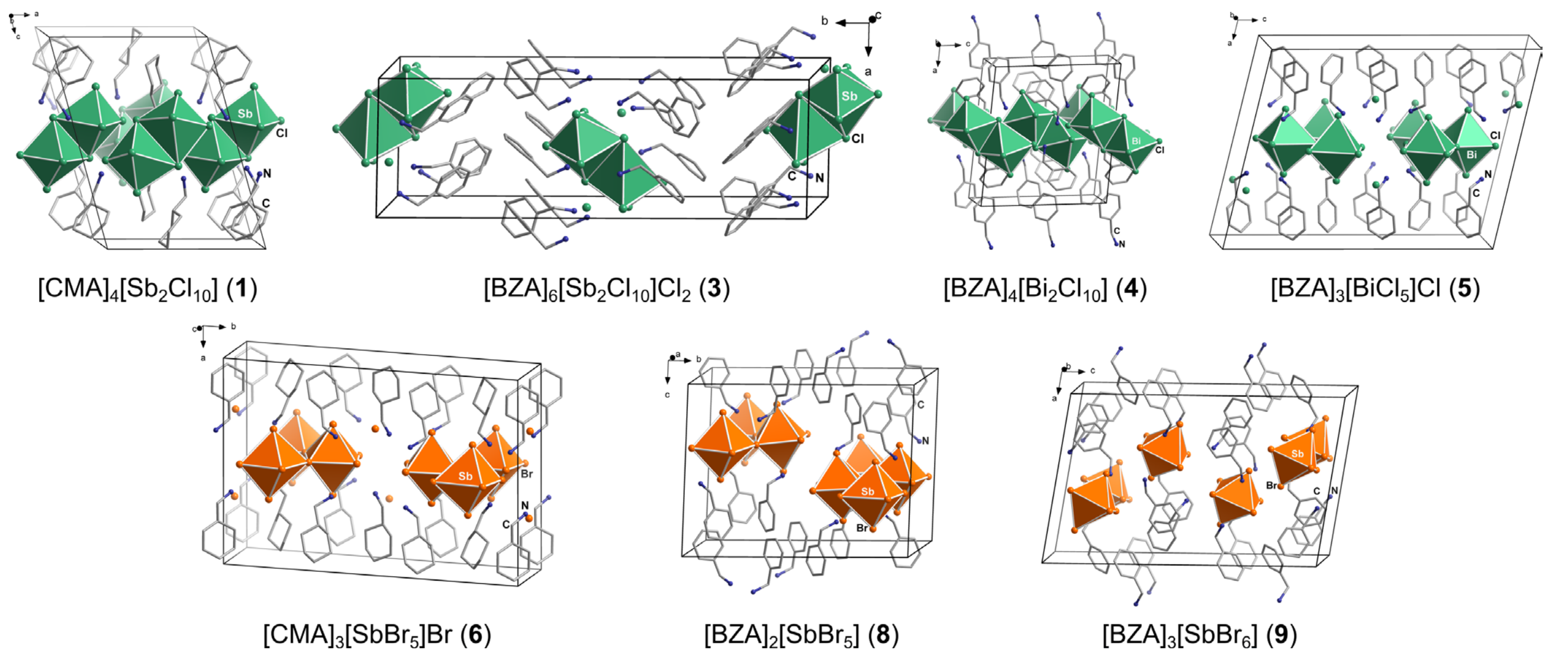

**Figure 2:** Excerpts of the single crystal structures of **1**, **3**, **4**, **5**, **6**, **8**, and **9** highlighting different packing arrangements. For isostructural pairs only the antimony compound is shown. Hydrogen atoms and disorder omitted for clarity, metal halide octahedra shown in polyhedral view.

## RESULTS

**Synthesis.** All compounds are available through a facile and well established[55] solution based route where the respective metal oxide $Sb_2O_3$ or $Bi_2O_3$ and the parent amine cyclohexylamine or benzylamine are dissolved in dilute or concentrated hydrohalic acid solution, heated to reflux for 30 minutes and left to cool and crystallize. Reagent and acid concentrations must be optimized for each synthesis to obtain phase-pure products (see Tables S1 and S2 in the SI). We note that reaction stoichiometry, acid concentration and overall concentration of the reaction mixture can influence the reaction outcome in non-trivial ways and often the reaction stoichiometry needed for phase pure crystalline products differ from the composition of the target compound. For example, $[BZA]_4[Bi_2Cl_{10}]$ (**4**) is obtained from 2M hydrochloric acid, while the synthesis of $[BZA]_3[BiCl_5]Cl$ (**5**)[43] requires the use of concentrated acid. Similarly, for the synthesis of $[BZA]_2[SbBr_5]$ (**8**)[44] and $[BZA]_3[SbBr_6]$ (**9**)[45] different reaction stoichiometries (equivalent ratio of Sb to amine: 1:1 for (**8**), 1:1.7 for (**9**)) are needed.

**Crystal Structure Description.** An overview of the single crystal structures discussed here is shown in Figure 2. Compounds **1** and **2** are isostructural, crystallizing in the triclinic crystal system in space group *P*-1 (No. 2) with two formula units per unit cell as colorless plates. The dinuclear $[E_2Cl_{10}]^{4-}$ anion found in **1** and **2** is composed of two edge-sharing $ECl_6$ octahedra, a common anion motif in the chemistry of halogenido pentelates, with a number of examples reported for antimony and bismuth.[56–59] The bond lengths in the $[E_2Cl_{10}]^{4-}$ anions in **1** and **2** display a typical pattern (see also Figures S7 and S9 in the SI): Bridging E-Cl bonds are longer than terminal ones and terminal E-Cl bonds *trans* to the bridging ones are shortest due to the structural *trans* effect in halogenido pentelates.[60] These differences in bond length are more pronounced in the antimony compound, with Sb-Cl distances of 2.409–3.117 Å, than in the bismuth compound, with Bi-Cl distances of 2.538–2.910 Å. For both compounds, *trans*-Cl-E-Cl angles deviate from the ideal 180° (**1**: 171–176°, **2**: 171–177°). One molecule of solvate water is found in the single crystal structures of **1** and **2** which is lost upon drying according to TGA and IR data. PXRD patterns confirm that no major structural changes occur upon desolvation. CMA cations and anions in **1** and **2** are packed such that charge-neutral cation-anion-cation tri-layers are generated which are held together only by van der Waals forces. This makes the compounds suitable for exfoliation towards few- or monolayer materials (Figures S8 and S10 in the SI).

Compound **6** and **7** are isostructural as well. They crystallize as light yellow plates in the monoclinic crystal system in space group $P2_1/c$ (No. 14) with four formula units per unit cell. The anions in **6** and **7** are chains composed of *cis*-corner-sharing $EBr_6$ octahedra, another common anion motif among halogenido pentelates.[55,61] The bond length pattern of longer bridging and shorter terminal metal-halogen bonds is observed in **6** and **7** as well, although less pronounced than in **1** and **2** (E-Br distances: 2.649–3.022 Å in **6**, 2.747–2.987 Å in **7**, see Figures S11 and S13). Similarly, the range of *trans*-Br-E-Br angles observed is smaller (**6**: 175–176°, **7**: 174°). Again, cations and anions pack in a pseudo-layered fashion, generating potentially exfoliable tri-layers (Figures S12 and S14 in the SI). Remarkably, the particular packing $[CMA]^+$ cations, $[EBr_5]^{2-}$ chains and additional bromide ions allows for a direct derivation of the structure from the cubic perovskite aristotype which has only been reported twice before in the chemistry of halogenido pentelates.[43,62] These packing motifs underscore the ability of the cylcohexylmethylammonium ion to template pseudo-layered arrangements. In the chemistry of 2D lead halide perovskites many different primary ammonium ions $[RNH_2]^+$, ranging from $[EtNH_3]^+$[35] to large organic semiconductors[56], can be used to template the formation of layered anions, especially <100> perovskites. In contrast, the formation of layered anions is much scarcer in the chemistry of halogenido penetelates[63–65] unless they are derived from double perovskites.[66] As a consequence, pseudo-layered, potentially exfoliable arrangements as observed in **1**, **2**, **6** and **7** have remained rare as well.

Compound **3** crystallizes in the monoclinic crystal system in space group $P2_1$ (No. 4) with two formula units per unit cell as colorless fragments. The crystal structure was refined as an inversion twin. The anion motif in **3** is the same as in **1**, a dinuclear $[Sb_2Cl_{10}]^{4-}$ unit,. The sum formula of $[BZA]_6[Sb_2Cl_{10}]Cl_2$ can be rewritten as $[BZA]_4[Sb_2Cl_{10}]\cdot 2[BZA]Cl$ to highlight the relationship between **1** and **3**. However, the bond length differences in the Sb-Cl bonds are more pronounced in **3**, here ranging 2.390–3.255 Å (Figure S15). Similarly, *trans*-Cl-Sb-Cl angles deviate more strongly from 180° and feature a broader range with values 169–175° in **3**. Despite the difference in composition between **1** and **3**, the packing in **3** can be viewed as pseudo-layered, featuring cation-anion-cation trilayers (Figure S16).

Compound **4** crystallizes in the monoclinic crystal system in space group $P2_1/c$ (No. 14) with two formula units per unit cell as colorless plates. **4** features dinuclear $[Bi_2Cl_{10}]^{4-}$ anions (Figure S17), here with Bi-Cl bond length of 2.530–2.970 Å and *trans*-Cl-Bi-Cl angles in a range of 172–179°. A pseudo-layered packing of anions and cations is observed, although a small amount of interpenetration of the benzyl moieties of adjacent layers are seen (Figure S18). While not isostructural with **2**, the two compounds are closely related regarding their composition and packing, allowing for a facile comparison of the influence of the organic cation on these compounds' properties (see below). The literature-known compound $[BZA]_3[BiCl_5]Cl$ (**5**)[43] features a similar composition, anion motif, and packing as **6** and **7**. In contrast, compound $[BZA]_2[SbBr_5]$ (**8**)[44] also displays chains composed of *cis*-corner-sharing $SbBr_6$ octahedra as its anion motif, but no pseudo-layered packing seen in the previously discussed compounds. Similarly, the two isostructural compounds $[BZA]_3[EBr_6]$ (E = Sb (**9**), Bi (**10**))[45], which feature simple mononuclear $[EBr_6]^{3-}$ units as their anion motif, do not show pseudo-layered packing (see Figures S19 for packing diagrams of **5**, **8**–**10**).

**Organizing Principles.** The compounds **1-10** can be categorized in several different ways. One approach is to differentiate by the templating cations cyclohexylmethylammonium and benzylammonium. Both are similar in the way they can template the formation of halogenido metalates, featuring a flexible methylammonium group extended either via a cyclohexyl or a phenyl moiety. However, clear differences in molecular volume exist: The parent hydrocarbons cyclohexane and benzene feature molecular volumes of 99 Å$^3$ and 85 Å$^3$, respectively.[67] Additionally, only the phenyl group is able to provide C-H…π interactions.[68] This is realized in the two related compounds $[CMA]_4[Bi_2Cl_{10}]$ (**2**) and $[BZA]_4[Bi_2Cl_{10}]$ (**4**), with **2** featuring a larger cell volume per formula unit (1141 Å$^3$ versus 1020 Å$^3$ in **4**), even when considering that the additional water molecules in **2** will have a small contribution (two times 17 Å$^3$)[67]. Also, a greater thickness of the charge-neutral trilayers in both compounds is found in **2** (15.3versus 13.3 Å in **4**) resulting from a lesser degree of interdigitation is observed in **2**. As a result of this difference in the features of the cationic template only **2** and **4** show the same composition $A_4E_2X_{10}$, while different compositions are observed for CMA and BZA for all other combinations of E and X.

Another approach toward the organization of the compounds focuses on the three different types of anion motifs: A dinuclear unit of edge-sharing octahedra $[E_2X_{10}]^{4-}$ in **1**–**4**, a flat chain of *cis*-corner-sharing octahedra $[EX_5]^{2-}$ in **5**–**8**, and isolated octahedral units $[EX_6]^{3-}$ in **9**–**10**. Notably, these motifs can be found either as simple salts, for example $[CMA]_4[Sb_2Cl_{10}]$ (**1**), or double salts like $[BZA]_4[Sb_2Cl_{10}]\cdot[BZA]Cl_2$ (**3**), althoughthis does not seem to have a significant influence on ground state distortion or the presence or absence of luminescence in our series of samples.

A third approach lies in organizing the series according to the constituent metal or halogen. The halogen appears to have a more drastic influence on what kind of anion motif and overall compound is observed. In contrast, three pairs of isostructural antimonates and bismuthates, **1**/**2**, **6**/**7** and **9**/**10** are observed, pointing to a smaller influence of the metal, at least on the anion motifs in our series.

**Distortion.** Ground state distortion has been discussed as an important factor for the observation of efficient photoluminescence based on self-trapped excitons.[41] Several ways have been put forth to quantify how much a given motif deviates from an ideal octahedron. These include the variance $\sigma^2$ of the *cis*-metal-halogen angles θ calculated according to the following equation[69]:

$$\sigma^2 = \frac{1}{11}\sum_{n=1}^{12}(\theta_n - 90)^2$$

To describe the variation in metal-halogen bond length, two measures have been established: Δd,[70] a measure of how much deviation from the average bond length is observed, and D,[71] which is used to described the off-centering of the metal atom within the octahedron from the center of gravity, according to the following equations, with individual M-X bond length $d_n$, average M-X bond length $d_0$ and long and short *trans*-bond length $a_n$ and $b_n$:

$$\Delta d = \frac{1}{6}\sum_{n=1}^{6}\left(\frac{d_n - d_0}{d_0}\right)^2$$

$$D = \sum_{n=1}^{3}\frac{|a_n - b_n|}{a_n + b_n}$$

Values for $\sigma^2$, Δd, and D for our series of compounds are shown in Table 1. Overall, significant distortions both in angles and bond length are observed for the combination Sb/Cl in **1** and **3**. These are higher than what is typically seen in mononuclear Sb/Cl anions (for example in $[EtPPh_3]_2[SbCl_5]\cdot EtOH$ with $\sigma^2 = 2.76$ and $\Delta d = 16.12 \times 10^{-4}$),[72] but similar to other chlorido antimonates with the same anion motif (for example in $(C_3H_{12}N_2)_2Sb_2Cl_{10}$ with $\sigma^2 = 51.29$ and $\Delta d = 118 \times 10^{-4}$).[73] Beyond this, no clear picture emerges across the entire series. We note that comparisons across multiple E/X combinations and different anion motifs will mostly show well-known trends, with compounds with lighter halogens and central group 15 elements displaying more pronounced ground state distortions.[74] Additionally, the presence of bridging halogen atoms induces a structural *trans*-effect,[60] evidenced by the observation of low values for Δd and D in **9** and **10**, where no bridging occurs. Finally, crystallographic symmetry may obscure the presence of ground state distortion to some degree as well.

**Stability.** The thermal analysis data of **1**–**10** is summarized in Table S9, with detailed TGA and DSC data shown in Figures S35-S43. The use of CMA cations produces thermally robust compounds with a decomposition range of 315–383 °C. In contrast, thermal stability in the BZA compounds only ranges between 184 °C and 280 °C.

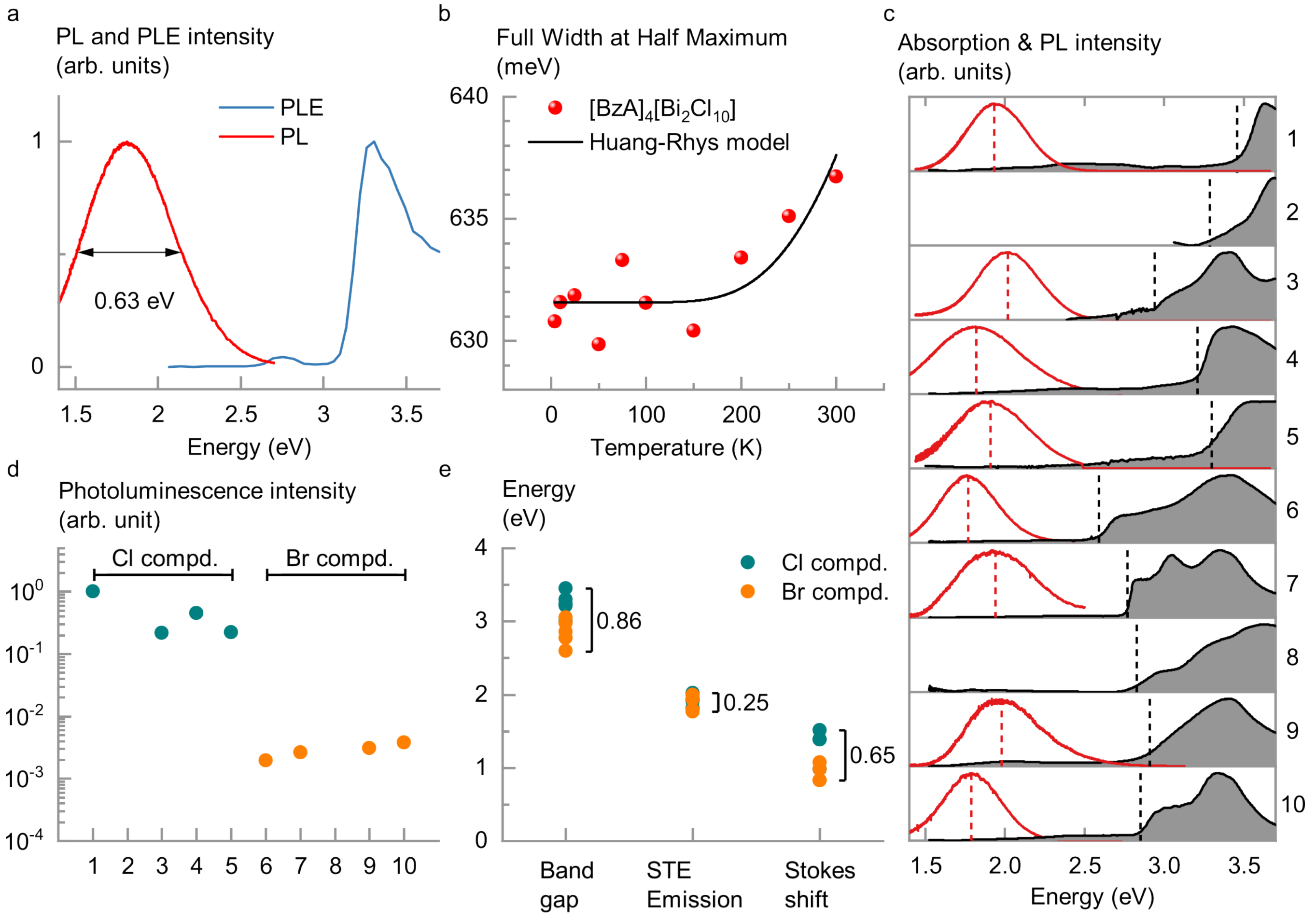

**Figure 4: Self-trapped exciton emission in hybrid Bi and Sb halides.** a) Photoluminescence (PL) and PL excitation spectra of $[BZA]_4[Bi_2Cl_{10}]$ (**4**) showing a broad STE emission with a large Stokes shift. b) Corresponding temperature-dependent full width at half maximum (FWHM) and a Huang-Rhys-model yielding $S = 4.9 \pm 0.5$ and $\hbar\Omega_e = 120$ meV. c) Absorption (black) and PL spectra (red) of all synthesized compounds. The vertical black (red) dashed line indicates the band-gap energy (emission center). d) Comparison of relative PL intensities of all synthesized compounds. e) Band-gap energy, emission center and Stokes shift of all compounds showing a narrow emission energy range.

**Table 1:** Measures of distortion in **1-10**, derived from the bond lengths and angles of the single crystal structures recorded at 100 K according to the equations given in the text. Highest and lowest values for each measure are shown in green and red, respectively.

| Compound | $\sigma^2$ | Δd (x10$^{-4}$) | D |
|---|---|---|---|
| $[CMA]_4[Sb_2Cl_{10}]$ (**1**) | 23.54 | 115.1 | **0.308** |
| $[CMA]_4[Bi_2Cl_{10}]$ (**2**) | 16.04 | 29.5 | 0.155 |
| $[CMA]_3[SbBr_5]Br$ (**6**) | 8.77 | 22.7 | 0.134 |
| $[CMA]_3[BiBr_5]Br$ (**7**) | 12.70 | 12.6 | 0.103 |
| $[BZA]_6[Sb_2Cl_{10}]Cl_2$ (**3**) | **35.80** | **119.8** | 0.303 |
| $[BZA]_4[Bi_2Cl_{10}]$ (**4**) | 8.18 | 24.6 | 0.110 |
| $[BZA]_3[BiCl_5]Cl$ (**5**) | 12.18 | 12.7 | 0.089 |
| $[BZA]_2[SbBr_5]$ (**8**) | **7.53** | 42.2 | 0.164 |
| $[BZA]_3[SbBr_6]$ (**9**) | 14.60 | 7.1 | 0.073 |
| $[BZA]_3[BiBr_6]$ (**10**) | 14.57 | **2.1** | **0.040** |

**Self-trapped exciton.** We first examine the fundamental optical properties of all compounds, focusing on the confirmation of their STE emission characteristics. We then analyze how variations in the metal, halide, organic cation, and anion connectivity influence the STE emission. We present steady-state photoluminescence (PL) and photoluminescence excitation (PLE) spectra of compound **4** (Figure 4a) as a representative example, with additional optical data provided in the SI. Compound **4** exhibits a bright, broadband emission across the visible spectrum. The PL peak centers at 1.83 eV (677 nm) with a large full width at half maximum (FWHM) of 0.63 eV (240 nm) making it a promising white-light-emitting material. PLE of the white-light emission allows us to determine the Stokes shift of the PL and confirm the absence of deep defects within the band gap of the material. The PLE spectrum reveals a dominant peak at 3.29 eV, which we attribute to the band gap. This results in a Stokes shift of 1.44 eV (293 nm), which is among the highest reported for white-light-emitting materials.[75] Importantly, the PLE intensity drops to nearly zero for excitation energies below the band-gap energy, and no PLE intensity is observed near the PL energy. This indicates the absence of deep defects or a negligible density of states within the band gap. Consequently, these results confirm that the material's emission arises from STEs. We further quantify the strength of the electron-phonon coupling by determining the Huang-Rhys factor $S$ from temperature-dependent PL measurements, analyzed within the framework of a configuration coordinate model (CCM). The PL FWHM shows a non-linear increase from 0.630 to 0.635 eV as the sample temperature increases from 4 to 300 K (Figure 4b). The CCM analysis yields a Huang-Rhys factor $S = 4.9$, indicating a medium-to-strong electron-phonon coupling, with an excited-state phonon energy of $\hbar\Omega_e = 120$ meV. Additional data is provided in the SI.

**Absorption.** We measured the absorption spectra of all compounds using micro-reflectance spectroscopy and determined the corresponding band-gap energies through Tauc analyses assuming direct allowed transitions (Figures S44). The absorption spectra extend from the blue region of the visible spectrum to the near-ultraviolet (Figure 4c), with band-gap energies ranging from 2.60 eV (480 nm) in $[CMA]_3[SbBr_5]Br$ (**6**) to 3.51 eV (353 nm) in $[CMA]_4[Sb_2Cl_{10}]$ (**1**). Table 2 summarizes the band-gap energies of all compounds. It reveals three robust trends: First, the halogen substitution of Cl for Br leads to higher band-gap energies shifting the absorption from the blue to the near-ultraviolet region (compounds **1**–**5** vs. **6**–**10**). Second, the effect of metal substitution is motif dependent. In the CMA chloride, edge-sharing compound pair, the optical gap is larger for Sb than for Bi (**1**= = 3.51 eV vs **2** = 3.46 eV). In contrast, in the chain-like arrangement of the isostructural compound pair **6**/**7**, the Bi analogue exhibits the larger optical gap. Third, the substitution of BZA for CMA of the organic cation the compound pair **2** and **4** slightly decreases the band gap, indicating a comparatively minor role of the organic cation in setting the absorption edge.

These observations can be understood from the electronic structures from density functional theory (DFT) (Figure 5). Compound **1** yields a band-gap of 3.72 eV, which is in good agreement with the optical gap. The density-of-states confirms that the band-edge states are confined to the $[Sb_2Cl_{10}]^{2-}$ anions, and mainly contributed by Sb *5p* electron orbitals at the conduction band minimum and Cl *3p* electron orbitals at the valence band maximum. The band structure (Figure 5e) exhibits very narrow bandwidth, and an indirect gap located between the Γ-point (VBM) and the Z-point (CBM). However, the direct Γ–Γ transition is higher by only 0.7 meV, which makes it energetically accessible. The flat bands indicate that the electronic states are highly localized. These results are qualitatively in excellent agreement with earlier results on compound **5**.[43] This explains our observations: As the bands near the band gap are dominated by the metal-halide-anion electron orbitals, substitutions in the anion strongly influence the absorption, while changes in the organic cation have a smaller effect. In isostructural metal pairs, the gap response is motif-dependent. Within our computational level, substituting Bi for Sb tends to widen the fundamental gap by primarily perturbing the p-derived CBM, which is consistent with the chain-like compounds of **6** and **7**, where the Bi analogue exhibits the larger gap. In the CMA chloride, edge-shar-ing pair **1**/**2**, the measured optical gaps invert, which can be understood because absorption probes the optical gap, rather than the fundamental one. The fundamental band gap is modified by the finite-temperature electron-phonon renormalization energy and the exciton binding energy. If either or both exceed 50 meV, the optical ordering can reverse even when the fundamental Bi-Sb trend favors a larger gap for Bi. Halide substitution (Cl vs Br) follows the expected direction as well: deeper Cl *p* levels and stronger metal-halide bonding lower the VBM and widen the gap relative to Br, which our calculations reproduce across the series.

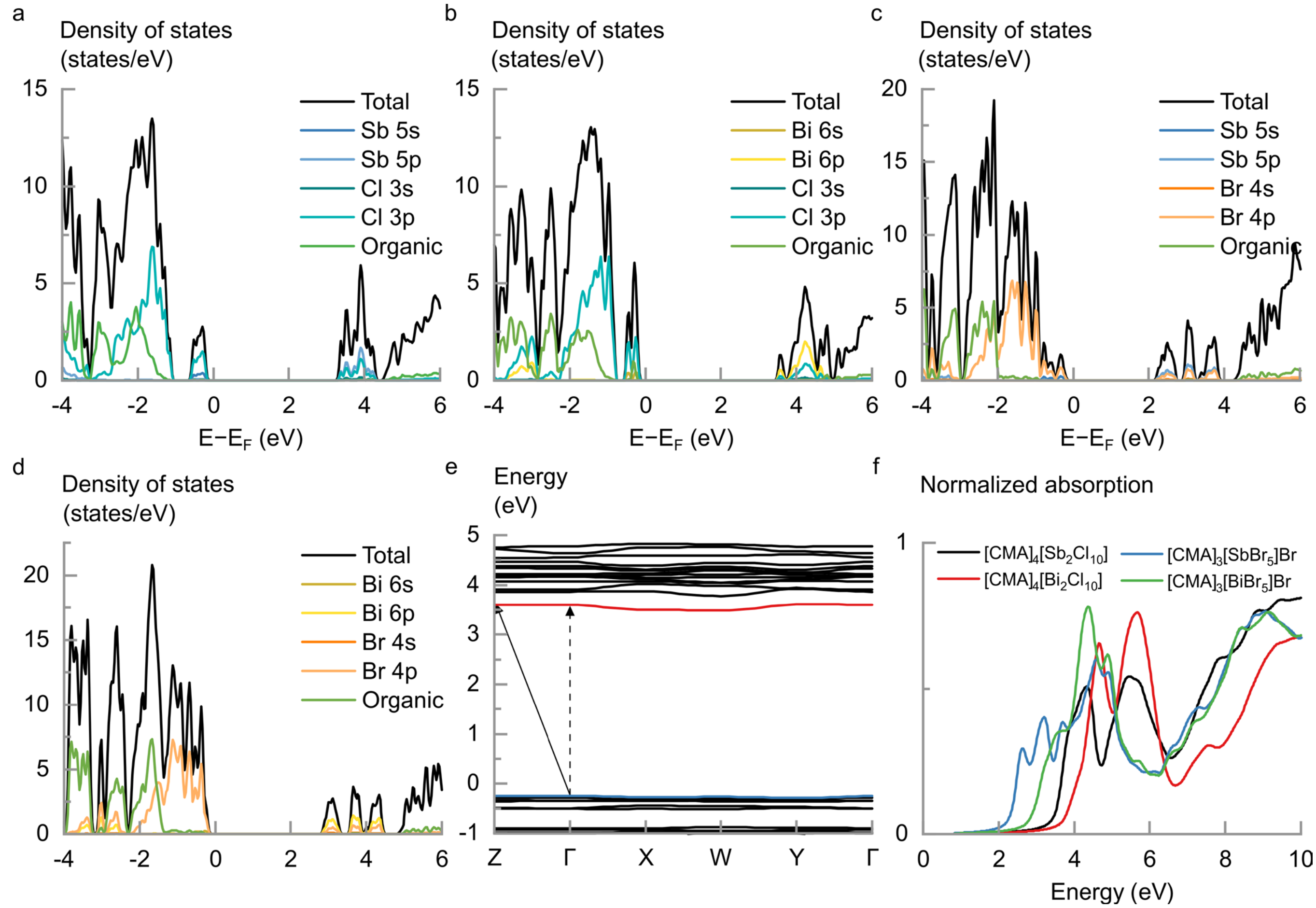

**Figure 5**: Calculated density of states (DOS), band structure, and absorption spectra of compound **1**, **2**, **6**, **7**. a) DOS of compound **1**. b) DOS of compound **2**. c) DOS of compound **6**. d) DOS of compound **7**. e) Band structure of compound **2**. f) Absorption spectra of compounds **1**, **2**, **6**, **7**.

**Photoluminescence.** Steady-state PL measurements confirm broad emission from STEs in 8 out of 10 compounds (Figure 4c). PL is observed in compounds **1**, **3-7**, **9**, and **10**; **2** shows no detectable PL. Compound **8** exhibits PL only at excitation powers that induce irreversible changes and is excluded from quantitative comparisons. Generally, the PL spectra exhibit a Gaussian shape, centered between 1.77 and 2.02 eV, with large FWHMs of 0.42 to 0.63 eV. These broad FWHMs enable the PL to cover a significant part of the visible spectrum, making all emissive compounds potential candidates for broadband emission materials, suitable as phosphors for solid-state lighting applications. Table 2 summarizes the PL properties of all emissive compounds, highlighting several key observations: First, by comparing halogen substitution of Cl for Br across connectivity-matched families, chlorides exhibit higher PL peak energies and modestly larger FWHMs, which is consistent with their larger optical gaps from absorption. The mean PL energy of all Br-compounds is 1.88 eV compared to 1.93 eV of all Cl-compounds. Similarly, the mean FWHM of all Br-compounds is 0.43 eV compared to 0.52 eV of all Cl-compounds. These observations can be exemplified by comparing $[BZA]_3[BiCl_5]Cl$ (**5**) and $[CMA]_3[BiBr_5]Br$ (**7**), which feature the same anion motif: The PL energy increases from 2.77 to 3.30 eV upon the halogen substitution of Cl for Br, while the FWHM increases slightly from 0.54 to 0.56 eV. Further, the Cl-containing compounds **1** and **3-5** exhibit bright photoluminescence emission compared to the Br-containing compounds, which are rather dim emitters (Figure 4d). Second, the influence of the metal is motif-dependent. The CMA chloride edge-sharing pair, $[CMA]_4[Sb_2Cl_{10}]$ (**1**) is strongly emissive whereas $[CMA]_4[Bi_2Cl_{10}]$ (**2**) is non-emissive under identical, non-destructive conditions; in the CMA bromide chains (**6**,**7**) both components are weak emitters but the Bi analogue (**7**) shows a higher PL energy and a broader FWHM; in the BZA bromides (**9**/**10**) both emit weakly with the Bi analogue only marginally stronger. Third, while cation effects are generally secondary across the series, a clean cation-only comparison exists in the Bi–chloride edge-sharing pair (**2**/**4**): $[BZA]_4[Bi_2Cl_{10}]$ (**4**) is bright whereas $[CMA]_4[Bi_2Cl_{10}]$ (**2**) is dark under identical, non-destructive conditions. This indicates that, within a fixed inorganic motif, the organic cation can tune lattice dynamics sufficiently to switch emissivity on/off (see Electron–phonon coupling and Table 2), even though halide and connectivity remain the primary drivers across the full dataset. Interestingly, the PL energies of all emissive compounds are confined to a relatively narrow range of 0.25 eV, while the corresponding band-gap energies span a large range of 0.86 eV (Figure 4e). This is also evident in the associated Stokes shifts, which are rather large and range from 0.8 eV in $[CMA]_3[SbBr_5]Br$ (**6**) to 1.51 eV in $[CMA]_4[Sb_2Cl_{10}]$ (**1**), with the latter being among the highest reported Stokes shifts in 1D.[76] Changes in band-gap energy appear to be offset by corresponding changes in Stokes shift, maintaining a narrow range of STE energies. This behavior arises from the nature of the STEs. The Stokes shift is governed by octahedral distortions in both the ground and excited states, along with the structural rearrangement between them.[14,77,78] In Cl-compounds, a more compact structure leads to greater structural rearrangement in the excited state compared to Br-compounds, resulting in larger Stokes shifts.[14] At the same time, Cl-compounds have larger band gaps than Br-compounds, reinforcing the narrow STE energy range. This trend is further supported by literature-reported 0D and 1D Sb and Bi chlorides and bromides, which align with our data (Figure S47). Notably, such a correlation is only evident due to the extensive dataset presented in this study.

**Electron-phonon coupling.** We further quantify the strength of the electron-phonon coupling in all emissive compounds by determining the Huang-Rhys factor *S* from temperature-dependent PL measurements, analyzed using a CCM. All emissive compounds exhibit medium-to-strong electron-phonon coupling, with *S* ranging from 4.9 in $[BZA]_4[Bi_2Cl_{10}]$ (**4**) to 22 in $[CMA]_3[SbBr_5]Br$ (**6**) (Table 2). The excited-state phonon energy $\hbar\Omega_e$ ranges from 38 meV in **6** to 120 meV in **4**. Detailed temperature-dependent data are provided in the SI (Figure S46). Interestingly, while previous studies proposed *S* values in the range 10–30 as optimal for efficient self-trapping in hybrid metal-halide materials,[79] compounds **1, 3 and 4** exhibit bright photoluminescence with *S* values 4.9–6.8. This suggests that the range for efficient self-trapping may be shifted in these hybrid metal-halide materials.

Several important observations arise when we compare *S* and $\hbar\Omega_e$ across different structural factors: First, the halogen substitution of Cl for Br generally reduces *S* and increases $\hbar\Omega_e$, aligned with enhanced PL. Specifically, the mean *S* and $\hbar\Omega_e$ of all Br-compounds are 12 and 60 meV, compared to 6 and 90 meV of all Cl-compounds. This trend can be attributed to the inverse relationship between the phonon energy and atom mass, where lighter Cl leads to higher phonon energies. Second, in the isostructural $[CMA]_3[SbBr_5]Br$ (**6**) and $[CMA]_3[BiBr_5]Br$ (**7**), substituting Bi for Sb decreases *S* from 22 to 13 and increases $\hbar\Omega_e$ from 38 to 71 meV. In this case, the above mentioned inverse relationship between phonon energy and atomic mass does not hold. A direct cation-only control appears in the Bi-chloride edge-sharing pair (**2**/**4**): **4** is bright with low *S* and high $E_{Ph}$, whereas **2** is non-emissive (no *S* and $E_{Ph}$ extractable). Across chloride emitters, both CMA and BZA land in the moderate *S* (5-7) and higher $E_{Ph}$ (80-120 meV) regime associated with robust STE PL. Across bromides, both cations correspond to higher *S* and lower $E_{Ph}$, consistent with weak emission. Thus, the cation can steer lattice dynamics within a fixed motif, but the halide and connectivity choice sets the dominant baseline.

## DISCUSSION

Our finding suggest several structural and electronic features to favor efficient STE-based emission: the presence of $s^2$ ions, particularly $5s^2$, large ground-state distortions, light halogens, 0D anion motifs, extended M…M distances, low coordination numbers, and intermediate electron-phonon coupling. Such coupling must be strong enough to enable the self-trapping process but not so excessive as to promote non-radiative recombination. Overall, our findings reveal that the electron-phonon coupling strength and excited-state phonon energy in hybrid metal-halide materials are significantly influenced by structural factors such as halogen substitution, metal choice, organic cation variation, and anion connectivity. Halogen substitution of Cl for Br consistently reduces the Huang-Rhys factor $S$ while increasing the phonon energy, which correlates with enhanced PL. Similarly, Bi substitution for Sb and changes in the organic cation induce comparable trends, highlighting the complex interplay between structural elements and optical properties. Notably, the edge-sharing octahedral anion motif emerges as a key structural feature for achieving strong PL, driven by its tendency to yield lower $S$ and higher $\hbar\Omega_e$. These insights establish a clearer understanding of the structure-property relationships governing efficient emission in hybrid metal-halide materials and provide guidance for the rational design of compounds with tailored optoelectronic properties.

Many of these factors are interdependent. For example, chlorido antimonates often exhibit pronounced ground-state distortions, while 0D anion motifs typically result in large M…M distances. Despite these apparent correlations, exceptions challenge the universality of these guidelines. Non-luminescent or low photoluminescence quantum yield (PLQY) compounds exist despite fulfilling the proposed criteria, as seen in $[CMA]_4[Bi_2Cl_{10}]$ (**2**). Conversely, materials with bright luminescence have been reported that do not conform to these guidelines. These discrepancies highlight the need for deeper mechanistic studies to refine the proposed design principles and disentangle the interwoven contributions of structural and electronic factors to STE emission.

**Influence of the metal atom.** Metal substitution tunes band edges and lattice dynamics, but does so not uniformly enhance emissivity. In CMA chloride, edge-sharing compounds (**1**/**2**), $[CMA]_4[Sb_2Cl_{10}]$ (**1**) is strongly emissive, whereas $[CMA]_4[Bi_2Cl_{10}]$ (**2**) is non-emissive under identical, non-destructive conditions. In CMA bromide chains (**6**/**7**), both compounds are weak emitters. The Bi analogue (**7**) shows a higher PL energy and a broader FWHM than **6**, consistent with the absorption blue-shift. Here, S decreases and EPh increases upon Sb→Bi, indicating that mode character and local force constants govern the effective phonon. In BZA bromides with isolated octahedra emission is weak, with Bi (**10**) only marginally stronger than Sb (**9**). The metal's impact is therefore filtered by halide identity and anion connectivity. Sb is favored in the chloride edge-sharing framework, whereas in bromide chains and isolated bromides neither metal yields strong emission and Bi primarily shifts the transition energy without improving brightness. We thus avoid global claims and treat Bi→Sb substitution as a motif-dependent knob; within a fixed motif, organic cation can further modulate this balance (**2** vs **4**).

**Influence of the halogen atom.** The halogen atom plays a pivotal role in shaping the optical properties of hybrid metal-halide materials. Substituting Cl for Br significantly impacts both absorption and PL. Cl substitution increases the band-gap energies, shifting absorption from the blue toward the near-ultraviolet region. This substitution also enhances PL, as evidenced by increased PL energies and broader FWHMs. Notably, chlorido halides proved to be significantly brighter emitters than bromido compounds, with the brightest emissive compounds **1**, **3-5** being chlorides, while most bromides are poor emitters. The halide choice greatly influences self-trapping depth and free exciton energy levels, affecting trapping and de-trapping processes in low-dimensional perovskites.[35,80,81] Furthermore, Cl substitution reduces $S$ and increases $\hbar\Omega e$, indicative of improved electron-phonon coupling. These findings underscore the central influence of halogen substitution in tuning electronic and optical properties, offering critical insights for designing advanced optoelectronic devices.

**Influence of organic cation.** A direct cation effect can be isolated in the Bi-chloride edge-sharing pair (**2**/**4**), where the inorganic framework is conserved and only the organic cation differs. Under identical excitation conditions $[BZA]_4[Bi_2Cl_{10}]$ (**4**) is bright, whereas $[CMA]_4[Bi_2Cl_{10}]$ (**2**) is dark. This contrast shows that, within a fixed motif, the cation can tune lattice dynamics sufficiently to cross the threshold for efficient STE emission. We attribute this to cation-controlled packing and hydrogen bonding, which stiffen the local framework and reduce the effective lattice displacement upon self-trapping. Beyond this control pair, cation-only conclusions are not warranted because cation changes co-vary with stoichiometry and anion packing (e.g., **1** vs **3** is not an isostructural comparison). Within this sample series, organic cations act as secondary, context-dependent tuners relative to halide and connectivity. Within Cl-edge-sharing motifs both CMA (e.g., **1**) and BZA (e.g., **4**) can access the favorable moderate Huang-Rhys values associated with strong STE PL, whereas in bromides (**6**/**7**, **9**/**10**) either cation is associated with higher $S$, lower $E_{Ph}$, and weak emission. Thus the organic cation's role is best viewed as a fine-tuning knob within a given motif, capable of toggling emissivity by steering the lattice into the optimal electron-phonon-coupling window.

**Table 2:** Summary of optical properties of all synthesized compounds, including band-gap energy ($E_g$), photoluminescence energy ($E_{PL}$) and full width at half maximum (FWHM), and corresponding Huang-Rhys factor (S) and phonon energies ($E_{Ph}$) from temperature-dependent photoluminescence measurements.

| Compound | | $E_{g,EXP}$ (eV) | $E_{g,DFT}$ (eV) | $E_{PL}$ (eV) | FWHM (eV) | $S$ | $E_{Ph}$ (meV) |
|---|---|---|---|---|---|---|---|
| $[CMA]_4[Sb_2Cl_{10}]$ | **1** | 3.51 | 3.72 | 1.94 | 0.43 | 6.8 | 79 |
| $[CMA]_4[Bi_2Cl_{10}]$ | **2** | 3.46 | 3.56 | | | | |
| $[BZA]_6[Sb_2Cl_{10}]Cl_2$ | **3** | 3.00 | - | 2.02 | 0.44 | 5.2 | 96 |
| $[BZA]_4[Bi_2Cl_{10}]$ | **4** | 3.21 | - | 1.83 | 0.63 | 4.9 | 120 |
| $[BZA]_3[BiCl_5]Cl$ [43] | **5** | 3.30 | - | 1.91 | 0.56 | 6.9 | 87 |
| $[CMA]_3[SbBr_5]Br$ | **6** | 2.60 | 2.46 | 1.77 | 0.42 | 22 | 38 |
| $[CMA]_3[BiBr_5]Br$ | **7** | 2.77 | 3.12 | 1.94 | 0.54 | 13 | 71 |
| $[BZA]_2[SbBr_5]$ | **8** | 3.05 | - | | | - | - |
| $[BZA]_3[SbBr_6]$ | **9** | 2.98 | - | 2.00 | 0.49 | 25 | 42 |
| $[BZA]_3[BiBr_6]$ | **10** | 2.86 | - | 1.79 | 0.41 | 7.7 | 65 |

## CONCLUSION

In summary, combining absorption, steady-state PL, and temperature-dependent PL across ten Sb/Bi halides yields a consistent, motif-aware picture of STE emission. Substitution is motif-dependent. While Bi→Sb substitution generally widens the band gap and can blue-shift PL, it does not uniformly enhance emissivity (CMA chloride edge-sharing: **1** is bright, **2** is dark; bromide chains **6**/**7** and isolated bromides **9**/**10** remain weak). Halide identity is the primary lever: Chlorides consistently outperform bromides, showing higher PL intensities and slightly broader emission bands. Connectivity sets the stage: Edge-sharing motifs consistently favor bright broadband STE PL, whereas chain-like and isolated connectivities largely do not. Organic cations modulate but rarely dominate; they can be decisive within a fixed motif as demonstrated in the isostructural Bi/Cl edge-sharing pair (**2**/**4**), where the cation toggles emissivity. Despite ~0.86 eV variation in optical gaps, we observed that PL energies cluster within ~0.25 eV because large Stokes shifts (up to 1.51 eV in **1**) compensate band-gap changes.

Despite these trends, exceptions highlight the nuanced dependencies of STE-based emission on material-specific factors, requiring deeper mechanistic insights. The findings provide a framework for designing high-performance hybrid halides, leveraging the synergistic effects of halogen substitution, metal atom variation, organic cation modulation, and anion connectivity. This work establishes guiding principles for tailoring optoelectronic properties, paving the way for advanced materials with enhanced white-light emission for next-generation optoelectronic applications.

## ASSOCIATED CONTENT

**Supporting Information**.

Additional synthetic and crystallographic details, powder patterns, IR spectra, thermal analysis data, UV-Vis spectra and Tauc plots, details on the PL setup and additional PL and computational data. This material is available free of charge via the Internet at http://pubs.acs.org.

**Accession Codes**

CCDC 2479960 (**1**), 2479964 (**2**), 2479965 (**3**), 2479962 (**4**), 2479961 (**6**) and 2479963 (**7**) contain the supplementary crystallographic data for this paper. This data can be obtained free of charge via www.ccdc.cam.ac.uk/data_request/cif, or by emailing data_request@ccdc.cam.ac.uk, or by contacting The Cambridge Crystallographic Data Centre, 12 Union Road, Cambridge CB2 1EZ, UK; fax: +44 1223 336033.

## AUTHOR INFORMATION

### Corresponding Authors

* E-Mail: johanna.heine@uni-oldenburg.de

### Author Contributions

The manuscript was written through contributions of all authors. All authors have given approval to the final version of the manuscript.

### Notes

The authors declare no competing financial interests.

## ACKNOWLEDGMENT

The authors acknowledge financial support provided by the German Science Foundation (Deutsche Forschungsgemeinschaft, DFG), via project-ID 223848855-SFB 1083 “Structure and Dynamics of Internal Interfaces” within the project A15 and B2 and via the Heisenberg Programme (505757318) and contract (531281367)

### References

(1) Chouhan, L.; Ghimire, S.; Subrahmanyam, C.; Miyasaka, T.; Biju, V. Synthesis, optoelectronic properties and applications of halide perovskites. *Chem. Soc. Rev.* **2020**, *49* (10), 2869–2885. DOI: 10.1039/c9cs00848a.

(2) Kovalenko, M. V.; Protesescu, L.; Bodnarchuk, M. I. Properties and potential optoelectronic applications of lead halide perovskite nanocrystals. *Science* **2017**, *358* (6364), 745–750. DOI: 10.1126/science.aam7093.

(3) Dou, L.; Wong, A. B.; Yu, Y.; Lai, M.; Kornienko, N.; Eaton, S. W.; Fu, A.; Bischak, C. G.; Ma, J.; Ding, T.; Ginsberg, N. S.; Wang, L.-W.; Alivisatos, A. P.; Yang, P. Atomically thin two-dimensional organic-inorganic hybrid perovskites. *Science* **2015**, *349* (6255), 1518–1521. DOI: 10.1126/science.aac7660.

(4) Li, X.; Hoffman, J. M.; Kanatzidis, M. G. The 2D Halide Perovskite Rulebook: How the Spacer Influences Everything from the Structure to Optoelectronic Device Efficiency. *Chem. Rev.* **2021**, *121* (4), 2230–2291. DOI: 10.1021/acs.chemrev.0c01006.

(5) Straus, D. B.; Kagan, C. R. Electrons, Excitons, and Phonons in Two-Dimensional Hybrid Perovskites: Connecting Structural, Optical, and Electronic Properties. *J. Phys. Chem. Lett.* **2018**, *9* (6), 1434–1447. DOI: 10.1021/acs.jpclett.8b00201.

(6) Stoumpos, C. C.; Cao, D. H.; Clark, D. J.; Young, J.; Rondinelli, J. M.; Jang, J. I.; Hupp, J. T.; Kanatzidis, M. G. Ruddlesden–Popper Hybrid Lead Iodide Perovskite 2D Homologous Semiconductors. *Chem. Mater.* **2016**, *28* (8), 2852–2867. DOI: 10.1021/acs.chemmater.6b00847.

(7) Mitzi, D. B. Synthesis, Crystal Structure, and Optical and Thermal Properties of (C 4 H 9 NH 3 ) 2 MI 4 (M = Ge, Sn, Pb). *Chem. Mater.* **1996**, *8* (3), 791–800. DOI: 10.1021/cm9505097.

(8) Yang, S.; Niu, W.; Wang, A.-L.; Fan, Z.; Chen, B.; Tan, C.; Lu, Q.; Zhang, H. Ultrathin Two-Dimensional Organic-Inorganic Hybrid Perovskite Nanosheets with Bright, Tunable Photoluminescence and High Stability. *Angew. Chem. Int. Ed.* **2017**, *56* (15), 4252–4255. DOI: 10.1002/anie.201701134.

(9) Du, K.-Z.; Tu, Q.; Zhang, X.; Han, Q.; Liu, J.; Zauscher, S.; Mitzi, D. B. Two-Dimensional Lead(II) Halide-Based Hybrid Perovskites Templated by Acene Alkylamines: Crystal Structures, Optical Properties, and Piezoelectricity. *Inorg. Chem.* **2017**, *56* (15), 9291–9302. DOI: 10.1021/acs.inorgchem.7b01094.

(10) Dohner, E. R.; Hoke, E. T.; Karunadasa, H. I. Self-assembly of broadband white-light emitters. *J. Am. Chem. Soc.* **2014**, *136* (5), 1718–1721. DOI: 10.1021/ja411045r.

(11) Dohner, E. R.; Jaffe, A.; Bradshaw, L. R.; Karunadasa, H. I. Intrinsic white-light emission from layered hybrid perovskites. *J. Am. Chem. Soc.* **2014**, *136* (38), 13154–13157. DOI: 10.1021/ja507086b.

(12) Luo, J.; Wang, X.; Li, S.; Liu, J.; Guo, Y.; Niu, G.; Yao, L.; Fu, Y.; Gao, L.; Dong, Q.; Zhao, C.; Leng, M.; Ma, F.; Liang, W.; Wang, L.; Jin, S.; Han, J.; Zhang, L.; Etheridge, J.; Wang, J.; Yan, Y.; Sargent, E. H.; Tang, J. Efficient and stable emission of warm-white light from lead-free halide double perovskites. *Nature* **2018**, *563* (7732), 541–545. DOI: 10.1038/s41586-018-0691-0.

(13) Yuan, Z.; Zhou, C.; Tian, Y.; Shu, Y.; Messier, J.; Wang, J. C.; van de Burgt, L. J.; Kountouriotis, K.; Xin, Y.; Holt, E.; Schanze, K.; Clark, R.; Siegrist, T.; Ma, B. One-dimensional organic lead halide perovskites with efficient bluish white-light emission. *Nat. Commun.* **2017**, *8*, 14051. DOI: 10.1038/ncomms14051.
(14) Morad, V.; Yakunin, S.; Kovalenko, M. V. Supramolecular Approach for Fine-Tuning of the Bright Luminescence from Zero-Dimensional Antimony(III) Halides. *ACS Mater. Lett.* **2020**, *2* (7), 845–852. DOI: 10.1021/acsmaterialslett.0c00174.
(15) Zhou, C.; Lin, H.; Tian, Y.; Yuan, Z.; Clark, R.; Chen, B.; van de Burgt, L. J.; Wang, J. C.; Zhou, Y.; Hanson, K.; Meisner, Q. J.; Neu, J.; Besara, T.; Siegrist, T.; Lambers, E.; Djurovich, P.; Ma, B. Luminescent zero-dimensional organic metal halide hybrids with near-unity quantum efficiency. *Chem. Sci.* **2018**, *9* (3), 586–593. DOI: 10.1039/c7sc04539e.
(16) Zhou, C.; Lin, H.; Worku, M.; Neu, J.; Zhou, Y.; Tian, Y.; Lee, S.; Djurovich, P.; Siegrist, T.; Ma, B. Blue Emitting Single Crystalline Assembly of Metal Halide Clusters. *J. Am. Chem. Soc.* **2018**, *140* (41), 13181–13184. DOI: 10.1021/jacs.8b07731.
(17) Cortecchia, D.; Yin, J.; Petrozza, A.; Soci, C. White light emission in low-dimensional perovskites. *J. Mater. Chem. C* **2019**, *7* (17), 4956–4969. DOI: 10.1039/C9TC01036J.
(18) Li, M.; Xia, Z. Recent progress of zero-dimensional luminescent metal halides. *Chem. Soc. Rev.* **2021**, *50* (4), 2626–2662. DOI: 10.1039/d0cs00779j.
(19) Kundu, J.; Das, D. K. Low Dimensional, Broadband, Luminescent Organicganic halides. des. ers. ters. lusters. Assembly of Me *Eur. J. Inorg. Chem.* **2021**, *2021* (44), 4508–4520. DOI: 10.1002/ejic.202100685.
(20) Guo, Q.; Zhao, X.; Song, B.; Luo, J.; Tang, J. Light Emission of Self-Trapped Excitons in Inorganic Metal Halides for Optoelectronic Applications. *Adv. Mater.* **2022**, e2201008. DOI: 10.1002/adma.202201008.
(21) Mo, Q.; Shi, Y.; Cai, W.; Zhao, S.; Ying, Y.; Zang, Z. Opportunities and challenges of low-dimensional hybrid metal halides in white light-emitting diodes. *J. Phys. D: Appl. Phys.* **2022**, *55* (33), 333003. DOI: 10.1088/1361-6463/ac7264.
(22) Morad, V.; Shynkarenko, Y.; Yakunin, S.; Brumberg, A.; Schaller, R. D.; Kovalenko, M. V. Disphenoidal Zero-Dimensional Lead, Tin, and Germanium Halides: Highly Emissive Singlet and Triplet Self-Trapped Excitons and X-ray Scintillation. *J. Am. Chem. Soc.* **2019**, *141* (25), 9764–9768. DOI: 10.1021/jacs.9b02365.
(23) Zdražil, L.; Kalytchuk, S.; Langer, M.; Ahmad, R.; Pospíšil, J.; Zmeškal, O.; Altomare, M.; Osvet, A.; Zbořil, R.; Schmuki, P.; Brabec, C. J.; Otyepka, M.; Kment, Š. Transparent and Low-Loss Luminescent Solar Concentrators Based on Self-Trapped Exciton Emission in Lead-Free Double Perovskite Nanocrystals. *ACS Appl. Energy Mater.* **2021**, *4* (7), 6445–6453. DOI: 10.1021/acsaem.1c00360.
(24) Morad, V.; Yakunin, S.; Benin, B. M.; Shynkarenko, Y.; Grotevent, M. J.; Shorubalko, I.; Boehme, S. C.; Kovalenko, M. V. Hybrid 0D Antimony Halides as Air-Stable Luminophores for High-Spatial-Resolution Remote Thermography. *Adv. Mater.* **2021**, *33* (9), e2007355. DOI: 10.1002/adma.202007355.
(25) Yakunin, S.; Benin, B. M.; Shynkarenko, Y.; Nazarenko, O.; Bodnarchuk, M. I.; Dirin, D. N.; Hofer, C.; Cattaneo, S.; Kovalenko, M. V. High-resolution remote thermometry and thermography using luminescent low-dimensional tin-halide perovskites. *Nat. Mater.* **2019**, *18* (8), 846–852. DOI: 10.1038/s41563-019-0416-2.
(26) Luo, J.-B.; Wei, J.-H.; Zhang, Z.-Z.; Kuang, D.-B. Water-Molecule-Induced Emission Transformation of Zero-Dimension Antimony-Based Metal Halide. *Inorg. Chem.* **2022**, *61* (1), 338–345. DOI: 10.1021/acs.inorgchem.1c02871.
(27) Williams, R. T.; Song, K. S. The self-trapped exciton. *J. Phys. Chem. Solids.* **1990**, *51* (7), 679–716. DOI: 10.1016/0022-3697(90)90144-5.
(28) Peng, H.; Zou, B. Effects of Electron-Phonon Coupling and Spin-Spin Coupling on the Photoluminescence of Low-Dimensional Metal Halides. *J. Phys. Chem. Lett.* **2022**, *13* (7), 1752–1764. DOI: 10.1021/acs.jpclett.1c03849.
(29) Debever, J. M.; Bonnot, A.; Bonnot, A. M.; Coletti, F.; Hanus, J. Excitonic luminescence spectrum of electron excited solid xenon. *Solid State Commun.* **1974**, *14* (10), 989–992. DOI: 10.1016/0038-1098(74)90409-8.
(30) Kabler, M. N. Low-Temperature Recombination Luminescence in Alkali Halide Crystals. *Phys. Rev.* **1964**, *136* (5A), A1296-A1302. DOI: 10.1103/PhysRev.136.A1296.
(31) Matsui, A.; Iemura, M.; Nishimura, H. Dynamical behavior of excitons in pyrene. *J. Lumin.* **1981**, *24-25*, 445–448. DOI: 10.1016/0022-2313(81)90309-4.
(32) Lee; Mysyrowicz; Nurmikko; Fitzpatrick. Exciton self-trapping in ZnSe-ZnTe alloys. *Phys. Rev. Lett.* **1987**, *58* (14), 1475–1478. DOI: 10.1103/PhysRevLett.58.1475.
(33) Smith, M. D.; Jaffe, A.; Dohner, E. R.; Lindenberg, A. M.; Karunadasa, H. I. Structural origins of broadband emission from layered Pb-Br hybrid perovskites. *Chem. Sci.* **2017**, *8* (6), 4497–4504. DOI: 10.1039/c7sc01590a.
(34) Smith, M. D.; Karunadasa, H. I. White-Light Emission from Layered Halide Perovskites. *Acc. Chem. Res.* **2018**, *51* (3), 619–627. DOI: 10.1021/acs.accounts.7b00433.
(35) Mao, L.; Wu, Y.; Stoumpos, C. C.; Traore, B.; Katan, C.; Even, J.; Wasielewski, M. R.; Kanatzidis, M. G. Tunable White-Light Emission in Single-Cation-Templated Three-Layered 2D Perovskites (CH3CH2NH3)4Pb3Br10-xClx. *J. Am. Chem. Soc.* **2017**, *139* (34), 11956–11963. DOI: 10.1021/jacs.7b06143.
(36) Zhou, C.; Worku, M.; Neu, J.; Lin, H.; Tian, Y.; Lee, S.; Zhou, Y.; Han, D.; Chen, S.; Hao, A.; Djurovich, P. I.; Siegrist, T.; Du, M.-H.; Ma, B. Facile Preparation of Light Emitting Organic Metal Halide Crystals with Near-Unity Quantum Efficiency. *Chem. Mater.* **2018**, *30* (7), 2374–2378. DOI: 10.1021/acs.chemmater.8b00129.
(37) Benin, B. M.; Dirin, D. N.; Morad, V.; Wörle, M.; Yakunin, S.; Rainò, G.; Nazarenko, O.; Fischer, M.; Infante, I.; Kovalenko, M. V. Highly Emissive Self-Trapped Excitons in Fully Inorganic Zero-Dimensional Tin Halides. *Angew. Chem. Int. Ed.* **2018**, *57* (35), 11329–11333. DOI: 10.1002/anie.201806452.
(38) McCall, K. M.; Morad, V.; Benin, B. M.; Kovalenko, M. V. Efficient Lone-Pair-Driven Luminescence: Structure-Property Relationships in Emissive 5s2 Metal Halides. *ACS Mater. Lett.* **2020**, *2* (9), 1218–1232. DOI: 10.1021/acsmaterialslett.0c00211.
(39) Zhou, C.; Xu, L.-J.; Lee, S.; Lin, H.; Ma, B. Recent Advances in Luminescent Zero‐Dimensional Organic Metal Halide Hybrids. *Adv. Opt. Mater.* **2021**, *9* (18), 2001766. DOI: 10.1002/adom.202001766.
(40) Zhou, L.; Liao, J.-F.; Kuang, D.-B. An Overview for Zero‐Dimensional Broadband Emissive Metal‐Halide Single Crystals. *Adv. Opt. Mater.* **2021**, *9* (17), 2100544. DOI: 10.1002/adom.202100544.
(41) Molokeev, M. S.; Su, B.; Aleksandrovsky, A. S.; Golovnev, N. N.; Plyaskin, M. E.; Xia, Z. Machine Learning Analysis and Discovery of Zero-Dimensional ns 2 Metal Halides toward Enhanced Photoluminescence Quantum Yield. *Chem. Mater.* **2022**, *34* (2), 537–546. DOI: 10.1021/acs.chemmater.1c02725.
(42) Hu, X.; Zhu, Y.; Wang, J.; Zheng, G.; Yao, D.; Lin, B.; Tian, N.; Zhou, B.; Long, F. Stable organic-inorganic hybrid bismuth-halide:

Exploration of crystal-structural, morphological, thermal, spectroscopic and optoelectronic properties. *J. Mol. Struct.* **2022**, *1264*, 133102. DOI: 10.1016/j.molstruc.2022.133102.
(43) Klement, P.; Dehnhardt, N.; Dong, C.-D.; Dobener, F.; Bayliff, S.; Winkler, J.; Hofmann, D. M.; Klar, P. J.; Schumacher, S.; Chatterjee, S.; Heine, J. Atomically Thin Sheets of LeadLead, J. Atomically Thin Sheets of Leadof Leadf Leadof Leadtion of crystal-structural, morph *Adv. Mater.* **2021**, *33* (23), 2100518. DOI: 10.1002/adma.202100518.
(44) Anyfantis, G. C.; Ganotopoulos, N.-M.; Savvidou, A.; Raptopoulou, C. P.; Psycharis, V.; Mousdis, G. A. Synthesis and characterization of new organic–inorganic hybrid compounds based on Sb, with a perovskite like structure. *Polyhedron* **2018**, *151*, 299–305. DOI: 10.1016/j.poly.2018.05.024.
(45) Da Chen; Dai, F.; Hao, S.; Zhou, G.; Liu, Q.; Wolverton, C.; Zhao, J.; Xia, Z. Crystal structure and luminescence properties of lead-free metal halides (C 6 H 5 CH 2 NH 3 ) 3 MBr 6 (M = Bi and Sb). *J. Mater. Chem. C* **2020**, *8* (22), 7322–7329. DOI: 10.1039/D0TC00562B.
(46) Sheldrick, G. M. SHELXT - integrated space-group and crystal-structure determination. *Acta Crystallogr., Sect. A: Found. Adv.* **2015**, *71* (Pt 1), 3–8. DOI: 10.1107/S2053273314026370.
(47) Sheldrick, G. M. A short history of SHELX. *Acta Crystallogr., Sect. A: Found. Crystallogr.* **2008**, *64* (Pt 1), 112–122. DOI: 10.1107/S0108767307043930.
(48) Sheldrick, G. M. Crystal structure refinement with SHELXL. *Acta Crystallogr., Sect. C: Struct. Chem.* **2015**, *71* (Pt 1), 3–8. DOI: 10.1107/S2053229614024218.
(49) Dolomanov, O. V.; Bourhis, L. J.; Gildea, R. J.; Howard, J. A. K.; Puschmann, H. OLEX2 : a complete structure solution, refinement and analysis program. *J. Appl. Crystallogr.* **2009**, *42* (2), 339–341. DOI: 10.1107/S0021889808042726.
(50) Brandenburg, K. *Diamond*; Crystal Impact GbR, 2005.
(51) Groom, C. R.; Bruno, I. J.; Lightfoot, M. P.; Ward, S. C. The Cambridge Structural Database. *Acta Crystallogr., Sect. B: Struct. Sci., Cryst. Eng. Mater.* **2016**, *72* (Pt 2), 171–179. DOI: 10.1107/S2052520616003954.
(52) Kresse, G.; Hafner, J. Norm-conserving and ultrasoft pseudopotentials for first-row and transition elements. *J. Phys.: Condens. Matter* **1994**, *6* (40), 8245–8257. DOI: 10.1088/0953-8984/6/40/015.
(53) Blöchl, P. E. Projector augmented-wave method. *Phys. Rev. B Condens. Matter* **1994**, *50* (24), 17953–17979. DOI: 10.1103/PhysRevB.50.17953.
(54) Perdew, J. P.; Burke, K.; Ernzerhof, M. Generalized Gradient Approximation Made Simple. *Phys. Rev. Lett.* **1996**, *77* (18), 3865–3868. DOI: 10.1103/PhysRevLett.77.3865.
(55) Mousdis, G. A.; Papavassiliou, G. C.; Terzis, A.; Raptopoulou, C. P. Notizen: Preparation, Structures and Optical Properties of [H 3 N(CH 2 ) 6 NH 3 ]BiX 5 (X=I, Cl) and [H 3 N(CH 2 ) 6 NH 3 ]SbX 5 (X=I, Br). *Z. Naturforsch. B* **1998**, *53* (8), 927–932. DOI: 10.1515/znb-1998-0825.
(56) Adonin, S. A.; Sokolov, M. N.; Fedin, V. P. Polynuclear halide complexes of Bi(III): From structural diversity to the new properties. *Coord. Chem. Rev.* **2016**, *312*, 1–21. DOI: 10.1016/j.ccr.2015.10.010.
(57) Bigoli, F.; Lanfranchi, M.; Pellinghelli, M. A. The structures of bis(1H+,5H+-S-methylisothiocarbonohydrazidium) di-µ-chloro-octachlorodibismuthate(III) tetrahydrate and tris(1H+-S-methylisothiocarbonohydrazidium) esachlorobismuthate(III). *Inorg. Chim. Acta* **1984**, *90* (3), 215–220. DOI: 10.1016/S0020-1693(00)80749-5.
(58) Du Bois, A.; Abriel, W. Notizen: Darstellung und Kristallstruktur von [H 3 N(CH 2 ) 3 NH 3 ]SbCl 5 / Preparation and Crystal Structure of [H 3 N(CH 2 ) 3 NH 3 ]SbCl 5. *Z. Naturforsch. B* **1989**, *44* (9), 1151–1154. DOI: 10.1515/znb-1989-0925.
(59) Lipka, A. Chloroantimonate(III): Die Kristallstruktur des 4,4?-Dipyridylium-pentachloroantimonats, (C10H8N2H2)SbCl5. *Z. Anorg. Allg. Chem.* **1980**, *469* (1), 229–233. DOI: 10.1002/zaac.19804690130.
(60) Sheu, H.-L.; Laane, J. Trans effect in halobismuthates and haloantimonates revisited. Molecular structures and vibrations from theoretical calculations. *Inorg. Chem.* **2013**, *52* (8), 4244–4249. DOI: 10.1021/ic302082a.
(61) Buikin, P. A.; Rudenko, A. Y.; Baranchikov, A. E.; Ilyukhin, A. B.; Kotov, V. Y. 1D-Bromobismuthates of Dipyridinoalkane Derivatives. *Russ. J. Coord. Chem.* **2018**, *44* (6), 373–379. DOI: 10.1134/S1070328418060015.
(62) Mitzi, D. B. Organic-inorganic perovskites containing trivalent metal halide layers: the templating influence of the organic cation layer. *Inorg. Chem.* **2000**, *39* (26), 6107–6113. DOI: 10.1021/ic000794i.
(63) Wojciechowska, M.; Gągor, A.; Piecha-Bisiorek, A.; Jakubas, R.; Ciżman, A.; Zaręba, J. K.; Nyk, M.; Zieliński, P.; Medycki, W.; Bil, A. Ferroelectricity and Ferroelasticity in Organic Inorganic Hybrid (Pyrrolidinium) 3 [Sb 2 Cl 9 ]. *Chem. Mater.* **2018**, *30* (14), 4597–4608. DOI: 10.1021/acs.chemmater.8b00962.
(64) Li, M.-Q.; Hu, Y.-Q.; Bi, L.-Y.; Zhang, H.-L.; Wang, Y.; Zheng, Y.-Z. Structure Tunable Organic–Inorganic Bismuth Halides for an Enhanced Two-Dimensional Lead-Free Light-Harvesting Material. *Chem. Mater.* **2017**, *29* (13), 5463–5467. DOI: 10.1021/acs.chemmater.7b01017.
(65) Dehnhardt, N.; Luy, J.-N.; Szabo, M.; Wende, M.; Tonner, R.; Heine, J. Synthesis of a two-dimensional organic-inorganic bismuth iodide metalate through in situ formation of iminium cations. *Chem. Commun.* **2019**, *55* (98), 14725–14728. DOI: 10.1039/c9cc06625j.
(66) Schmitz, F.; Horn, J.; Dengo, N.; Sedykh, A. E.; Becker, J.; Maiworm, E.; Bélteky, P.; Kukovecz, Á.; Gross, S.; Lamberti, F.; Müller-Buschbaum, K.; Schlettwein, D.; Meggiolaro, D.; Righetto, M.; Gatti, T. Large Cation Engineering in Two-Dimensional Silver–Bismuth Bromide Double Perovskites. *Chem. Mater.* **2021**, *33* (12), 4688–4700. DOI: 10.1021/acs.chemmater.1c01182.
(67) Gavezzotti, A. The calculation of molecular volumes and the use of volume analysis in the investigation of structured media and of solid-state organic reactivity. *J. Am. Chem. Soc.* **1983**, *105* (16), 5220–5225. DOI: 10.1021/ja00354a007.
(68) Nishio, M. CH/? hydrogen bonds in crystals. *CrystEngComm* **2004**, *6* (27), 130. DOI: 10.1039/b313104a.
(69) Robinson, K.; Gibbs, G. V.; Ribbe, P. H. Quadratic elongation: a quantitative measure of distortion in coordination polyhedra. *Science* **1971**, *172* (3983), 567–570. DOI: 10.1126/science.172.3983.567.
(70) Lufaso, M. W.; Woodward, P. M. Jahn-Teller distortions, cation ordering and octahedral tilting in perovskites. *Acta Crystallogr., Sect. B: Struct. Sci.* **2004**, *60* (Pt 1), 10–20. DOI: 10.1107/S0108768103026661.
(71) Lü, X.; Stoumpos, C.; Hu, Q.; Ma, X.; Zhang, D.; Guo, S.; Hoffman, J.; Bu, K.; Guo, X.; Wang, Y.; Ji, C.; Chen, H.; Xu, H.; Jia, Q.; Yang, W.; Kanatzidis, M. G.; Mao, H.-K. Regulating off-centering distortion maximizes photoluminescence in halide perovskites. *Natl. Sci. Rev.* **2021**, *8* (9), nwaa288. DOI: 10.1093/nsr/nwaa288.
(72) Peng, Y.-C.; Jin, J.-C.; Gu, Q.; Dong, Y.; Zhang, Z.-Z.; Zhuang, T.-H.; Gong, L.-K.; Ma, W.; Wang, Z.-P.; Du, K.-Z.; Huang, X.-Y. Se-

lective Luminescence Response of a Zero-Dimensional Hybrid Antimony(III) Halide to Solvent Molecules: Size-Effect and Supramolecular Interactions. *Inorg. Chem.* **2021**, *60* (23), 17837–17845. DOI: 10.1021/acs.inorgchem.1c02445.

(73) Li, M.; Lin, J.; Liu, K.; Fan, L.; Wang, N.; Guo, Z.; Yuan, W.; Zhao, J.; Liu, Q. Light-Emitting 0D Hybrid Metal Halide (C3H12N2)2Sb2Cl10 with Antimony Dimers. *Inorg. Chem.* **2021**, *60* (15), 11429–11434. DOI: 10.1021/acs.inorgchem.1c01440.

(74) Atanasov, M.; Reinen, D. Predictive concept for lone-pair distortions - DFT and vibronic model studies of AXn-(n-3) molecules and complexes (A = NIII to BiIII; X = F-I to I-I; n = 3-6). *J. Am. Chem. Soc.* **2002**, *124* (23), 6693–6705. DOI: 10.1021/ja012408h.

(75) Mohd Yusof Chan, N. N.; Idris, A.; Zainal Abidin, Z. H.; Tajuddin, H. A.; Abdullah, Z. White light employing luminescent engineered large (mega) Stokes shift molecules: a review. *RSC Adv.* **2021**, *11* (22), 13409–13445. DOI: 10.1039/D1RA00129A.

(76) Zhou, C.; Lin, H.; Shi, H.; Tian, Y.; Pak, C.; Shatruk, M.; Zhou, Y.; Djurovich, P.; Du, M.-H.; Ma, B. A Zero‐Dimensional Organic Seesaw‐Shaped Tin Bromide with Highly Efficient Strongly Stokes‐Shifted Deep‐Red Emission. *Angew. Chem. Int. Ed.* **2018**, *130* (4), 1033–1036. DOI: 10.1002/ange.201710383.

(77) Li, X.; Peng, C.; Xiao, Y.; Xue, D.; Luo, B.; Huang, X.-C. Guest-Induced Reversible Phase Transformation of Organic–Inorganic Phenylpiperazinium Antimony (III) Chlorides with Solvatochromic Photoluminescence. *J. Phys. Chem. C* **2021**, *125* (45), 25112–25118. DOI: 10.1021/acs.jpcc.1c07725.

(78) Biswas, A.; Bakthavatsalam, R.; Mali, B. P.; Bahadur, V.; Biswas, C.; Raavi, S. S. K.; Gonnade, R. G.; Kundu, J. The metal halide structure and the extent of distortion control the photo-physical properties of luminescent zero dimensional organic-antimony(iii ) halide hybrids. *J. Mater. Chem. C* **2021**, *9* (1), 348–358. DOI: 10.1039/D0TC03440A.

(79) Liu, S.; Wu, Y.; Wu, J.; Lin, Z. Blue Emission from Metal Halide Perovskites: Strategies and Applications. *ChemPhotoChem* **2024**, *8* (11). DOI: 10.1002/cptc.202400139.

(80) Li, J.; Wang, H.; Li, D. Self-trapped excitons in two-dimensional perovskites. *Front. Optoelectron.* **2020**, *13* (3), 225–234. DOI: 10.1007/s12200-020-1051-x.

(81) Gautier, R.; Paris, M.; Massuyeau, F. Exciton Self-Trapping in Hybrid Lead Halides: Role of Halogen. *J. Am. Chem. Soc.* **2019**, *141* (32), 12619–12623. DOI: 10.1021/jacs.9b04262.

**Supporting Information on**

# Enhanced White-Light Emission from Self-Trapped Excitons in Antimony and Bismuth Halides through Structural Design

Philip Klement[1], Lukas Gümbel[1], Meng Yang[2], Jan-Heinrich Littmann[1], Tatsuhiko Ohto[3], Hirokazu Tada[4], Sangam Chatterjee[1], Johanna Heine[5,*].

[1] Institute of Experimental Physics I and Center for Materials Research (ZfM), Justus Liebig University Giessen, Heinrich-Buff-Ring 16, 35392 Giessen, Germany.

[2] Department of Chemistry and mar.quest|Marburg Center for Quantum Materials and Sustainable Technologies, University of Marburg, Hans-Meerwein-Straße, 35043 Marburg, Germany.

[3] Graduate School of Engineering, Nagoya University, Aichi 464-8603, Japan

[4] Graduate School of Engineering Science, Osaka University, Japan

[5] Institute of Chemistry, School V, University of Oldenburg, Germany.

*Corresponding author E-mail address: johanna.heine@uni-oldenburg.de

## Table of Contents

# 1 Synthetic Details

**General procedure:** Under aerobic conditions, the respective metal oxides were dissolved in hydrohalic acid solution, followed by the addition of amine. The resulting clear reaction solution was heated to reflux for 30 minutes, filtered and left undisturbed for crystallization, which occurred within one to three days. Crystals were isolated by filtration, washed with small amounts of cold glacial acetic acid and pentane and dried under vacuum. Reagent amounts for individual reactions are given below in tables S1 and S2. Synthesis and analysis data for compound **5** is provided in a previous publication. [1] Compounds **8**, **9** and **10** were prepared according to modified literature procedures to allow for the use of metal oxides and amine as starting material. [2] [3] The phase purity of the resulting compounds was confirmed via CHN and PXRD analysis.

Single crystals for measurement were directly obtained from the mother liquor. Thermal analysis and IR data indicate that **1** and **2** loose water upon drying.

**Table S1:** Individual amounts of reagents and solvent in the synthesis of **1**, **2**, **6** and **7**, optimized for crystallization of pure product.

| **Compound** | **$E_2O_3$** | **Cyclohexanmethylamine** | **HX solution (2M)** | **Yield** |
|---|---|---|---|---|
| $[CMA]_4[Sb_2Cl_{10}]$ (**1**, dried) | 31.5mg 0.1mmol | 53.1µL, 0.4mmol | 5ml | 76.0mg 72% |
| $[CMA]_4[Bi_2Cl_{10}]$ (**2**, dried) | 49.0mg 0.1mmol | 53.1µL, 0.4mmol | 10ml | 41.5mg 34% |
| $[CMA]_3[SbBr_5]Br$ (**6**) | 30.3mg 0.1mmol | 39.8µL, 0.3mmol | 5ml | 58.2mg 62% |
| $[CMA]_3[BiBr_5]Br$ (**7**) | 49.1mg 0.1mmol | 39.8µL, 0.3mmol | 5ml | 66.8mg 65% |

**Table S2:** Individual amounts of reagents and solvent in the synthesis of **3**, **4** and **8**-**10** optimized for crystallization of pure product.

| Compound | $E_2O_3$ | Benzylamine | HX solution | Yield |
|---|---|---|---|---|
| $[BzA]_6[Sb_2Cl_{10}]Cl_2$ (**3**) | 31.8mg<br>0.1mmol | 65.4µL, 0.6mmol | 10ml<br>(Conc.) | 73.2mg<br>56% |
| $[BzA]_4[Bi_2Cl_{10}]$ (**4**) | 46.6mg<br>0.1mmol | 65.4µL, 0.6mmol | 5ml<br>(2M) | 73mg<br>61% |
| $[BzA]_2[SbBr_5]$ (**8**) | 148.0mg<br>0.5mmol | 0.11mL, 1.0mmol | 5ml<br>(Conc.) | 217.5mg<br>59% |
| $[BzA]_3[SbBr_6]$ (**9**) | 82.3mg<br>0.3mmol | 0.11mL, 1.0mmol | 5ml<br>(Conc.) | 174.1mg<br>31% |
| $[BzA]_3[BiBr_6]$ (**10**) | 48.2mg<br>0.1mmol | 65.4µL, 0.6mmol | 5ml<br>(2M) | 173.3mg<br>86% |

## CHN Analysis

CHN analysis was carried out on an *Elementar* CHN-analyzer. It should be noted that loss of solvate water in **1** and **2** as indicated by thermal analysis and IR data cannot be unambiguously confirmed with CHN analysis due to the small fraction of mass of the $H_2O$ molecules.

$[CMA]_4[Sb_2Cl_{10}]$ (**1**, dried). Colorless crystal. CHN data for **1**: Anal. Calcd for $C_{28}H_{64}N_4Sb_2Cl_{10}$, (M = 1054.75 g mol$^{-1}$): C, 31.88; H, 6.12; N, 5.31%. Found: C, 31.51; H, 6.01; N, 5.23%.

$[CMA]_4[Bi_2Cl_{10}]$ (**2**, dried). Colorless crystals. CHN Data for **2**: Anal. Calcd for $C_{28}H_{64}N_4Bi_2Cl_{10}$, (M = 1229.21 gmol$^{-1}$): C, 27.36; H, 5.25; N, 4.56%. Found: C, 27.28; H, 5.11; N, 4.60%.

$[CMA]_3[SbBr_5]Br$ (**6**). Light yellow crystals. CHN data for **6**: Anal. Calcd for $C_{21}H_{48}N_3SbBr_6$, (M = 943.83 g mol$^{-1}$): C, 26.70; H, 5.09; N, 4.45%. Found: C, 26.53; H, 4.97; N, 4.45%.

$[CMA]_3[BiBr_5]Br$ (**7**). Very light yellow crystals. CHN data for **7**: Anal. Calcd for $C_{21}H_{48}N_3BiBr_6$, (M = 1031.06 g mol$^{-1}$): C, 24.44; H, 4.66; N, 4.07%. Found: C, 24.47; H, 4.59; N, 4.12%.

$[BzA]_6[Sb_2Cl_{10}]Cl_2$ (**3**). Colorless crystal. CHN data for **3**: Anal. Calcd for $C_{42}H_{60}N_6Sb_2Cl_{12}$, (M = 1317.86 g mol$^{-1}$): C, 38.24; H, 4.55; N, 6.37%. Found: C, 38.67; H, 4.72; N, 6.46%.

$[BzA]_4[Bi_2Cl_{10}]$ (**4**). Colorless crystals. CHN Data for **4**: Anal. Calcd for $C_{28}H_{40}N_4Bi_2Cl_{10}$, (M = 1205.10 g mol$^{-1}$): C, 27.91; H, 3.35; N, 4.65%. Found: C, 28.05; H, 3.24; N, 4.73%.

$[BzA]_2[SbBr_5]$ (**8**). Light yellow crystals. CHN data for **8**: Anal. Calcd for $C_{14}H_{20}N_2SbBr_5$, (M = 737.80 g mol$^{-1}$): C, 22.77; H, 2.71; N, 3.80%. Found: C, 22.92; H, 2.94; N, 3.84%.

$[BzA]_3[SbBr_6]$ (**9**). Light yellow crystals. CHN data for **9**: Anal. Calcd for $C_{21}H_{30}N_3SbBr_6$, (M = 925.80 g mol$^{-1}$): C, 27.19; H, 3.24; N, 4.54%. Found: C, 27.15; H, 3.40; N, 4.51%.

$[BzA]_3[BiBr_6]$ (**10**). Light yellow crystals. CHN data for **10**: Anal. Calcd for $C_{21}H_{30}N_3BiBr_6$, (M = 1012.98 g mol$^{-1}$): C, 24.88; H, 2.96; N, 4.15%. Found: C, 25.30; H, 3.25; N, 4.24%.

# 2 Crystallographic Details

Single crystal X-ray determination was performed at 100 K on a Bruker Quest D8 diffractometer with microfocus MoKα radiation and a Photon 100 (CMOS) detector or a STOE IPDS-2 diffractometer equipped with an imaging plate detector system using MoKα radiation with graphite monochromatization. The structures were solved using direct methods, refined by full-matrix least-squares techniques and expanded using Fourier techniques, using the ShelX software package [4–6] within the OLEX2 suite. [7] All non-hydrogen atoms were refined anisotropically unless otherwise indicated. Hydrogen atoms were assigned to idealized geometric positions and included in structure factors calculations unless otherwise indicated. Pictures of the crystal structures were created using DIAMOND. [8] Additional details on individual refinements, including CCDC deposition numbers, are given below.

**Table S3:** Crystallographic data for **1**.

| CCDC No. | 2479960 |
|---|---|
| Empirical formula | $C_{28}H_{64}Cl_{10}N_4OSb_2$ |
| Formula weight | 1070.83 |
| Temperature/K | 100 |
| Crystal system | triclinic |
| Space group | P-1 |
| a/Å | 11.7166(3) |
| b/Å | 12.4608(3) |
| c/Å | 16.8058(5) |
| α/° | 70.527(2) |
| β/° | 76.086(2) |
| γ/° | 88.766(2) |
| Volume/Å$^3$ | 2240.78(11) |
| Z | 2 |
| $\rho_{calc}$ g/cm$^3$ | 1.587 |
| μ/mm$^{-1}$ | 1.829 |
| Absorption correction ($T_{min}/T_{max}$) | numerical (0.676/ 0.834) |
| F(000) | 1080.0 |
| Crystal size/mm$^3$ | 0.232 × 0.163 × 0.103 |
| Radiation | MoKα (λ = 0.71073 Å) |
| 2Θ range for data collection/° | 3.9 to 63.848 |
| Index ranges | -17 ≤ h ≤ 16, -18 ≤ k ≤ 18, -24 ≤ l ≤ 24 |
| Reflections collected | 53177 |
| Independent reflections | 15390 [$R_{int}$ = 0.0773, $R_{sigma}$ = 0.0822] |
| Data/restraints/parameters | 15390/6/473 |
| Goodness-of-fit on F$^2$ | 0.988 |
| Final R indexes [I>=2σ (I)] | $R_1$ = 0.0395, $wR_2$ = 0.1293 |
| Final R indexes [all data] | $R_1$ = 0.0602, $wR_2$ = 0.1335 |
| Largest diff. peak/hole / e Å$^{-3}$ | 1.63/-1.78 |

**Details of crystal structure measurement and refinement:** Single crystal X-ray determination was performed at 100 K on a STOE IPDS-2diffractometer equipped with an imaging plate detector system using MoKα radiation with graphite monochromatization. One cyclohexylmethylammonium cation was found to be disordered over two positions, with the exception of the nitrogen atom. Occupations in the disordered units were first refined freely, then set to the nearest decimal value (0.4 and 0.6) to ensure a more stable refinement. An ISOR command had to be used on one of the disordered carbon atom to achieve reasonable atomic displacement parameters. Hydrogen atoms of the water molecule could not be reliably located and were thus omitted.

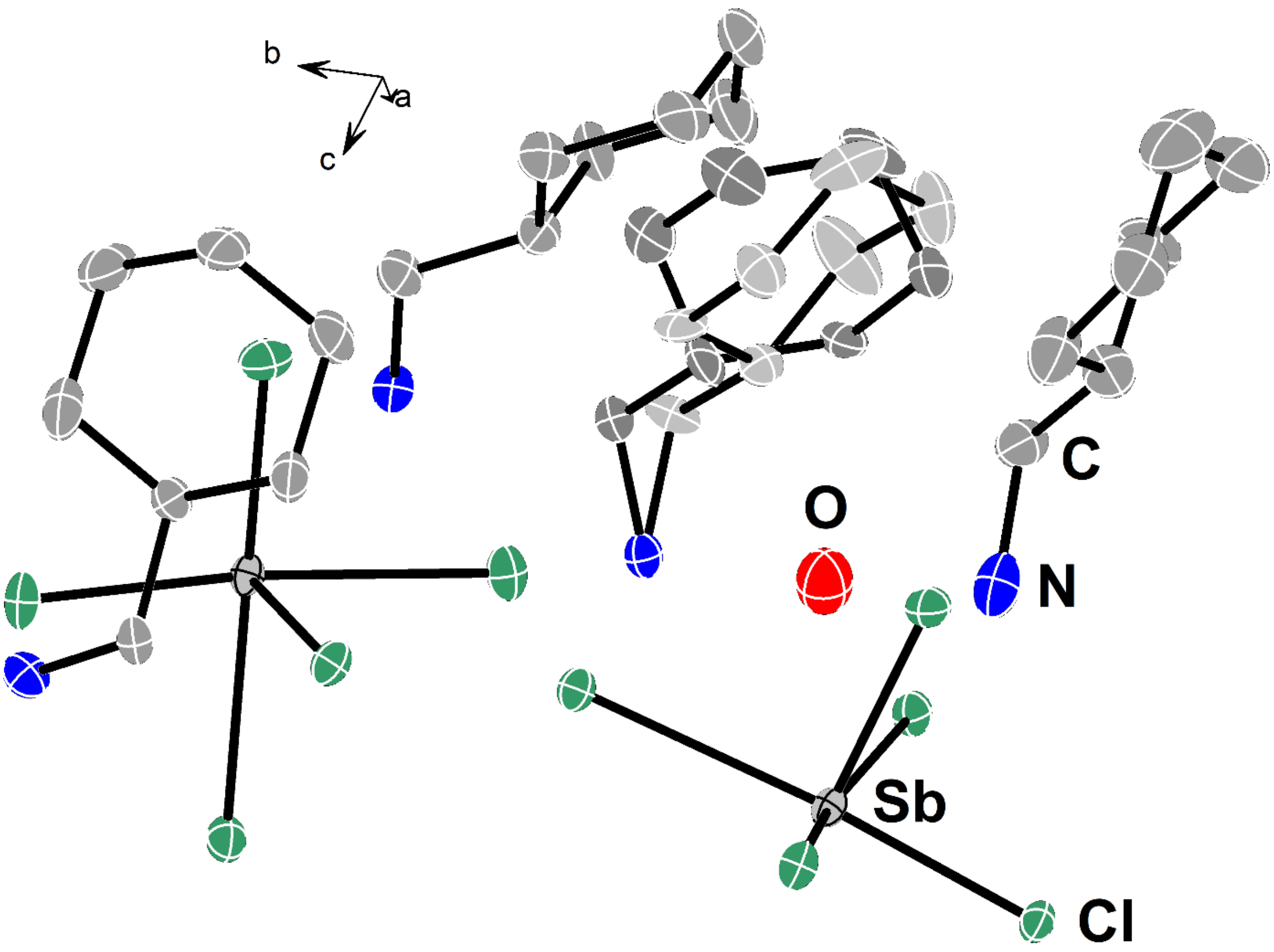

**Figure S1:** Asymmetric unit of **1**, ellipsoids at 50% probability. Hydrogen atoms are omitted for clarity. The two positions in the disordered part of the cyclohexylmethylammonium cation are shown in darker and lighter colour for 0.6 and 0.4 occupation, respectively.

**Table S4:** Crystallographic data for **2**.

| CCDC No. | 2479964 |
|---|---|
| Empirical formula | $C_{28}H_{66}Bi_2Cl_{10}N_4O$ |
| Formula weight | 1247.30 |
| Temperature/K | 100 |
| Crystal system | triclinic |
| Space group | P-1 |
| a/Å | 12.0090(3) |
| b/Å | 12.4027(3) |
| c/Å | 16.2752(4) |
| α/° | 106.775(2) |
| β/° | 99.311(2) |
| γ/° | 91.841(2) |
| Volume/Å$^3$ | 2282.37(10) |
| Z | 2 |
| $\rho_{calc}$g/cm$^3$ | 1.815 |
| μ/mm$^{-1}$ | 8.312 |
| Absorption correction ($T_{min}/T_{max}$) | numerical (0.0873/0.4485) |
| F(000) | 1212.0 |
| Crystal size/mm$^3$ | 0.235 × 0.147 × 0.103 |
| Radiation | MoKα (λ = 0.71073 Å) |
| 2Θ range for data collection/° | 3.442 to 58.432 |
| Index ranges | -16 ≤ h ≤ 16, -16 ≤ k ≤ 16, -22 ≤ l ≤ 22 |
| Reflections collected | 58866 |
| Independent reflections | 12294 [$R_{int}$ = 0.0659, $R_{sigma}$ = 0.0430] |
| Data/restraints/parameters | 12294/46/406 |
| Goodness-of-fit on F$^2$ | 1.066 |
| Final R indexes [I>=2σ (I)] | $R_1$ = 0.0316, $wR_2$ = 0.0856 |
| Final R indexes [all data] | $R_1$ = 0.0443, $wR_2$ = 0.0881 |
| Largest diff. peak/hole / e Å$^{-3}$ | 3.09/-1.83 |

**Details of crystal structure refinement:** Single crystal X-ray determination was performed at 100 K on a STOE IPDS-2diffractometer equipped with an imaging plate detector system using MoKα radiation with graphite monochromatization. Two cyclohexylmethylammonium cations were found to be disordered over two positions. In the first case, only part of the cyclohexyl ring are disordered, in the second case, all atoms except for nitrogen atom are disordered. Occupations in the disordered units were first refined freely, then set to the nearest decimal value (0.5 for all) to ensure a more stable refinement. Disordered atoms were refined isotropically. A number of SADI restraints had to be used to ensure reasonable carbon-carbon distances within the two disordered cyclohexyl moieties.

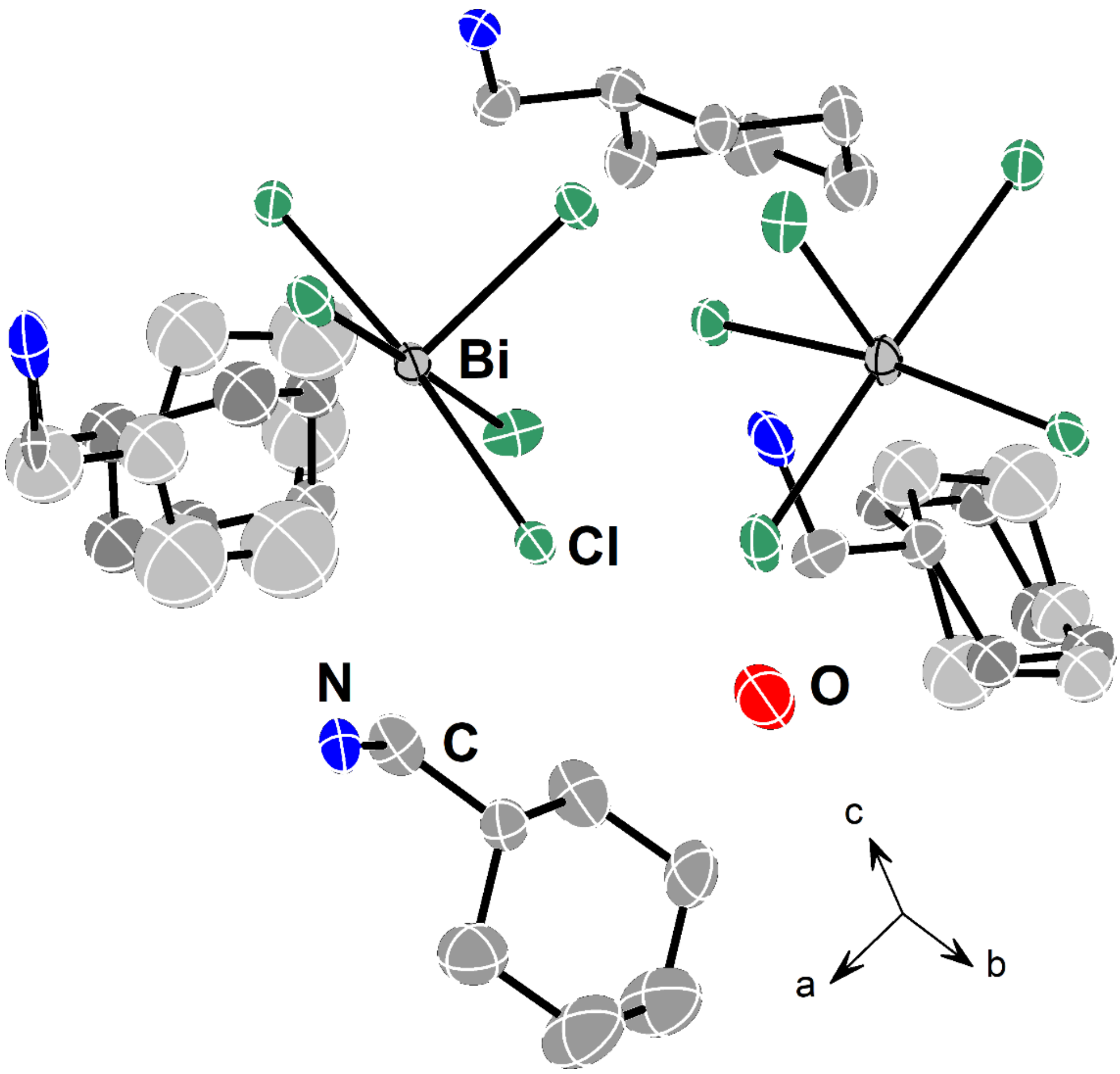

**Figure S2:** Asymmetric unit of **2**, ellipsoids at 50% probability. Hydrogen atoms are omitted for clarity. The two positions in the disordered parts of the cyclohexylmethylammonium cations are shown in darker and lighter colour for improved clarity although it should be noted that all feature an occupancy of 0.5.

**Table S5:** Crystallographic data for **6**.

| CCDC No. | 2479961 |
|---|---|
| Empirical formula | $C_{21}H_{48}Br_6N_3Sb$ |
| Formula weight | 943.83 |
| Temperature/K | 100 |
| Crystal system | monoclinic |
| Space group | $P2_1/c$ |
| a/Å | 17.0648(7) |
| b/Å | 23.7816(8) |
| c/Å | 7.9888(3) |
| α/° | 90 |
| β/° | 98.683(3) |
| γ/° | 90 |
| Volume/Å$^3$ | 3204.9(2) |
| Z | 4 |
| $\rho_{calc}$g/cm$^3$ | 1.956 |
| μ/mm$^{-1}$ | 8.356 |
| Absorption correction ($T_{min}/T_{max}$) | numerical (0.1974/0.6573) |
| F(000) | 1824.0 |
| Crystal size/mm$^3$ | 0.242 × 0.187 × 0.067 |
| Radiation | MoKα (λ = 0.71073 Å) |
| 2Θ range for data collection/° | 2.96 to 53.51 |
| Index ranges | -10 ≤ h ≤ 7, -30 ≤ k ≤ 28, -21 ≤ l ≤ 21 |
| Reflections collected | 26548 |
| Independent reflections | 6794 [$R_{int}$ = 0.0782, $R_{sigma}$ = 0.0467] |
| Data/restraints/parameters | 6794/6/283 |
| Goodness-of-fit on F$^2$ | 1.053 |
| Final R indexes [I>=2σ (I)] | $R_1$ = 0.0442, $wR_2$ = 0.1560 |
| Final R indexes [all data] | $R_1$ = 0.0614, $wR_2$ = 0.1786 |
| Largest diff. peak/hole / e Å$^{-3}$ | 1.72/-2.32 |

**Details of crystal structure measurement and refinement:** Single crystal X-ray determination was performed at 100 K on a STOE IPDS-2diffractometer equipped with an imaging plate detector system using MoKα radiation with graphite monochromatization. An ISOR restraint had to be used on one carbon atom to obtain reasonable atomic displacement parameters.

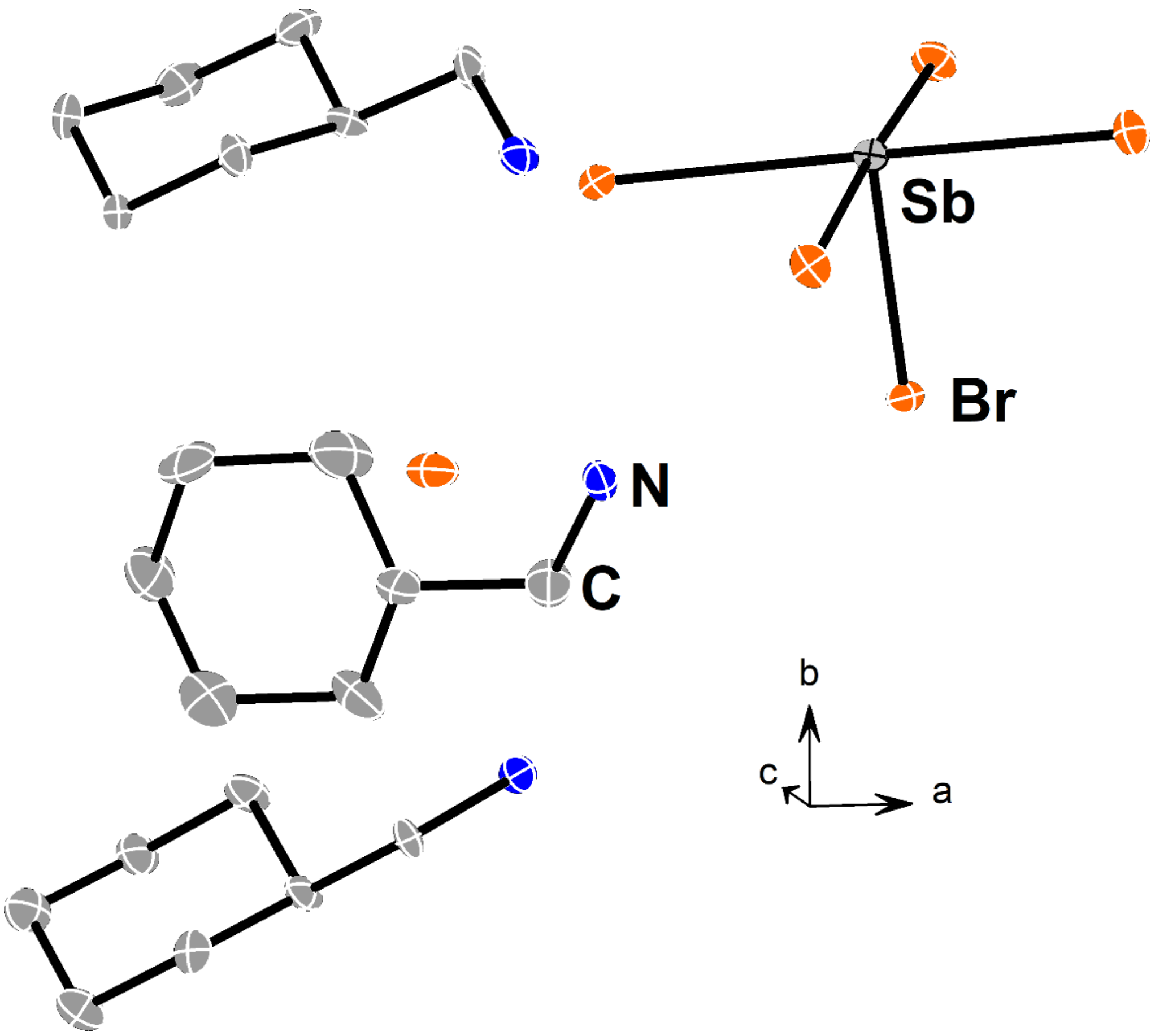

**Figure S3:** Asymmetric unit of **6**, ellipsoids at 50% probability. Hydrogen atoms omitted for clarity.

**Table S6:** Crystallographic data for **7**.

| CCDC No. | 2479963 |
|---|---|
| Empirical formula | $C_{21}H_{48}BiBr_6N_3$ |
| Formula weight | 1031.06 |
| Temperature/K | 100 |
| Crystal system | monoclinic |
| Space group | $P2_1/c$ |
| a/Å | 17.0824(10) |
| b/Å | 23.9227(10) |
| c/Å | 7.9868(5) |
| α/° | 90 |
| β/° | 99.081(5) |
| γ/° | 90 |
| Volume/Å$^3$ | 3223.0(3) |
| Z | 4 |
| $\rho_{calc}$g/cm$^3$ | 2.125 |
| μ/mm$^{-1}$ | 12.924 |
| Absorption correction ($T_{min}/T_{max}$) | numerical (0.2200/0.5062) |
| F(000) | 1952.0 |
| Crystal size/mm$^3$ | 0.118 × 0.088 × 0.07 |
| Radiation | MoKα (λ = 0.71073 Å) |
| 2Θ range for data collection/° | 3.404 to 58.384 |
| Index ranges | -23 ≤ h ≤ 23, -32 ≤ k ≤ 32, -10 ≤ l ≤ 10 |
| Reflections collected | 26797 |
| Independent reflections | 8688 [$R_{int}$ = 0.0949, $R_{sigma}$ = 0.0837] |
| Data/restraints/parameters | 8688/0/283 |
| Goodness-of-fit on $F^2$ | 1.010 |
| Final R indexes [I>=2σ (I)] | $R_1$ = 0.0428, $wR_2$ = 0.0961 |
| Final R indexes [all data] | $R_1$ = 0.0733, $wR_2$ = 0.1039 |
| Largest diff. peak/hole / e Å$^{-3}$ | 2.00/-2.31 |

**Details of crystal structure measurement and refinement:** Single crystal X-ray determination was performed at 100 K on a STOE IPDS-2diffractometer equipped with an imaging plate detector system using MoKα radiation with graphite monochromatization.

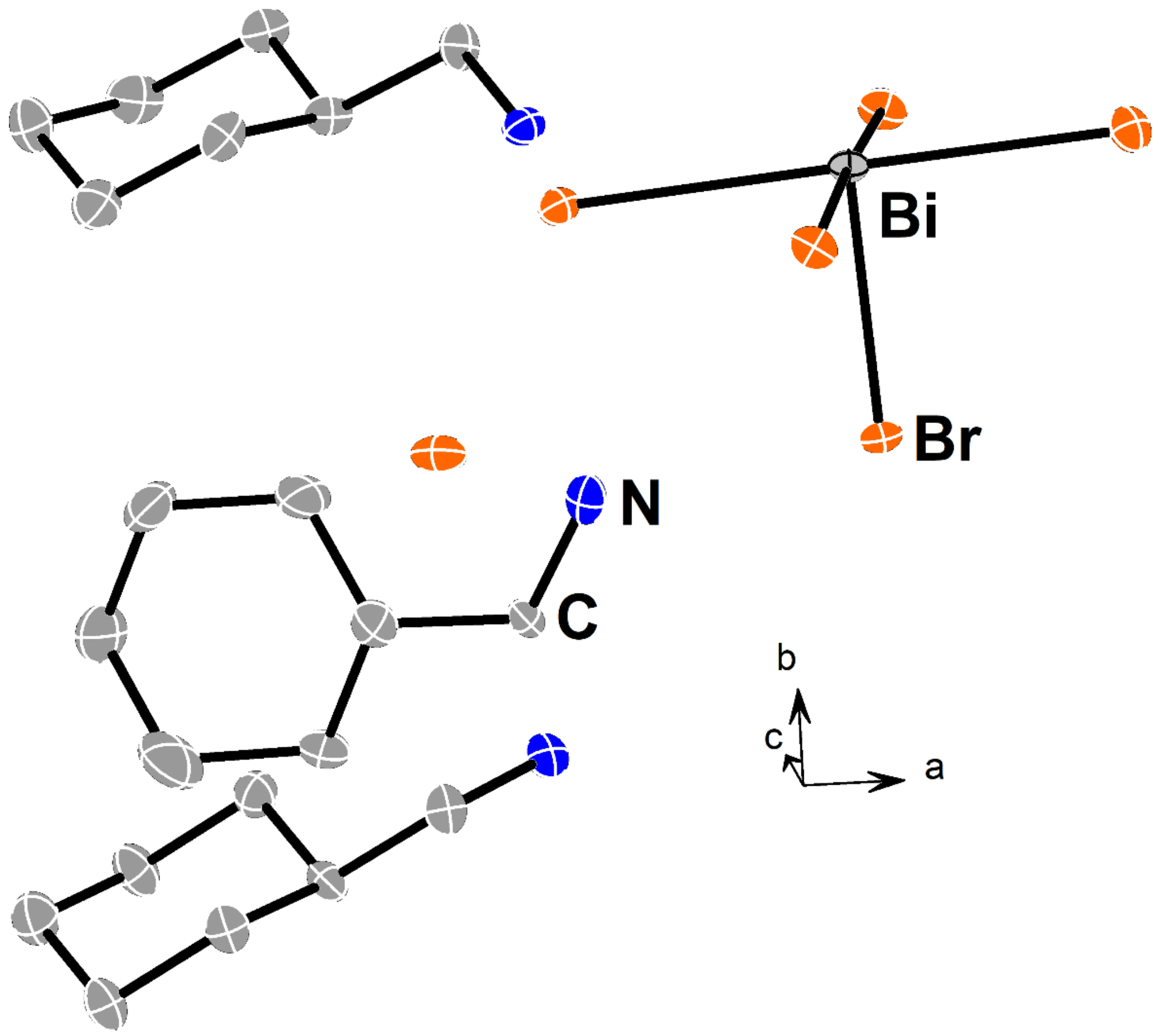

**Figure S4:** Asymmetric unit of **7**, ellipsoids at 50% probability. Hydrogen atoms are omitted for clarity.

**Table S7:** Crystallographic data for **3**.

| CCDC | 2479965 |
|---|---|
| Empirical formula | $C_{42}H_{60}Cl_{12}N_6Sb_2$ |
| Formula weight | 1317.86 |
| Temperature/K | 100.0 |
| Crystal system | monoclinic |
| Space group | $P2_1$ |
| a/Å | 9.5168(7) |
| b/Å | 29.602(2) |
| c/Å | 10.0137(7) |
| α/° | 90 |
| β/° | 104.803(2) |
| γ/° | 90 |
| Volume/Å$^3$ | 2727.4(3) |
| Z | 2 |
| $\rho_{calc}$g/cm$^3$ | 1.605 |
| μ/mm$^{-1}$ | 1.614 |
| Absorption correction ($T_{min}/T_{max}$) | Multiscan (0.944/0.766) |
| F(000) | 1320.0 |
| Crystal size/mm$^3$ | 0.174 × 0.039 × 0.036 |
| Radiation | MoKα (λ = 0.71073 Å) |
| 2Θ range for data collection/° | 4.426 to 50.592 |
| Index ranges | $-11 \le h \le 11$, $-35 \le k \le 35$, $-11 \le l \le 12$ |
| Reflections collected | 67462 |
| Independent reflections | 9905 [$R_{int}$ = 0.1791, $R_{sigma}$ = 0.1044] |
| Data/restraints/parameters | 9905/79/547 |
| Goodness-of-fit on $F^2$ | 1.075 |
| Final R indexes [I>=2σ (I)] | $R_1$ = 0.0748, $wR_2$ = 0.1463 |
| Final R indexes [all data] | $R_1$ = 0.1046, $wR_2$ = 0.1574 |
| Largest diff. peak/hole / e Å$^{-3}$ | 2.22/-1.39 |
| Flack parameter | 0.22(6) |

**Details of crystal structure measurement and refinement:** Single crystal X-ray determination was performed at 100 K on a Bruker Quest D8 diffractometer with microfocus MoKα radiation and a Photon 100 (CMOS) detector. The structure was refined as an inversion twin. One benzylammonium cation was disordered over two positions with occupancies first refined freely and then set to 2/3 and 1/3 to provide a more stable refinement. The nitrogen and one carbon atom of the unit were not disordered. Disordered atoms were refined isotropically. A number of ISOR restraints had to be used to achieve reasonable atomic displacement parameters.

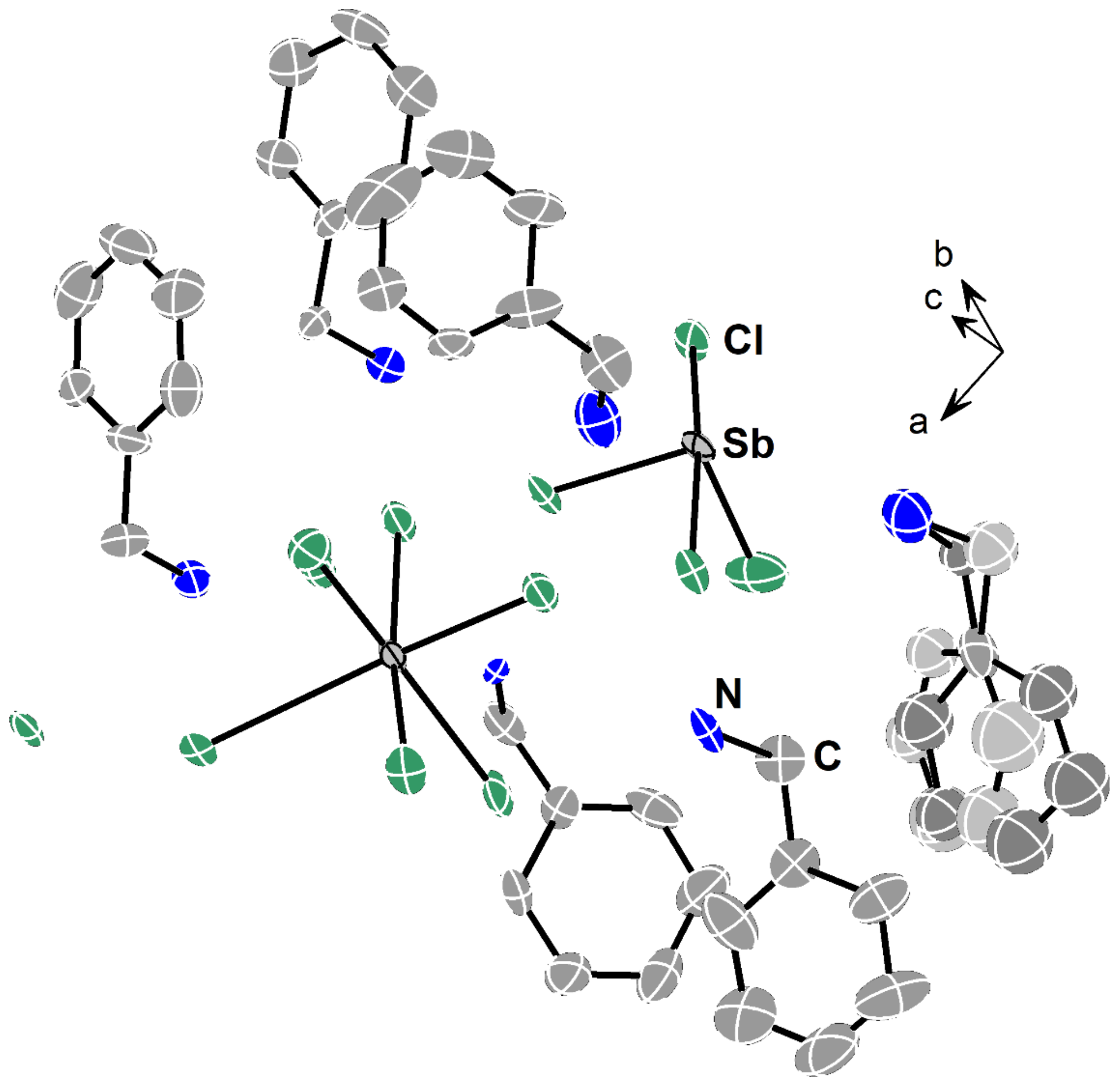

**Figure S5**: Asymmetric unit of **3**, ellipsoids at 80% probability. Hydrogen atoms omitted for clarity. The two positions in the disordered part of the benzylammonium cation are shown in darker and lighter colour for 2/3 and 1/3 occupation, respectively.

**Table S8:** Crystallographic data for **4**.

| CCDC No. | 2479962 |
|---|---|
| Empirical formula | $C_{28}H_{40}Bi_2Cl_{10}N_4$ |
| Formula weight | 1205.10 |
| Temperature/K | 100.0 |
| Crystal system | monoclinic |
| Space group | $P2_1/c$ |
| a/Å | 13.4356(19) |
| b/Å | 12.6845(17) |
| c/Å | 12.0570(16) |
| α/° | 90 |
| β/° | 97.175(4) |
| γ/° | 90 |
| Volume/Å$^3$ | 2038.7(5) |
| Z | 2 |
| $\rho_{calc}$g/cm$^3$ | 1.963 |
| μ/mm$^{-1}$ | 9.301 |
| Absorption correction ($T_{min}$/$T_{max}$) | multiscan (0.3615/0.7452) |
| F(000) | 1144.0 |
| Crystal size/mm$^3$ | 0.343 × 0.279 × 0.136 |
| Radiation | MoKα (λ = 0.71073 Å) |
| 2Θ range for data collection/° | 4.432 to 50.774 |
| Index ranges | -16 ≤ h ≤ 16, -15 ≤ k ≤ 15, -14 ≤ l ≤ 14 |
| Reflections collected | 41243 |
| Independent reflections | 3740 [$R_{int}$ = 0.0924, $R_{sigma}$ = 0.0393] |
| Data/restraints/parameters | 3740/6/201 |
| Goodness-of-fit on F$^2$ | 1.279 |
| Final R indexes [I>=2σ (I)] | $R_1$ = 0.0506, $wR_2$ = 0.1314 |
| Final R indexes [all data] | $R_1$ = 0.0641, $wR_2$ = 0.1376 |
| Largest diff. peak/hole / e Å$^{-3}$ | 2.58/-2.68 |

**Details of crystal structure measurement and refinement:** Single crystal X-ray determination was performed at 100 K on a Bruker Quest D8 diffractometer with microfocus MoKα radiation and a Photon 100 (CMOS) detector. An ISOR restraint had to be used on one carbon atom to obtain reasonable atomic displacement parameters.

We note that a room temperature measurement of the same compound has recently been reported by Long [9] under CCDC deposition number 1886951 / CSD entry code NAXXAP indicating that no phase transition occurs between 100 and 293 K.

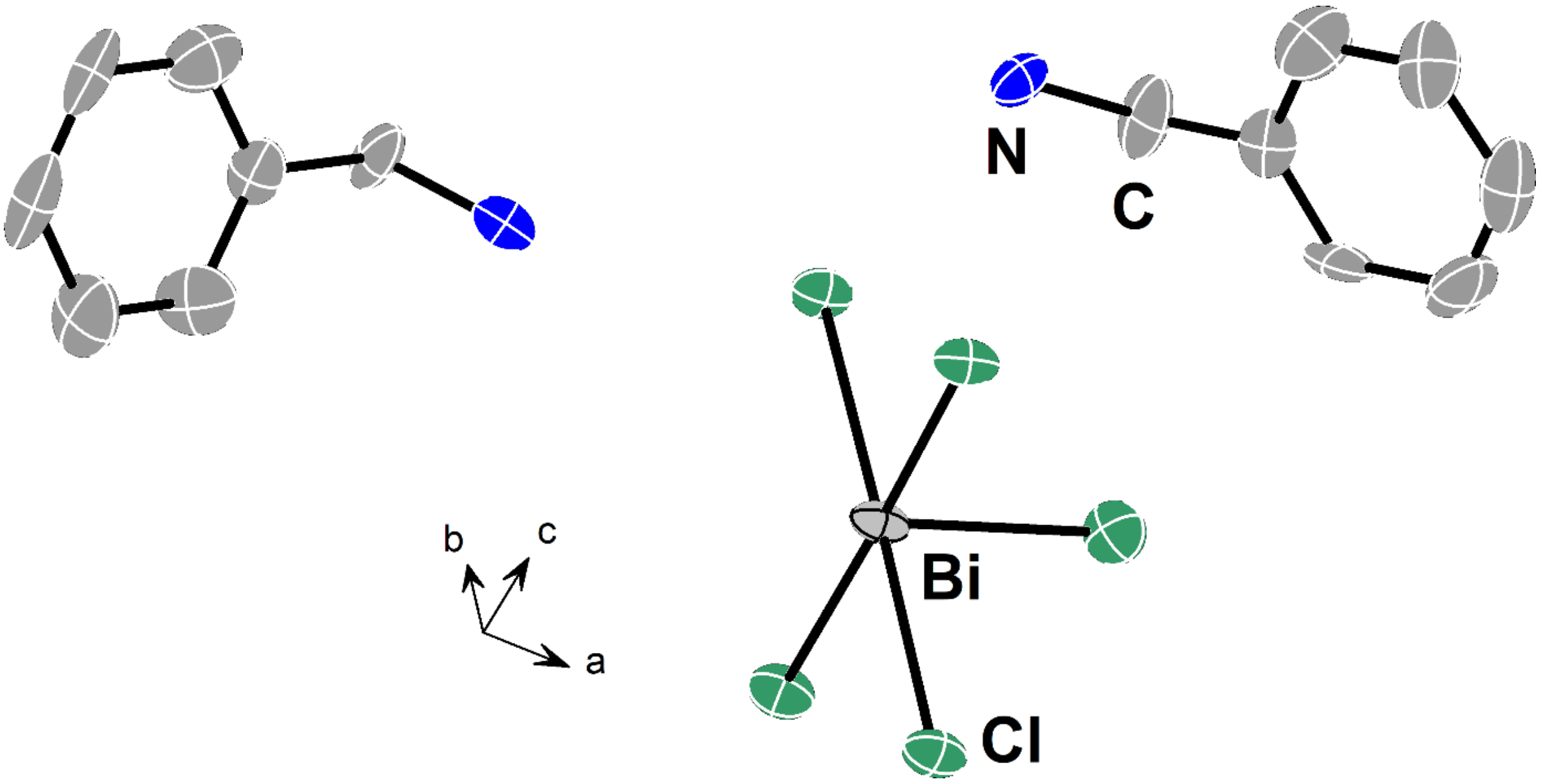

**Figure S6**: Asymmetric unit of **4**, ellipsoids at 80% probability.

# 3 Additional Crystallographic Figures & Tables

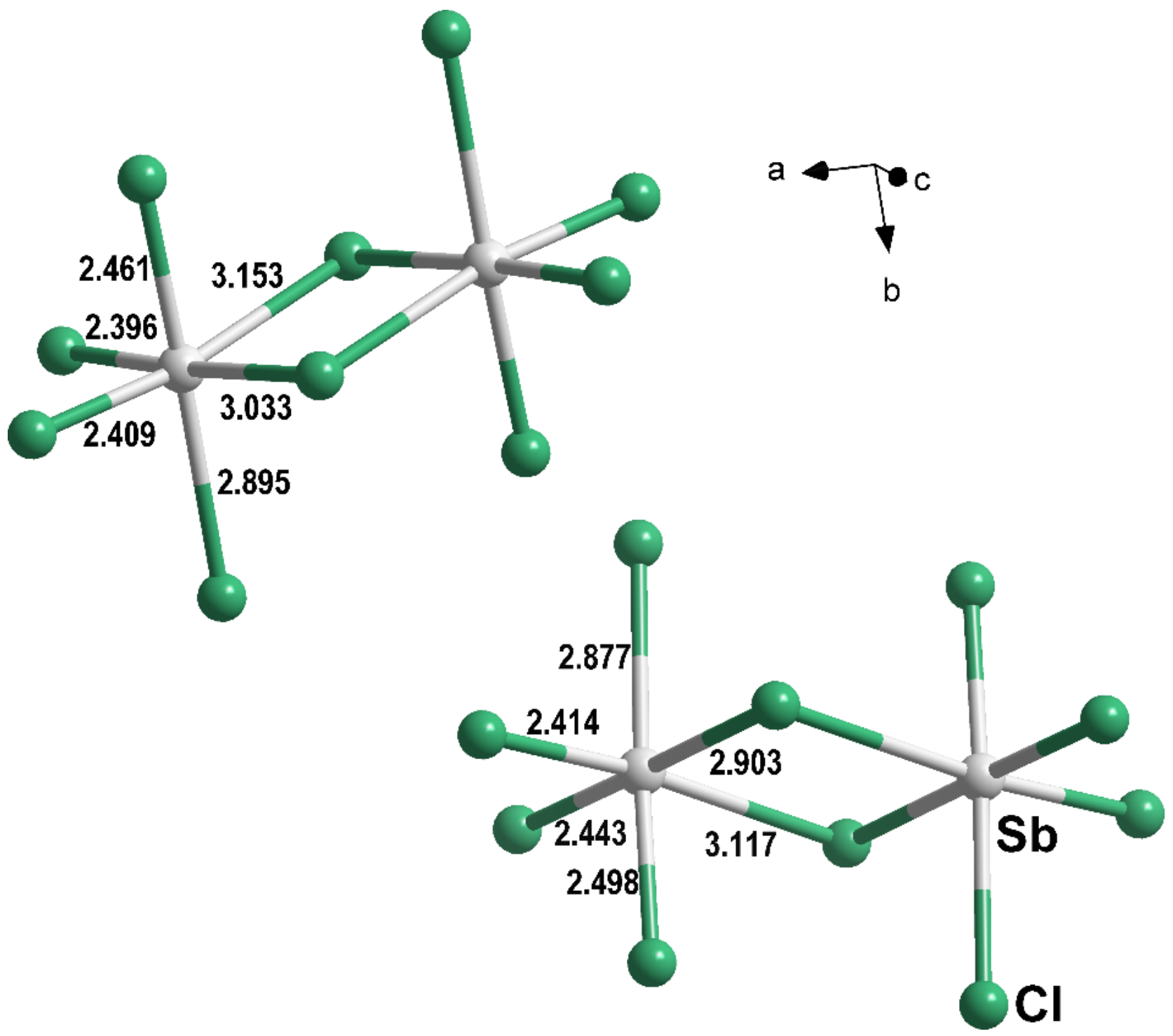

**Figure S7**: Bond length in Ångström in the $[Sb_2Cl_{10}]^{4-}$ anion in **1** with the second half of the anions generated via the centre of inversion.

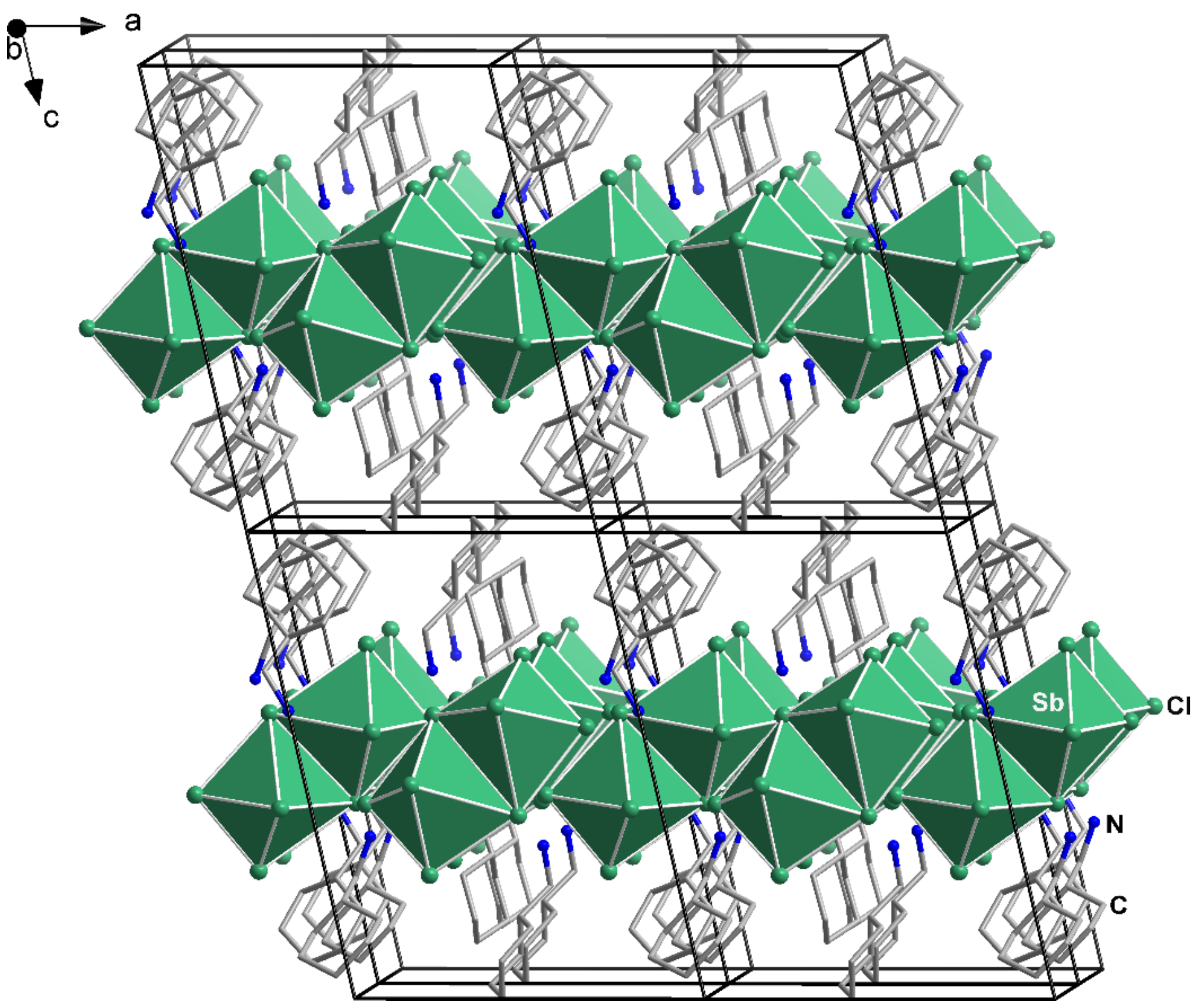

**Figure S8:** Packing diagram of **1** highlighting the pseudo-layered arrangement of cations and anions. Hydrogen atoms, cation disorder and solvate water atoms are omitted for clarity.

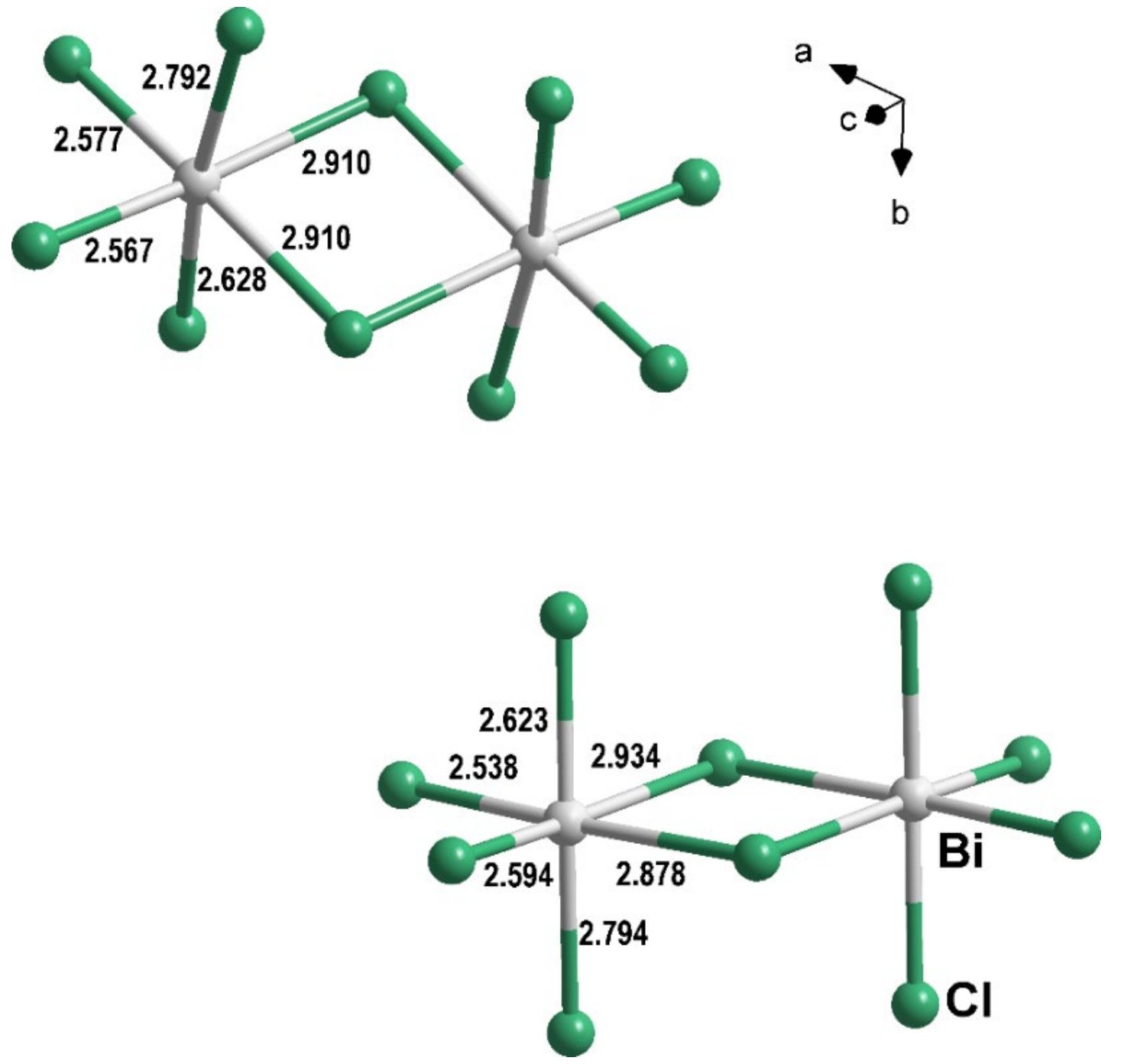

**Figure S9**: Bond length in Ångström in the $[Bi_2Cl_{10}]^{4-}$ anion in **2** with the second half of the anions generated via the centre of inversion.

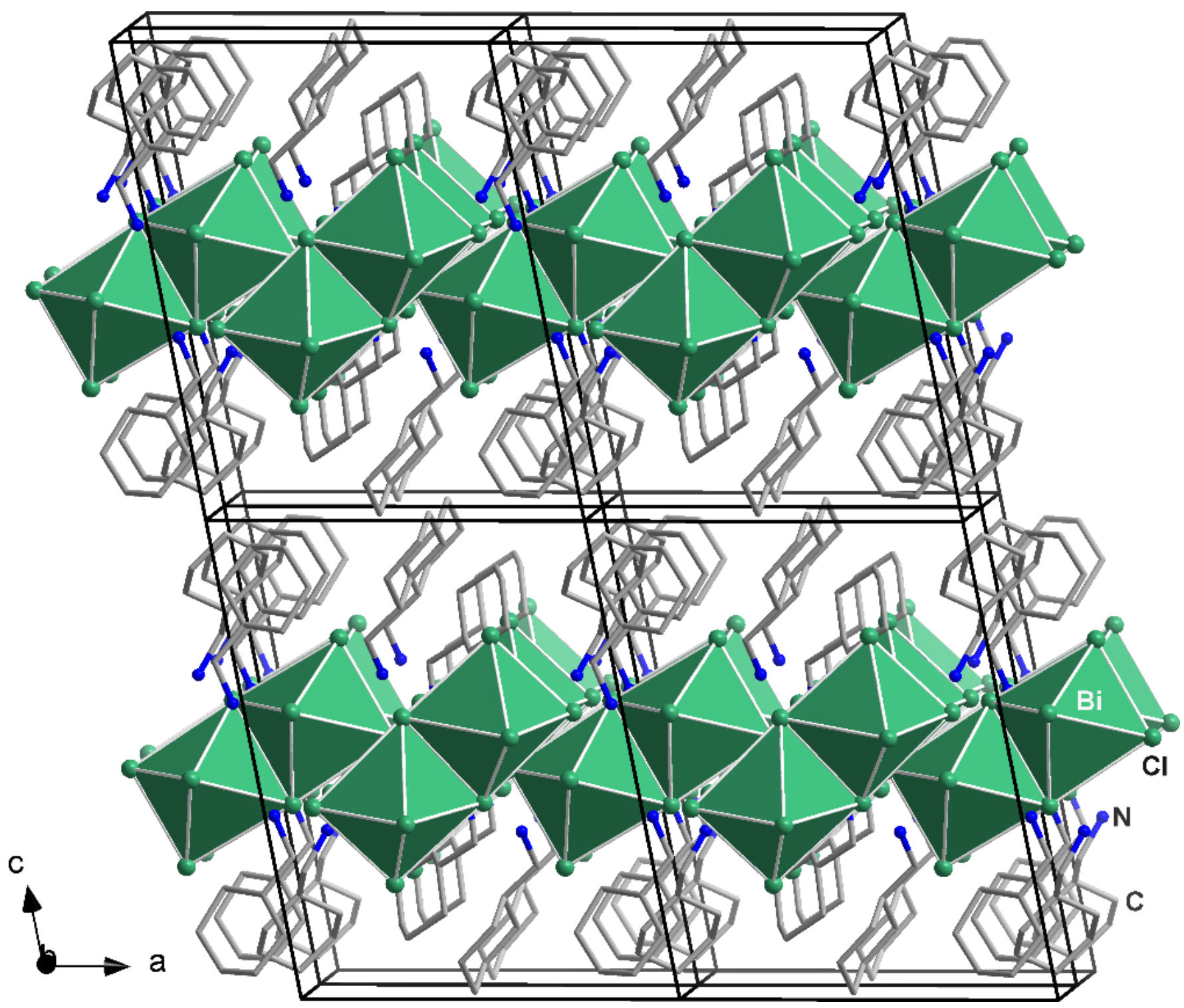

**Figure S10:** Packing diagram of **2** highlighting the pseudo-layered arrangement of cations and anions. Hydrogen atoms, cation disorder and solvate water atoms are omitted for clarity.

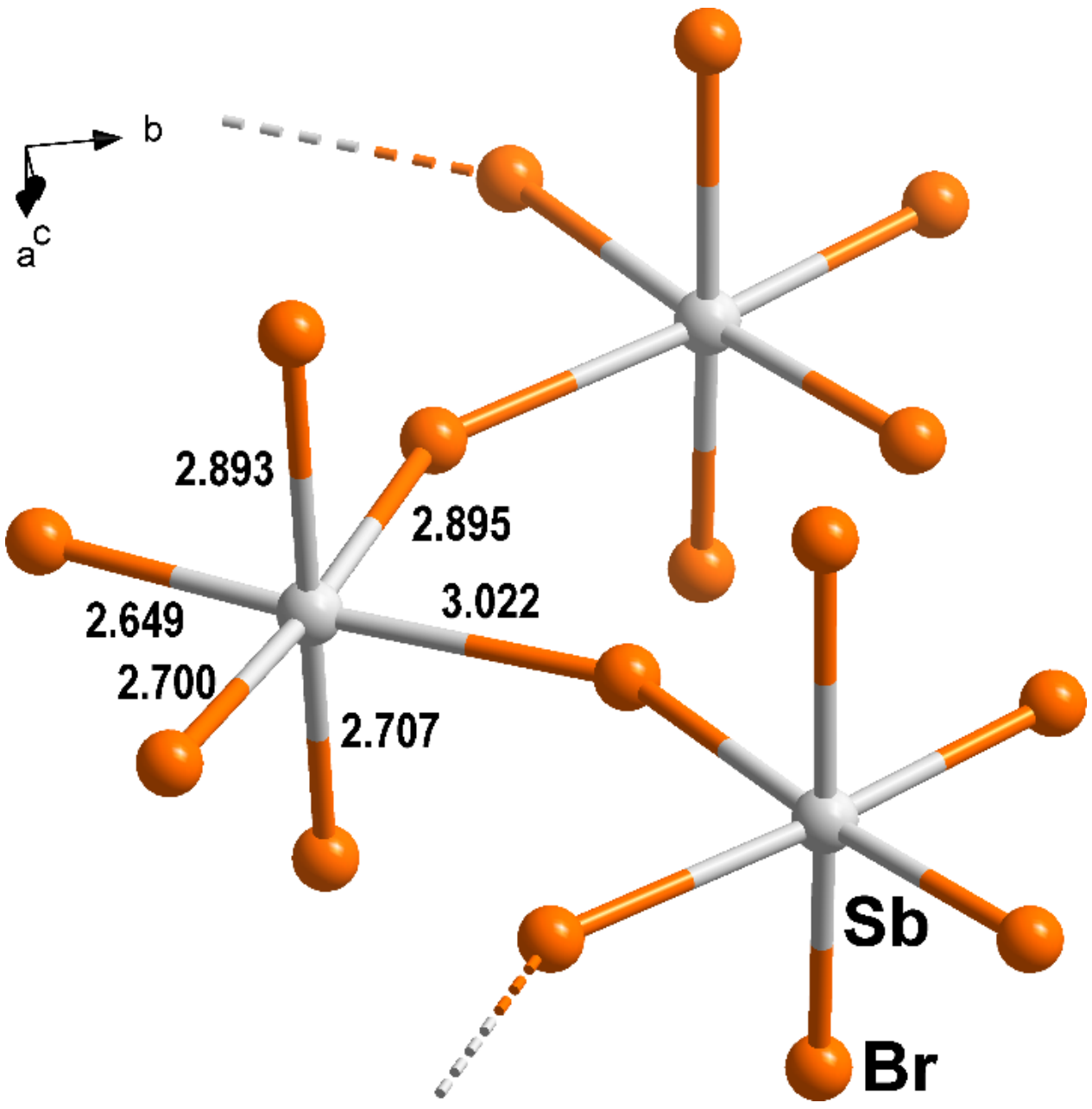

**Figure S11**: Bond length in Ångström in the $[SbBr_5]^{2-}$ anion in **6**.

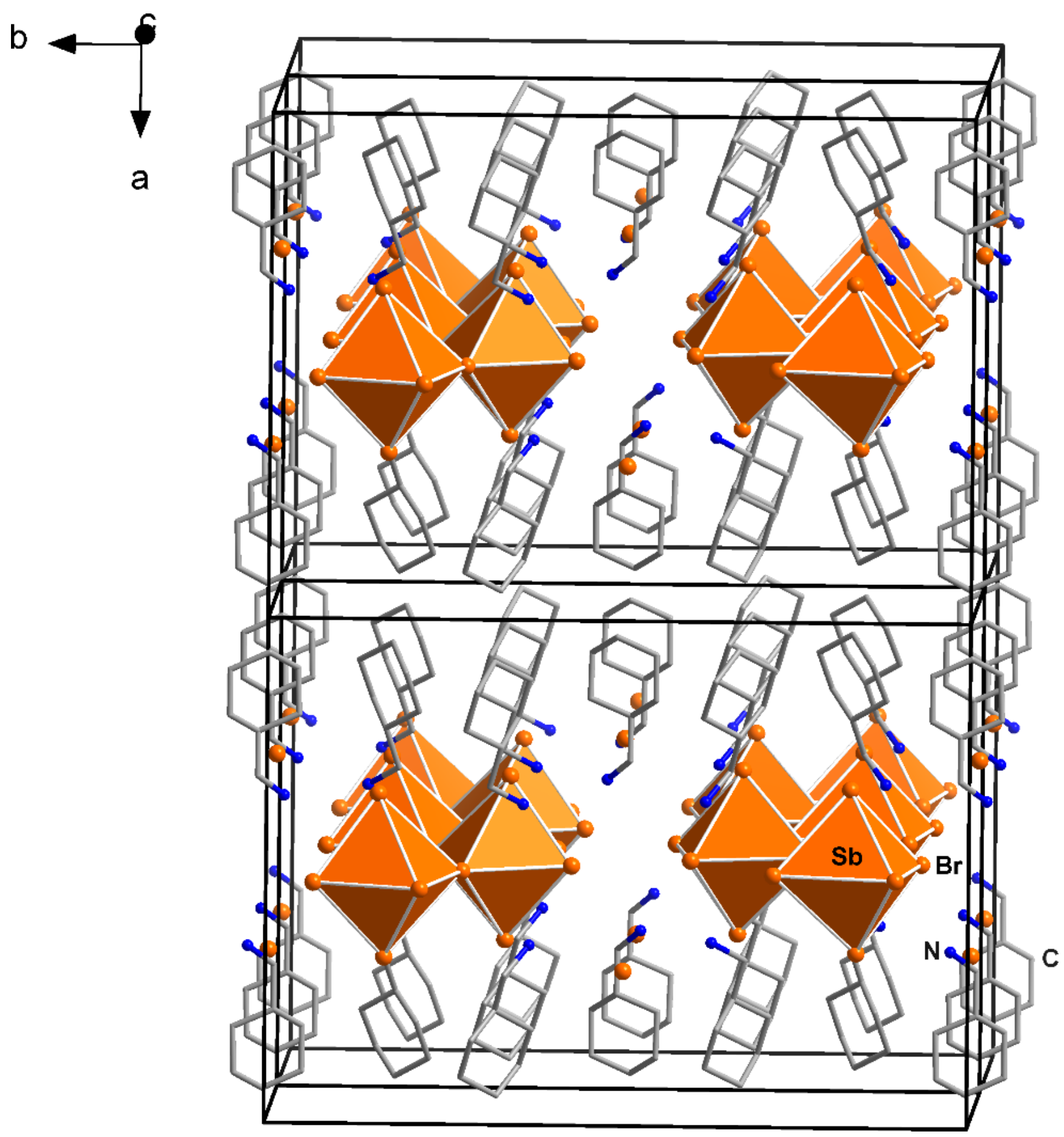

**Figure S12:** Packing diagram of **6** highlighting the pseudo-layered arrangement of cations and anions. Hydrogen atoms are omitted for clarity.

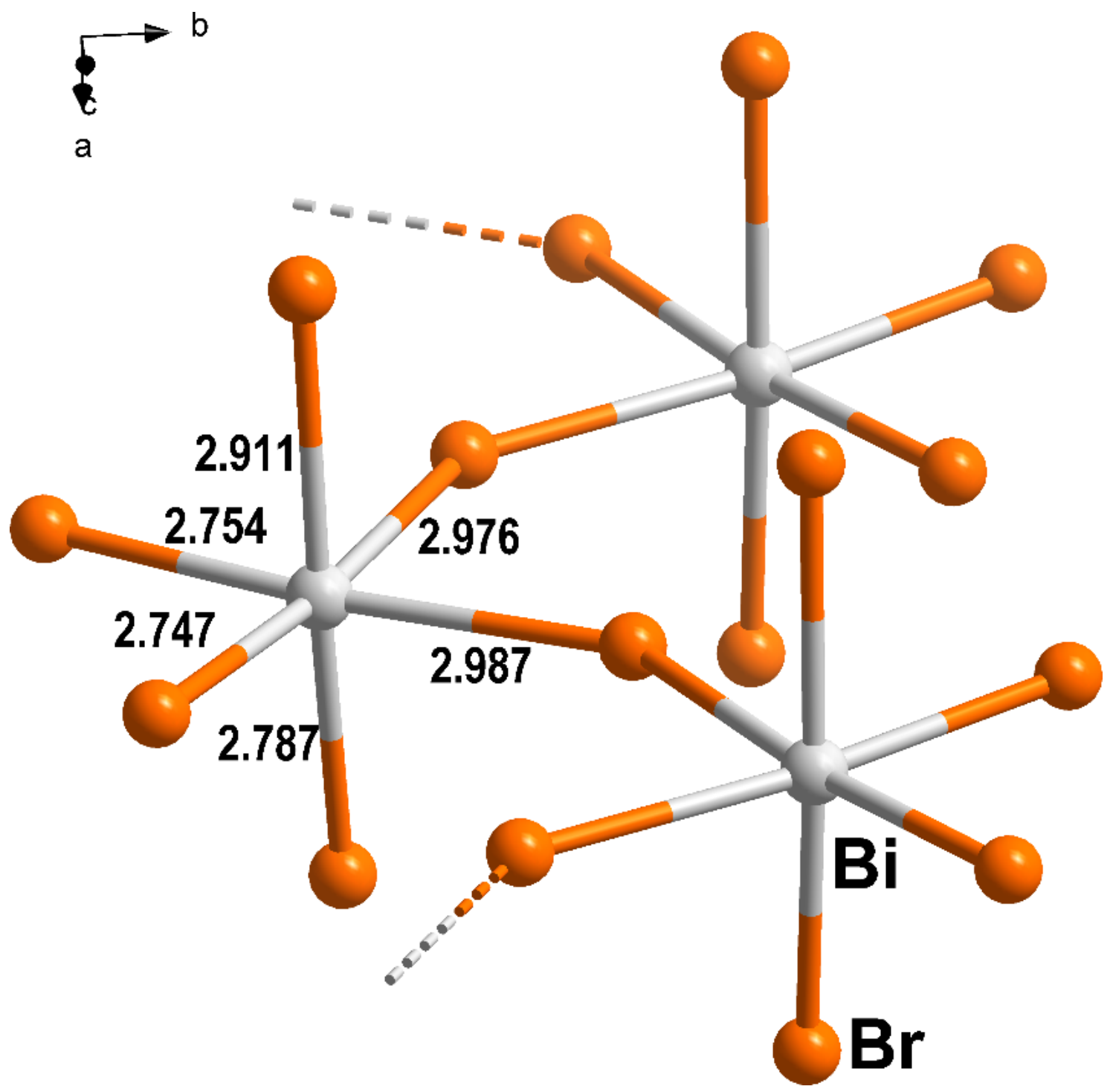

**Figure S12**: Bond length in Ångström in the $[BiBr_5]^{2-}$ anion in **7**.

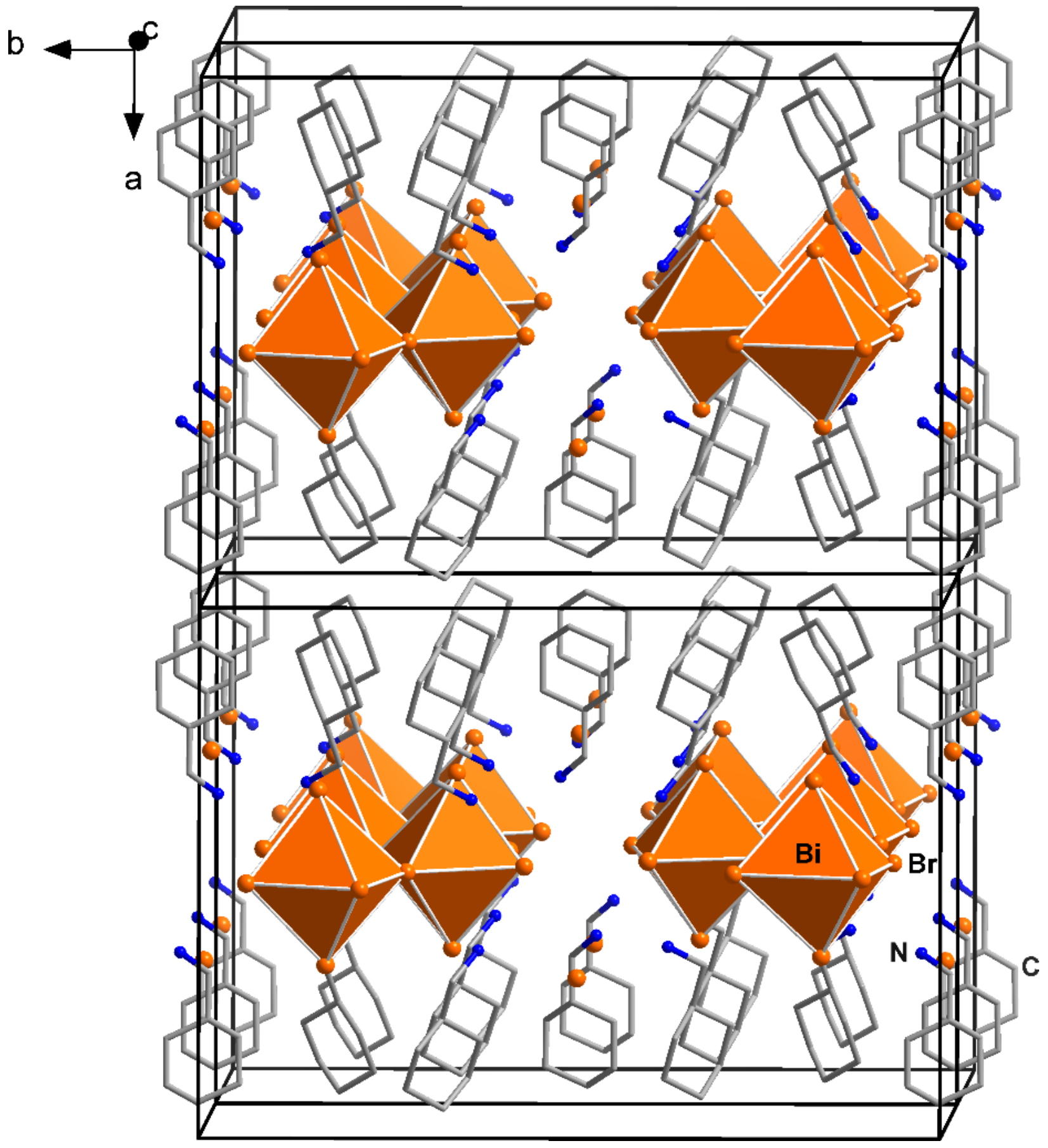

**Figure S13:** Packing diagram of **7** highlighting the pseudo-layered arrangement of cations and anions. Hydrogen atoms are omitted for clarity.

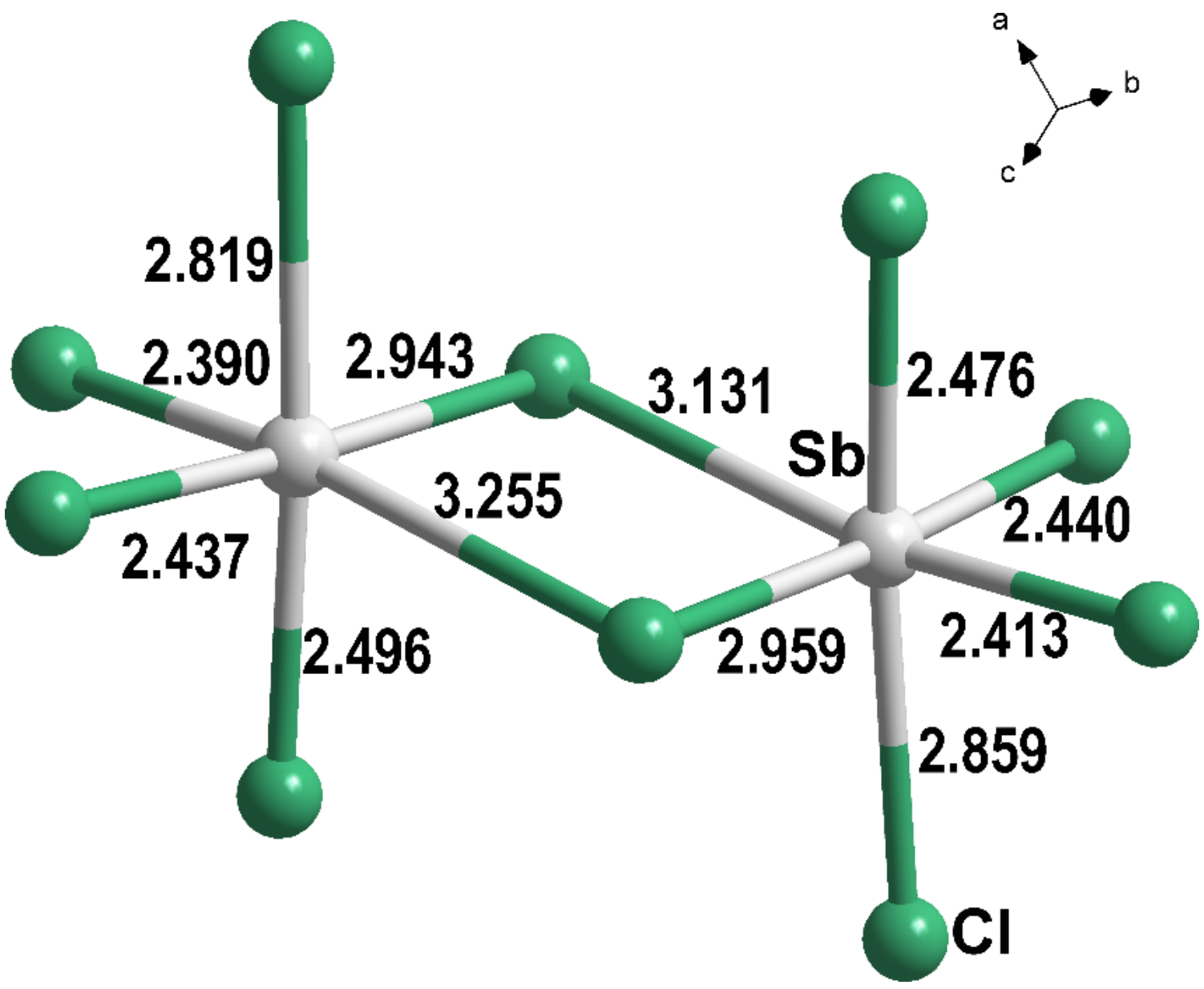

**Figure S14**: Bond length in Ångström in the $[Sb_2Cl_{10}]^{4-}$ anion in **3**.

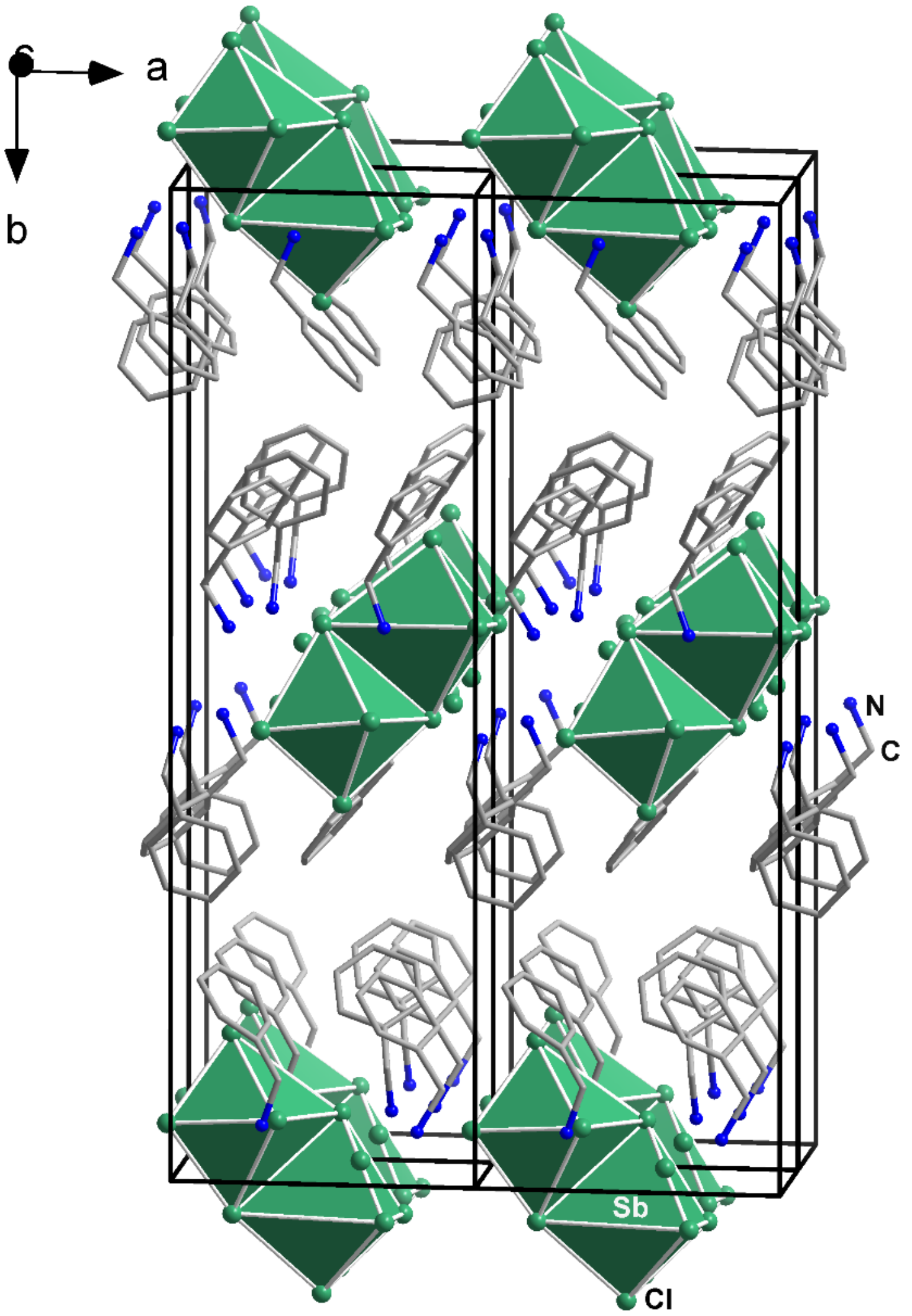

**Figure S15:** Packing diagram of **3** highlighting the pseudo-layered arrangement of cations and anions. Hydrogen atoms and cation disorder are omitted for clarity.

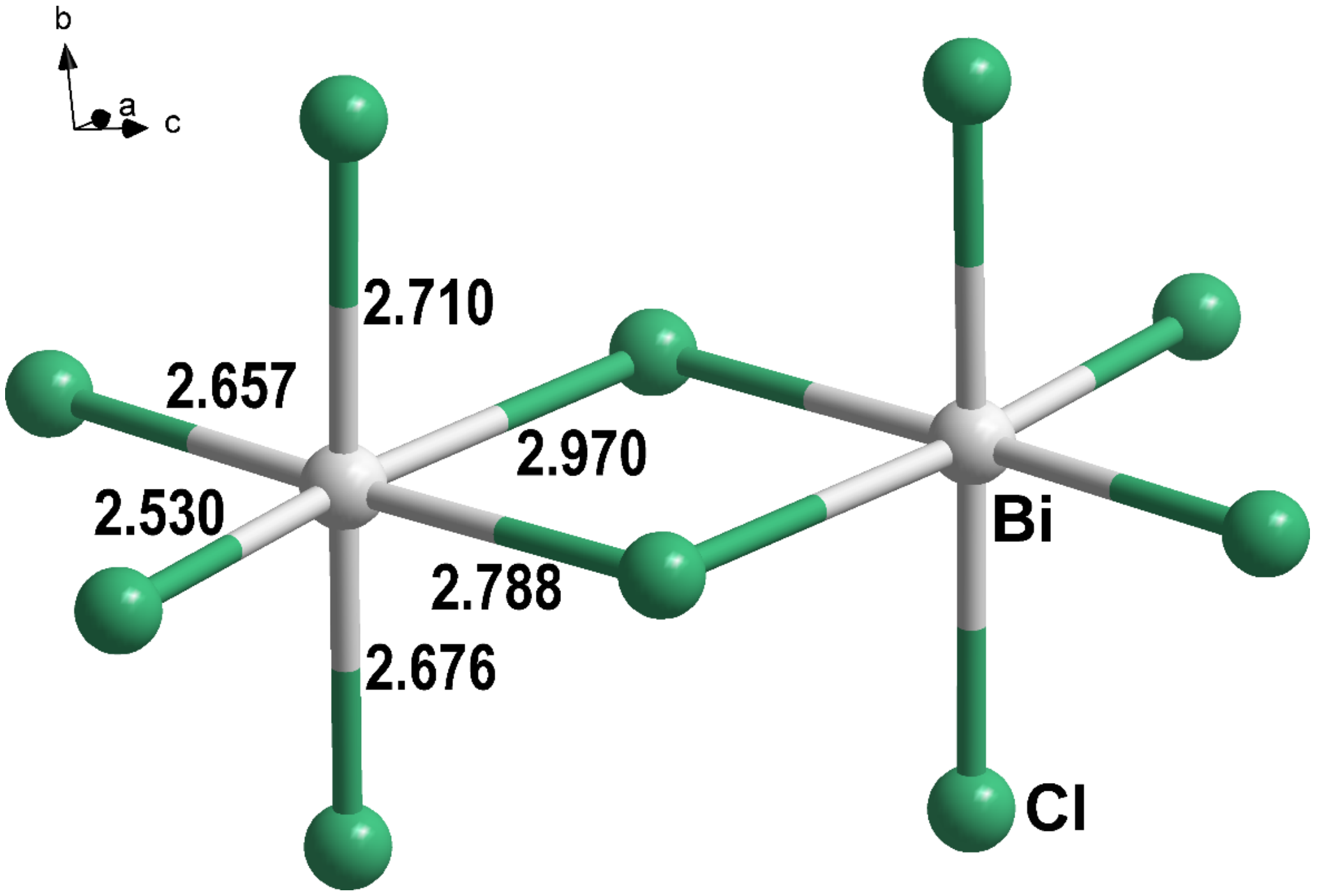

**Figure S16**: Bond length in Ångström in the $[Bi_2Cl_{10}]^{4-}$ anion in **4**.

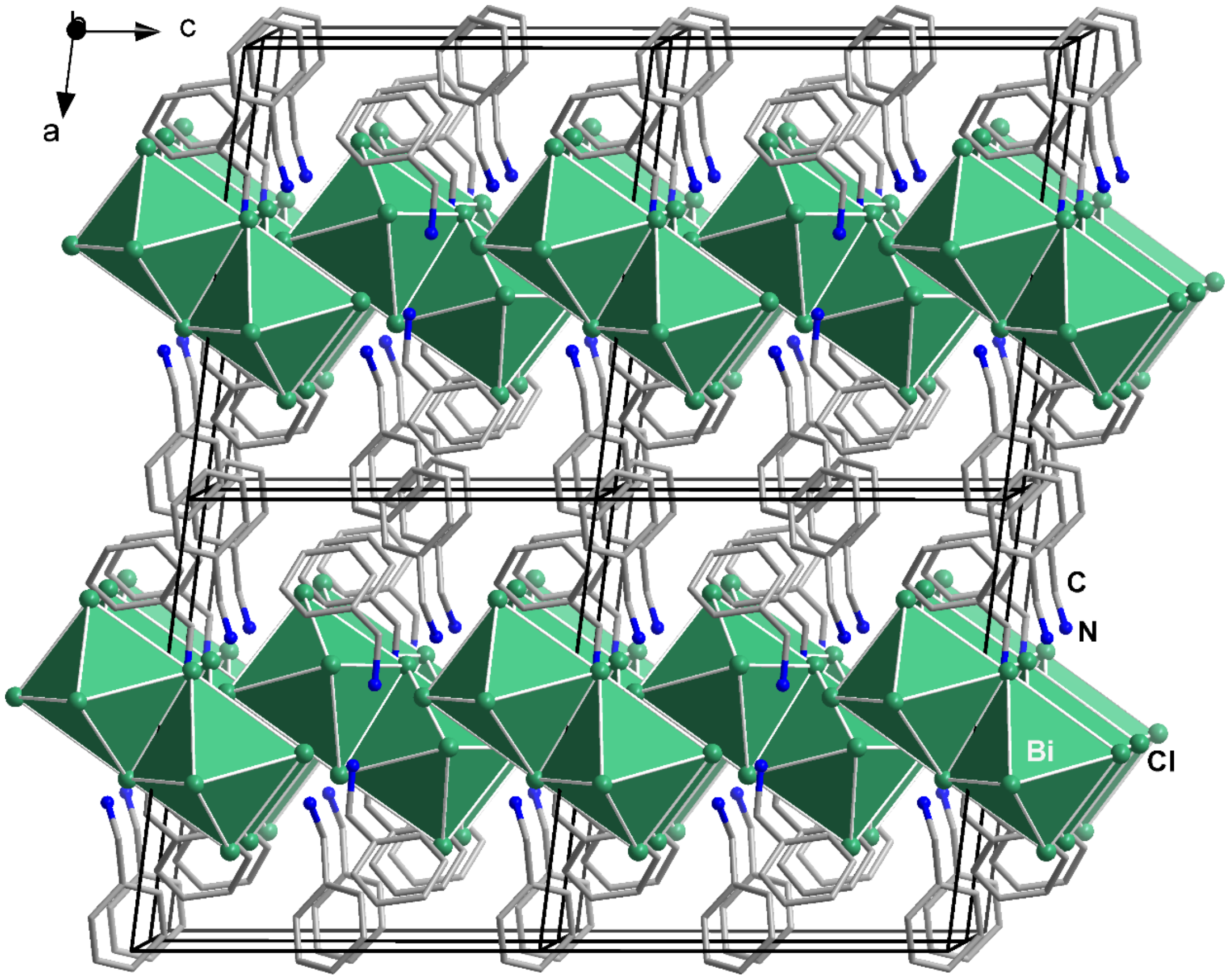

**Figure S17:** Packing diagram of **4** highlighting the pseudo-layered arrangement of cations and anions. Hydrogen atoms and cation disorder are omitted for clarity.

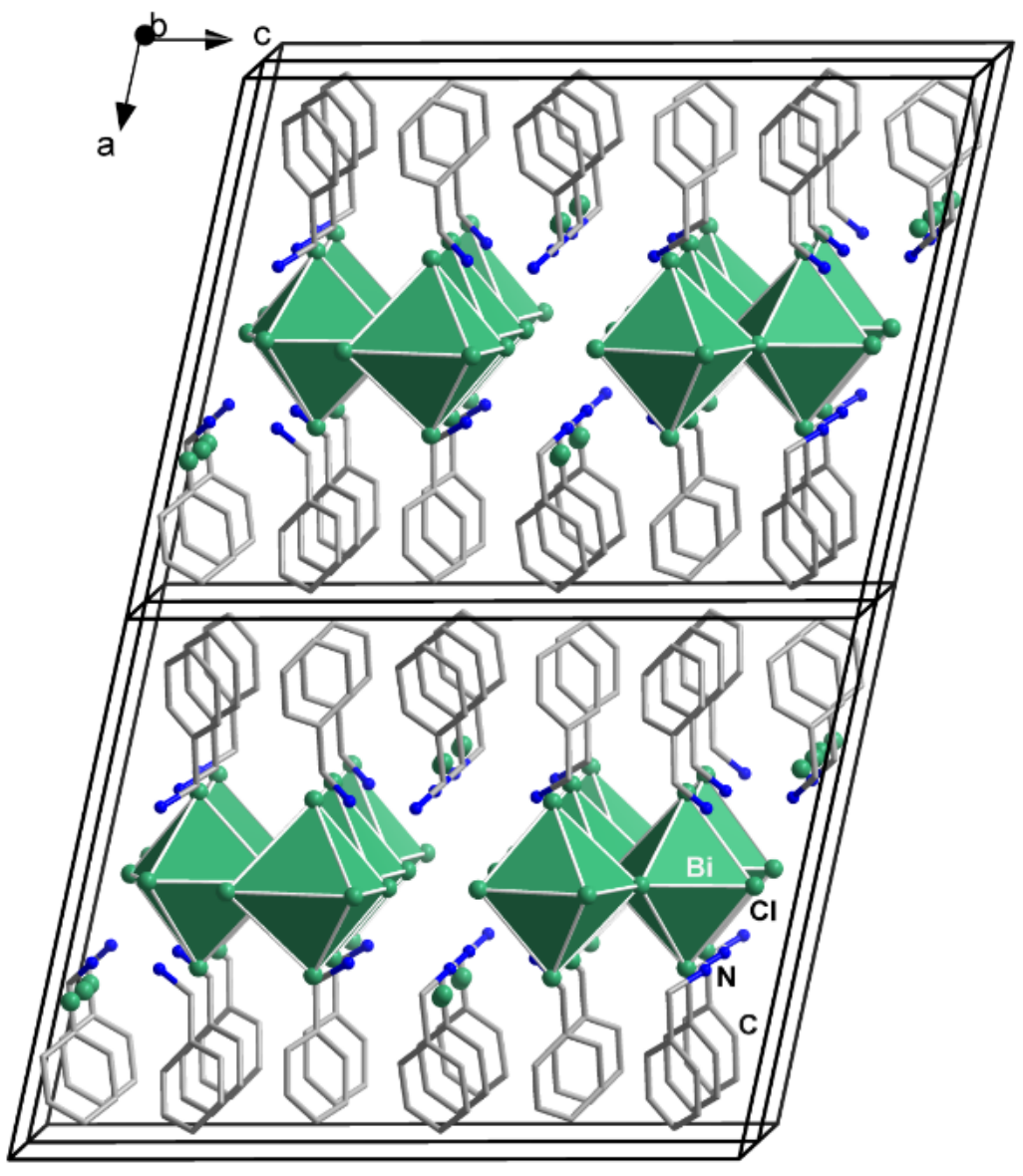

$[BZA]_3[BiCl_5]Cl$ (**5**)

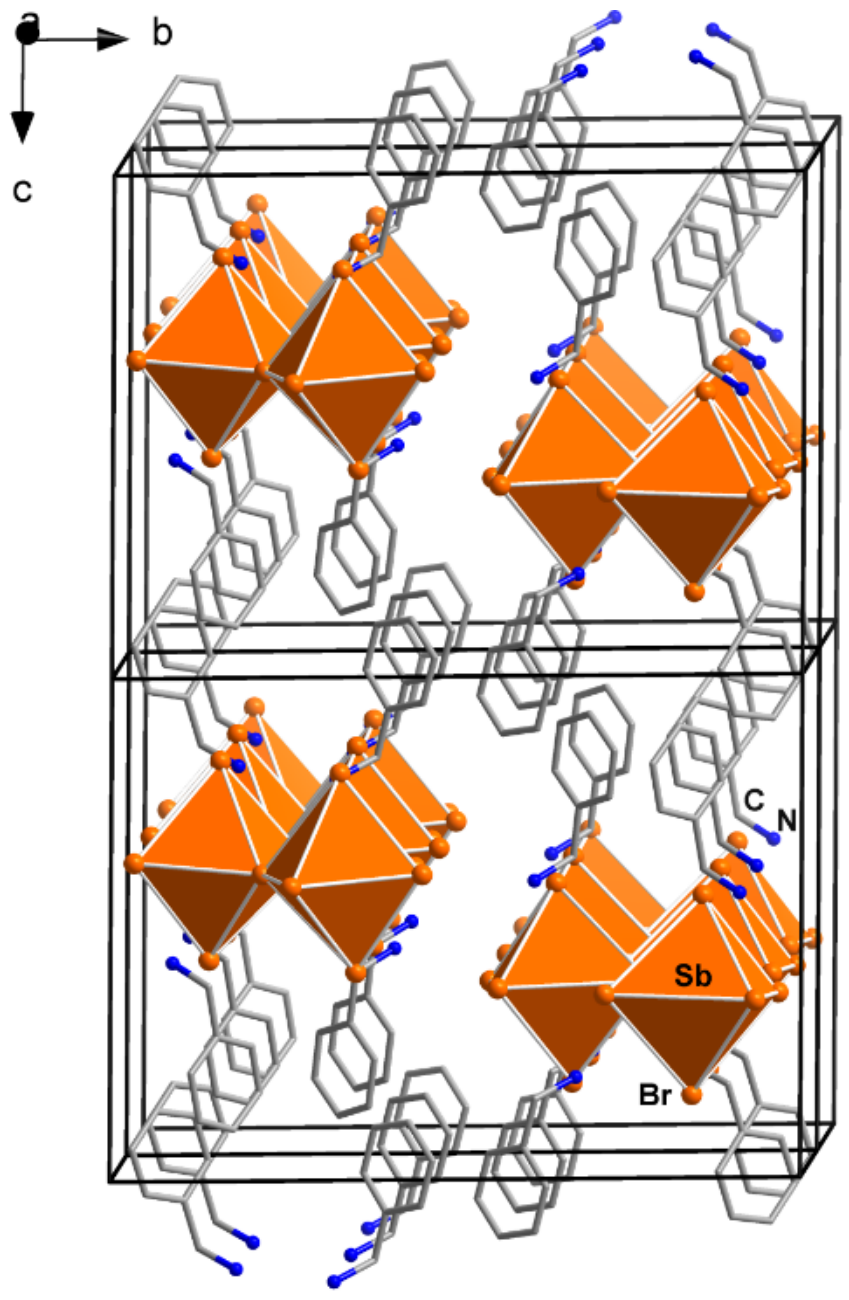

$[BZA]_2[SbBr_5]$ (**8**)

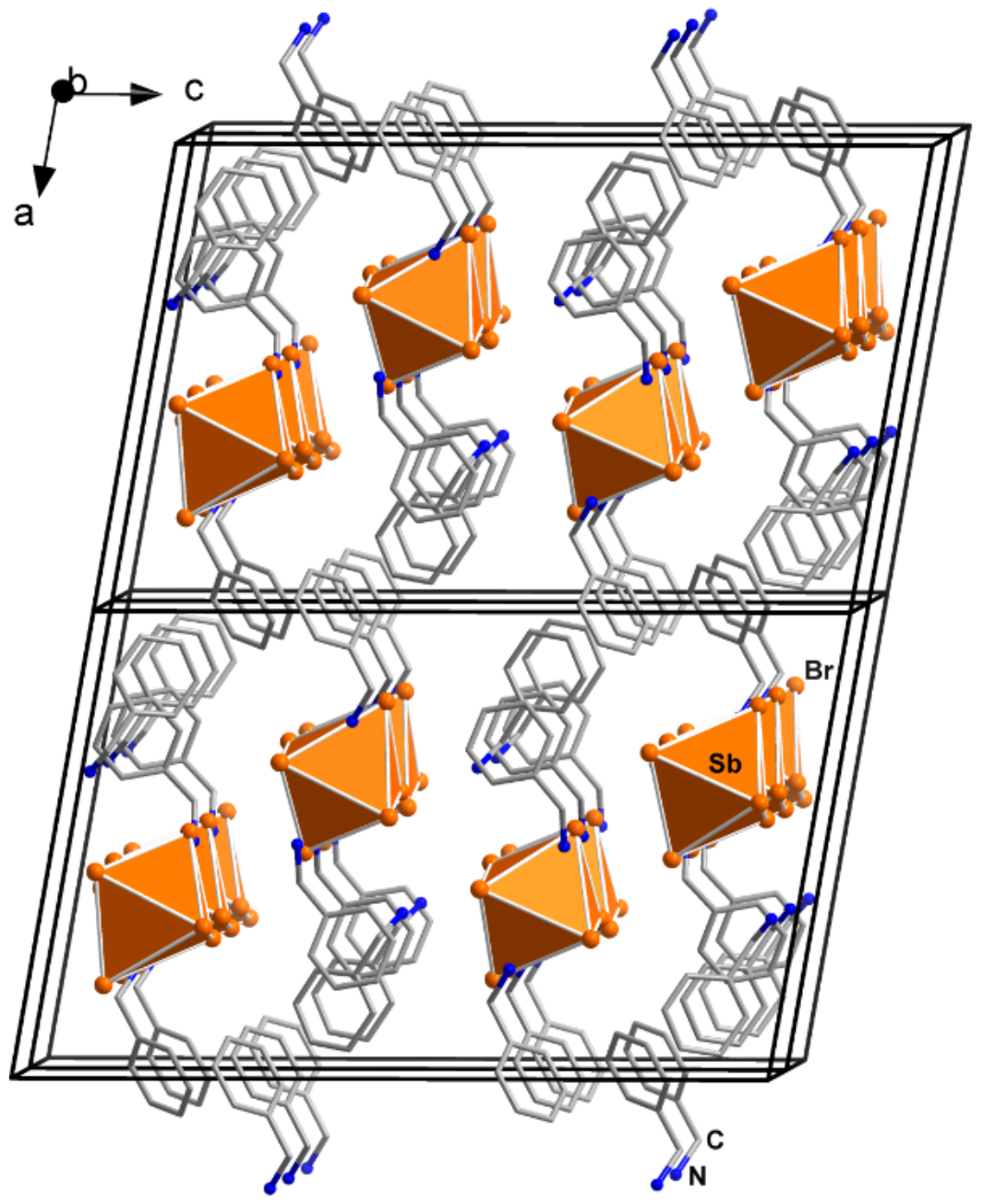

$[BZA]_3[SbBr_6]$ (**9**)

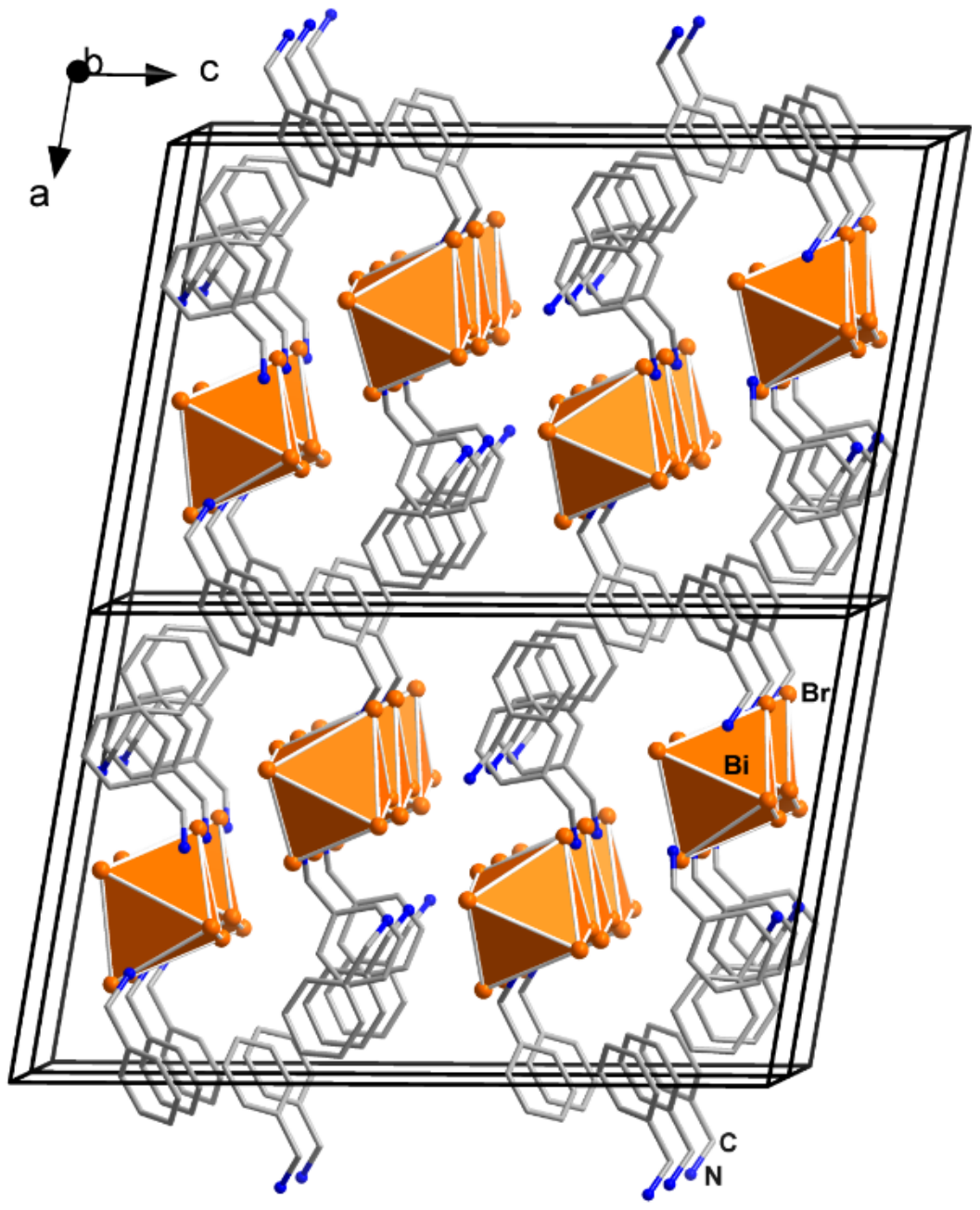

$[BZA]_3[BiBr_6]$ (**10**)

**Figure S18:** Overview of the packing diagrams of **5**,[1] **8**-**10**.[2,3] Hydrogen atoms are omitted for clarity.

# 4 Powder Diffraction

Powder patterns were recorded on a STADI MP (STOE Darmstadt) powder diffractometer, with CuKα1 radiation with λ= 1.54056 Å at room temperature in transmission mode from 3 to 70° in 2θ, although in the patterns shown below no major reflections are found beyond 50° in 2θ. The patterns confirm the presence of the phase determined by SCXRD measurement and the absence of any major crystalline by-product.

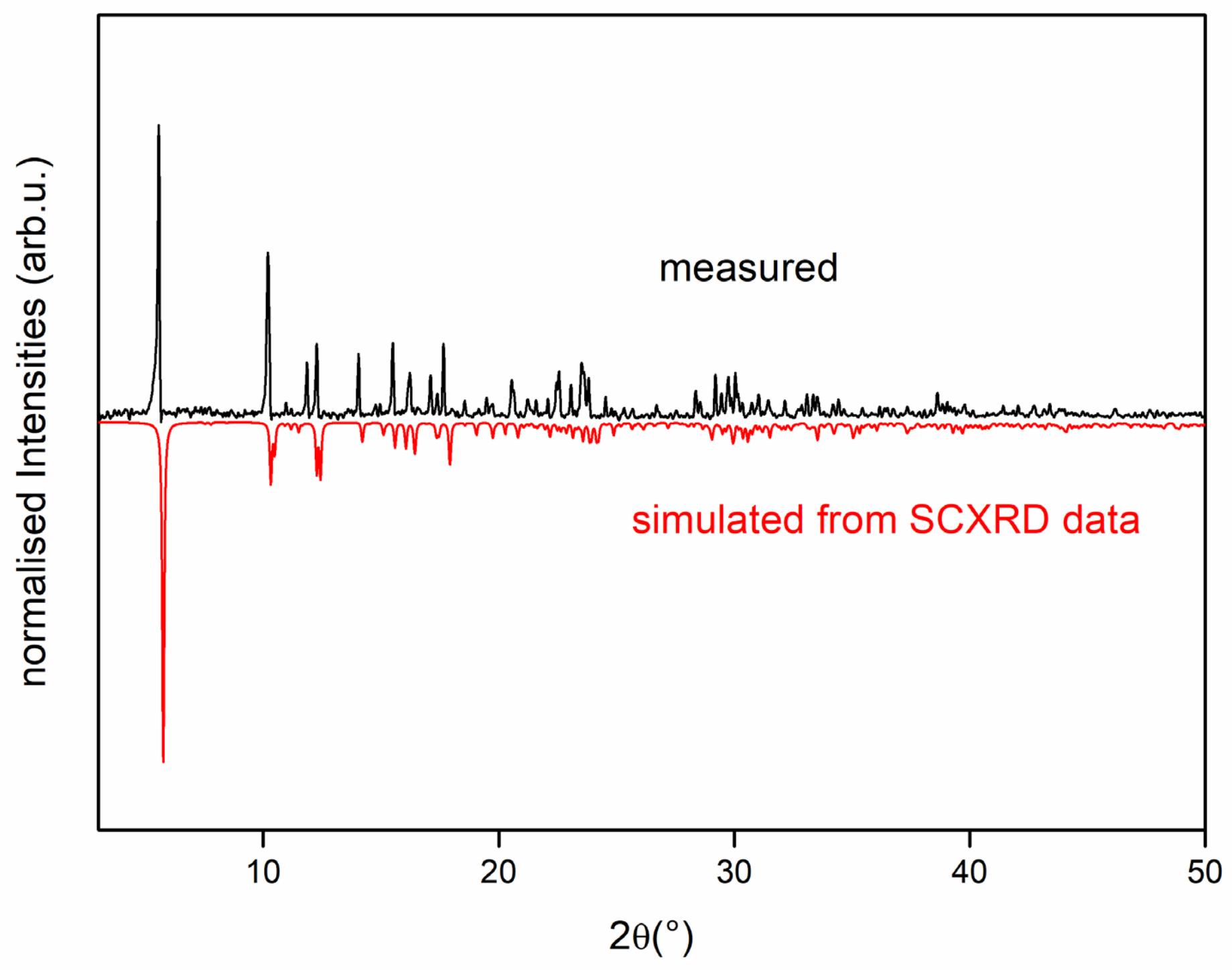

**Figure S20.** Powder diffraction pattern of $[CMA]_4[Sb_2Cl_{10}]$ (**1**).

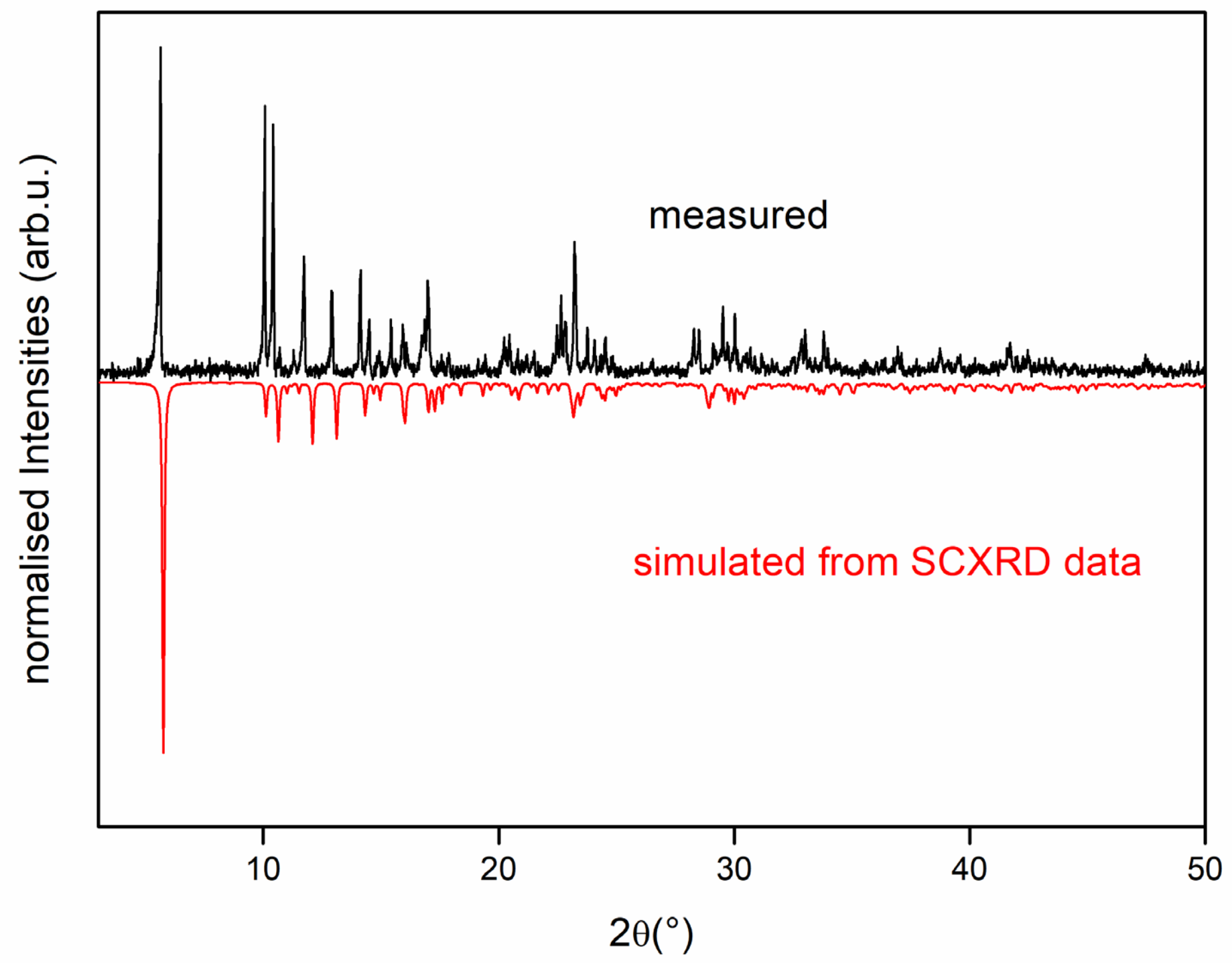

**Figure S21.** Powder diffraction pattern of $[CMA]_4[Bi_2Cl_{10}]$ (**2**).

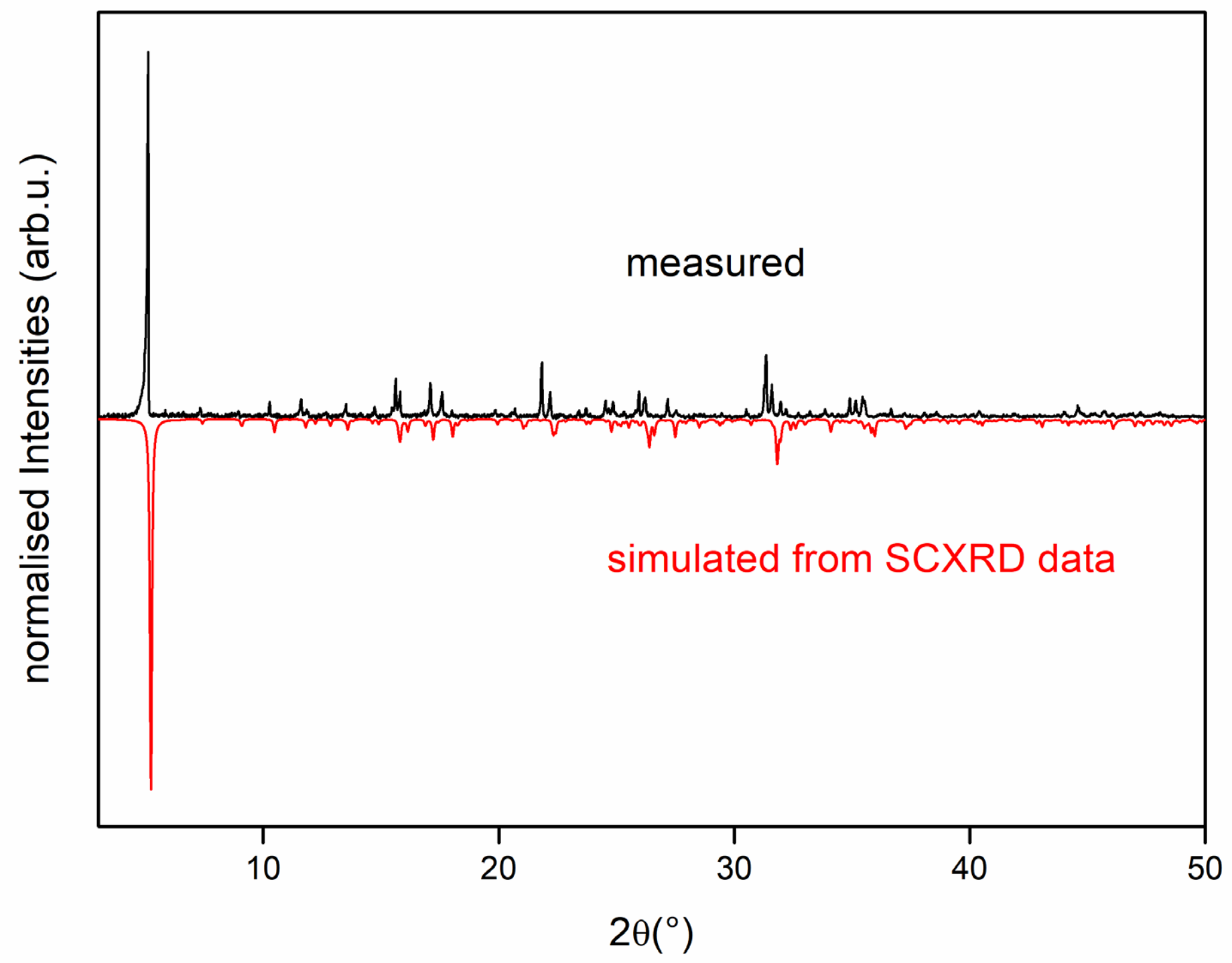

**Figure S22.** Powder diffraction pattern of $[CMA]_3[SbBr_5]Br$ (**6**).

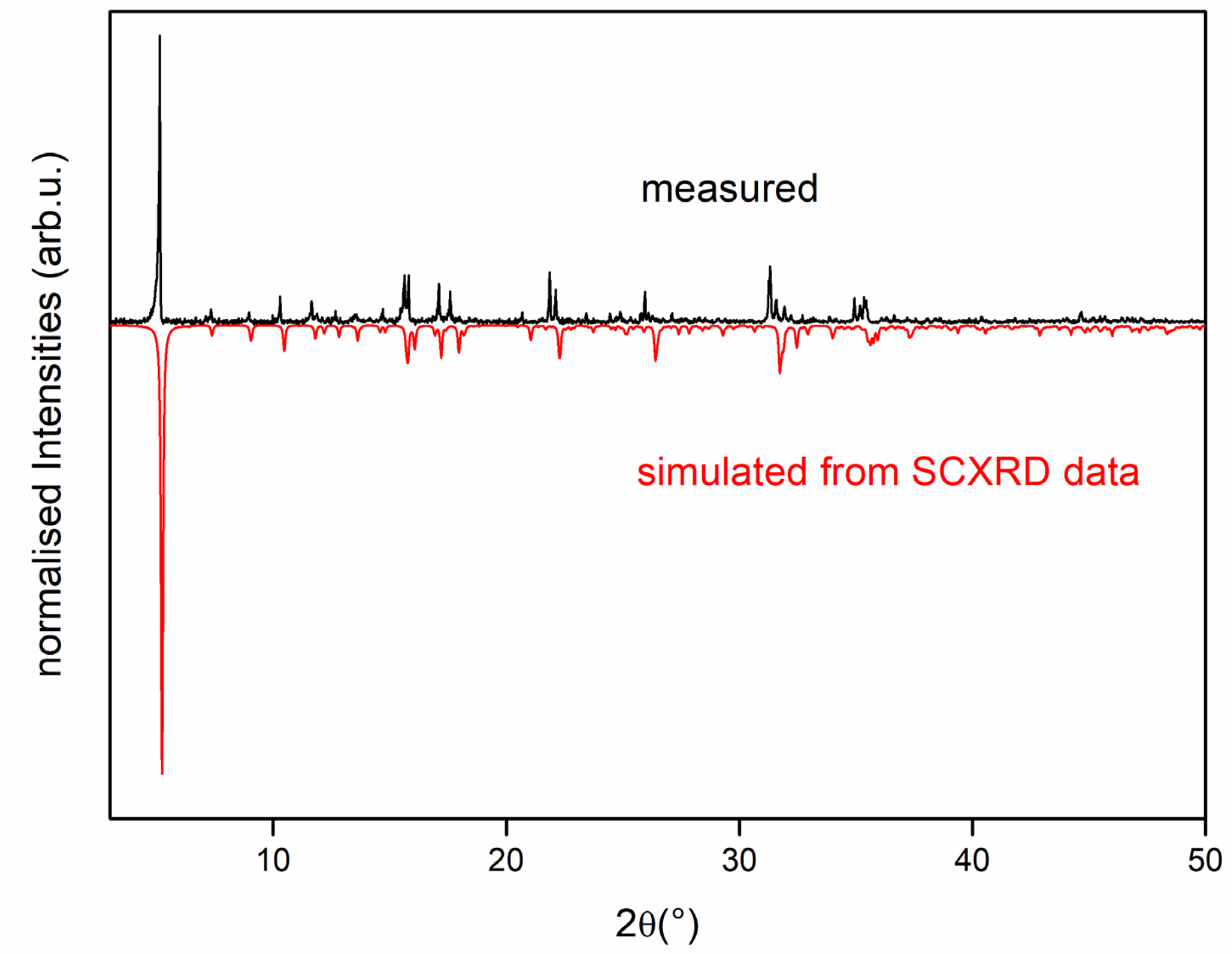

**Figure S23:** Powder diffraction pattern of $[CMA]_3[BiBr_5]Br$ (**7**).

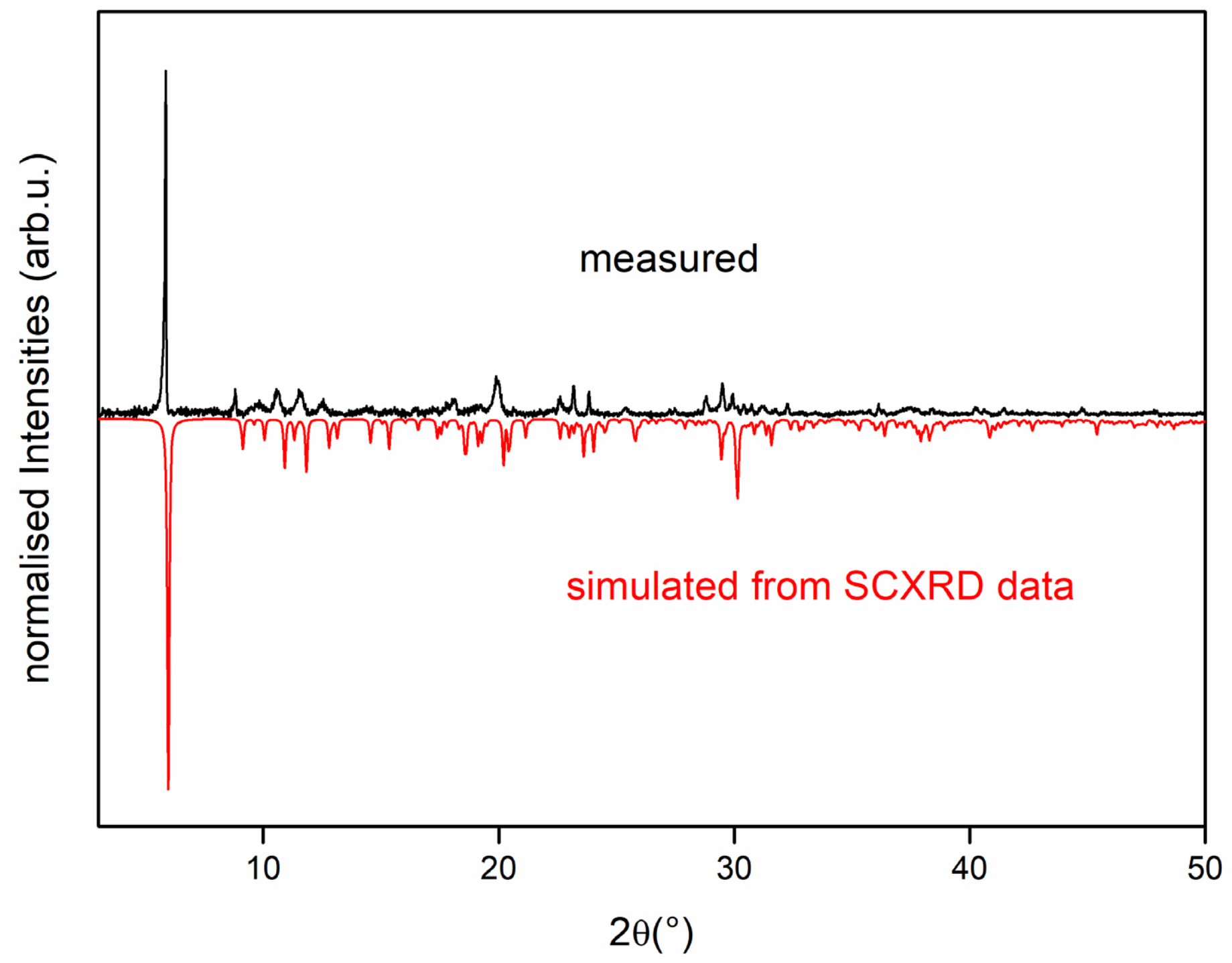

**Figure S24:** Powder diffraction pattern of $[BZA]_6[Sb_2Cl_{10}]Cl_2$ (**3**).

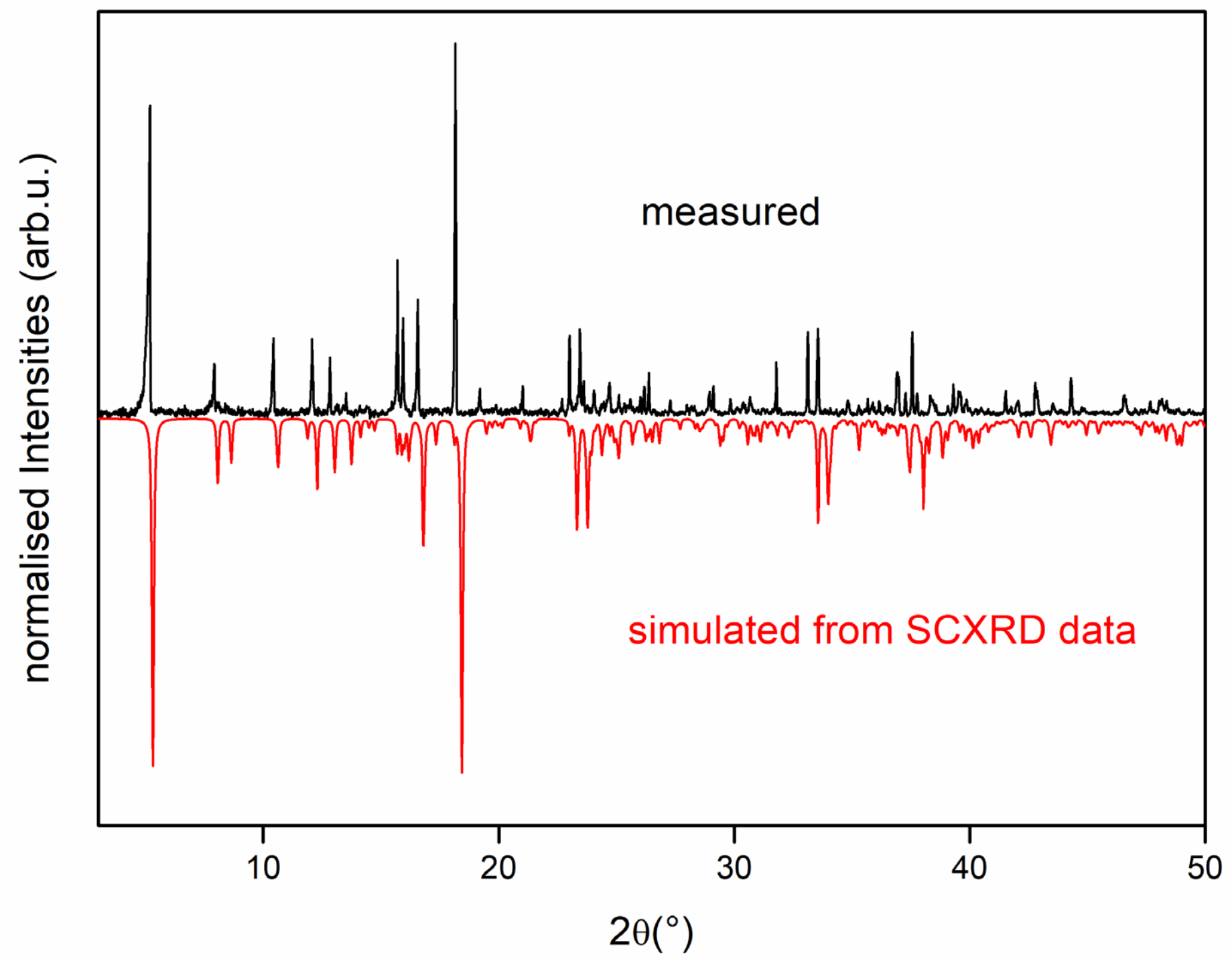

**Figure S25**: Powder diffraction pattern of $[BZA]_4[Bi_2Cl_{10}]$ (**4**).

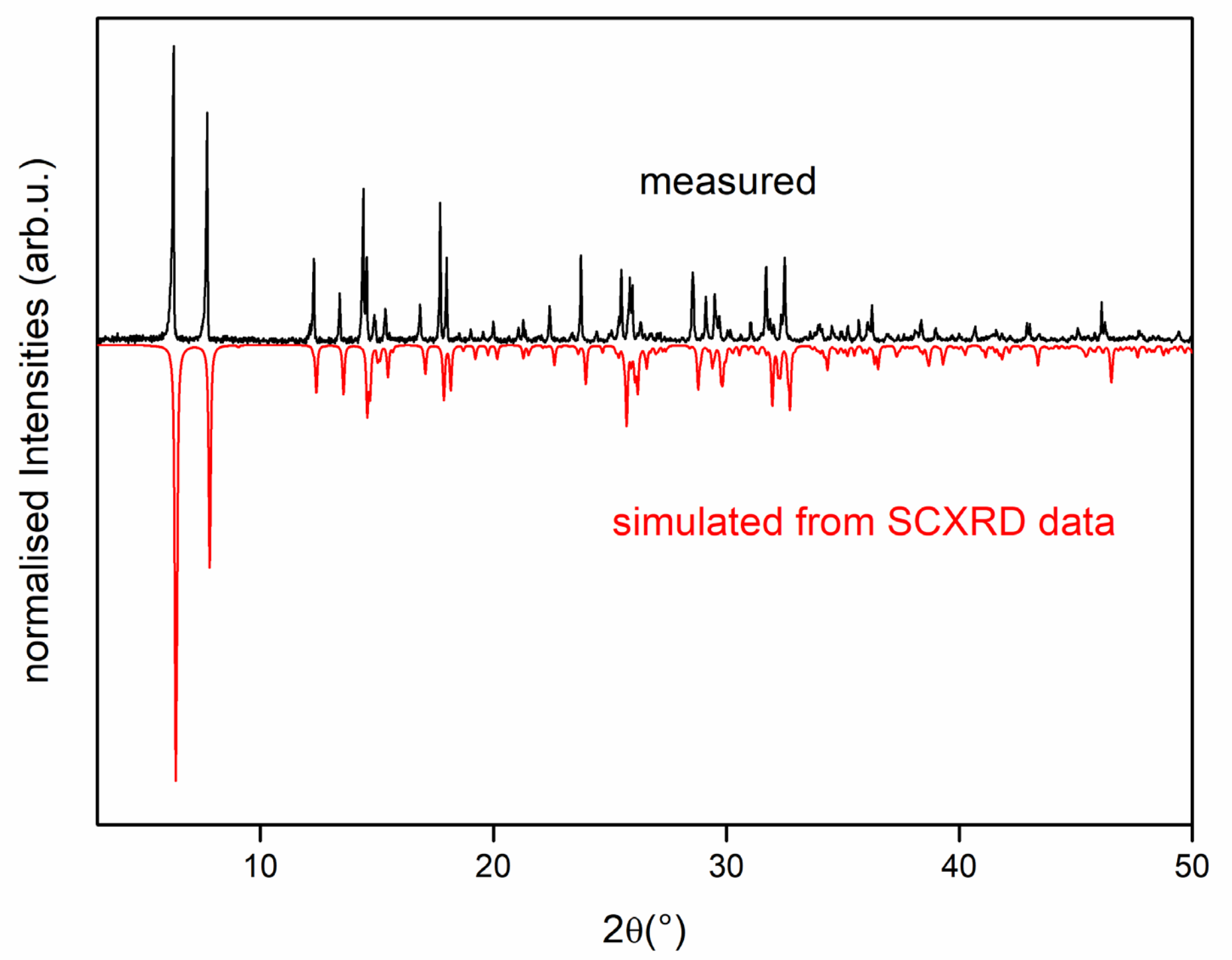

**Figure S26:** Powder diffraction pattern of $[BZA]_2[SbBr_5]$ (**8**) compared with the CSD entry MIFQUQ. [2]

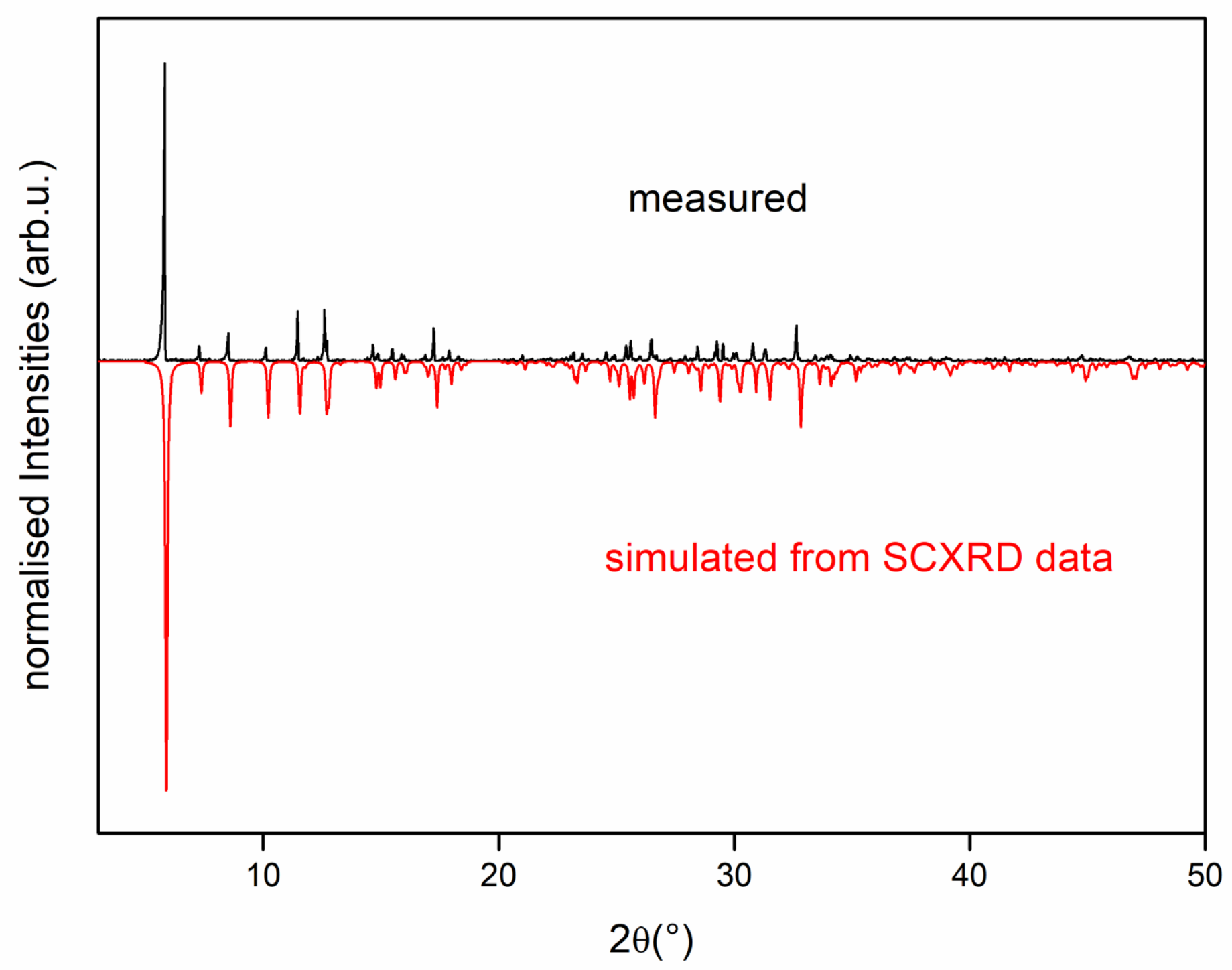

**Figure S27:** Powder diffraction pattern of $[BZA]_3[SbBr_6]$ (**9**) compared with the CSD entry OHUBIF. [3]

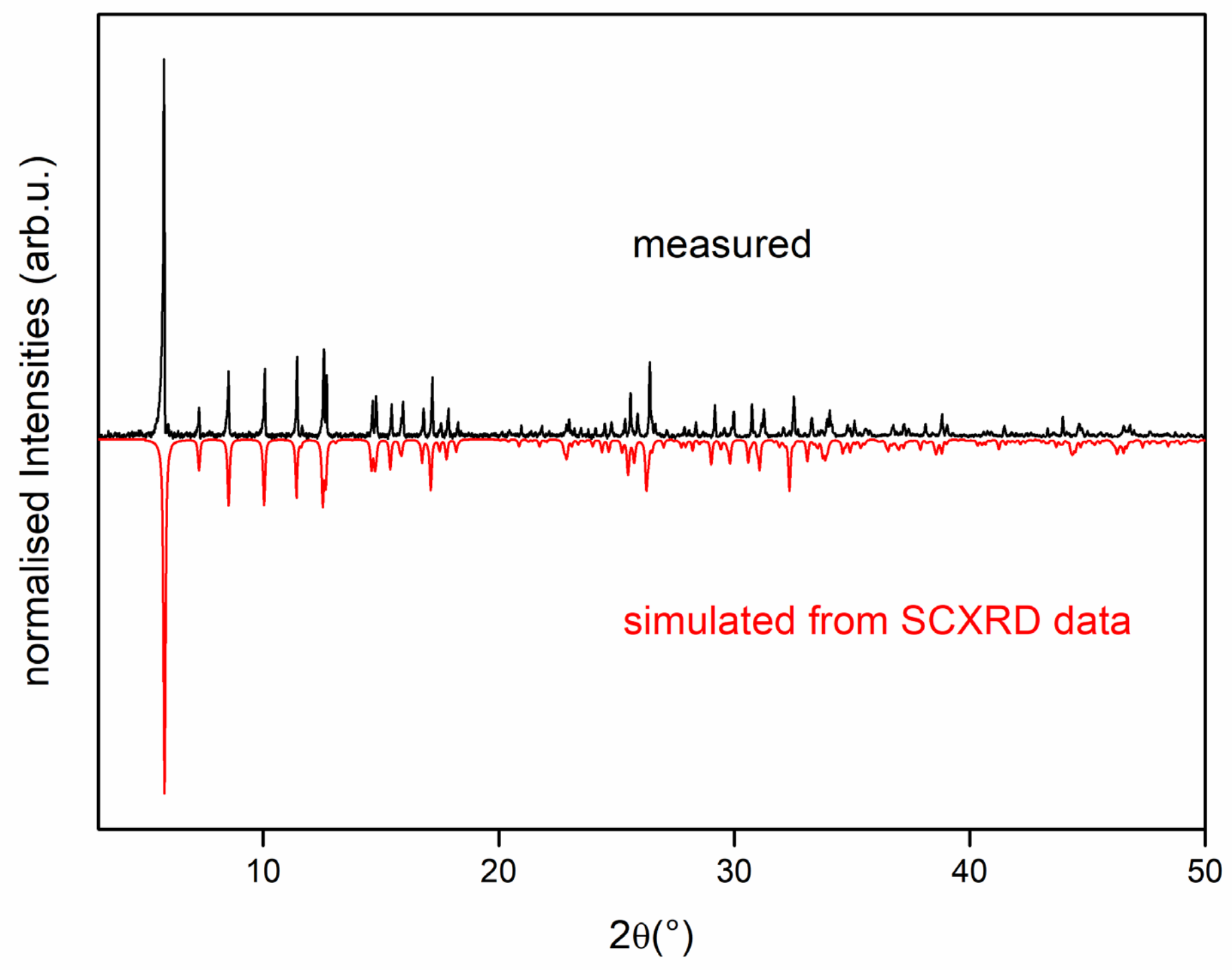

**Figure S28**: Powder diffraction pattern of $[BZA]_3[BiBr_6]$ (**10**)compared with the CSD entry OHUBEB. [3]

# 5 IR Spectroscopy

IR spectra were recorded on a Bruker Tensor 37 FT-IR spectrometer equipped with an ATR Platinum measuring unit.

In the spectra shown below, typical bands of the organic cation are observed, similar to findings for other cyclohexylmethylammonium and benzylammonium halide and metal halide compounds. [10–12] Structured N-H bands are observed around 3000 cm$^{-1}$. In the spectra of **1** and **2** no O-H bands appear between 4000 and 3500 cm$^{-1}$, indicating a loss of water in these compounds upon drying, as also indicated by the thermal analysis (see below).

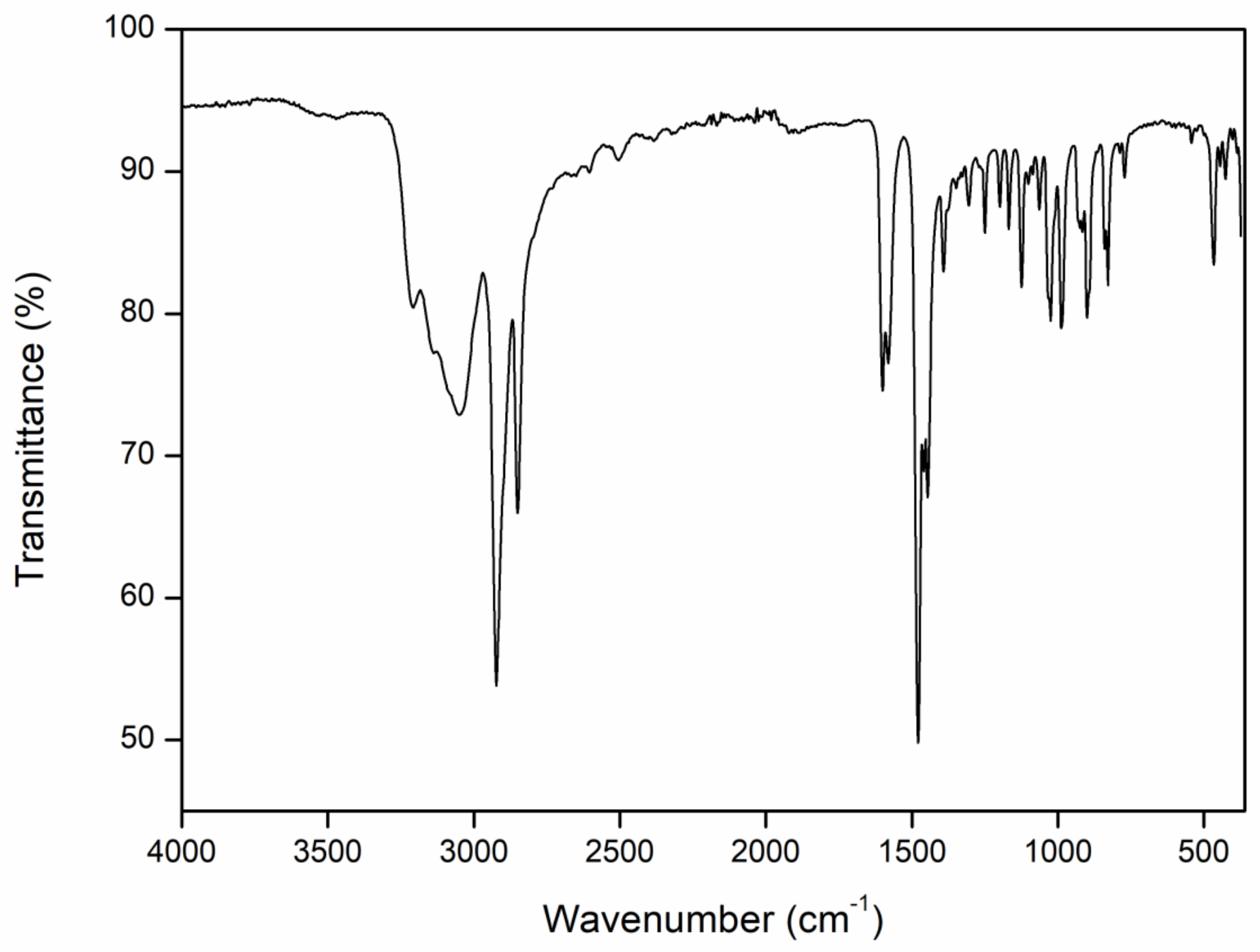

**Figure S29:** IR spectra of compound $[CMA]_4[Sb_2Cl_{10}]$ (**1**).

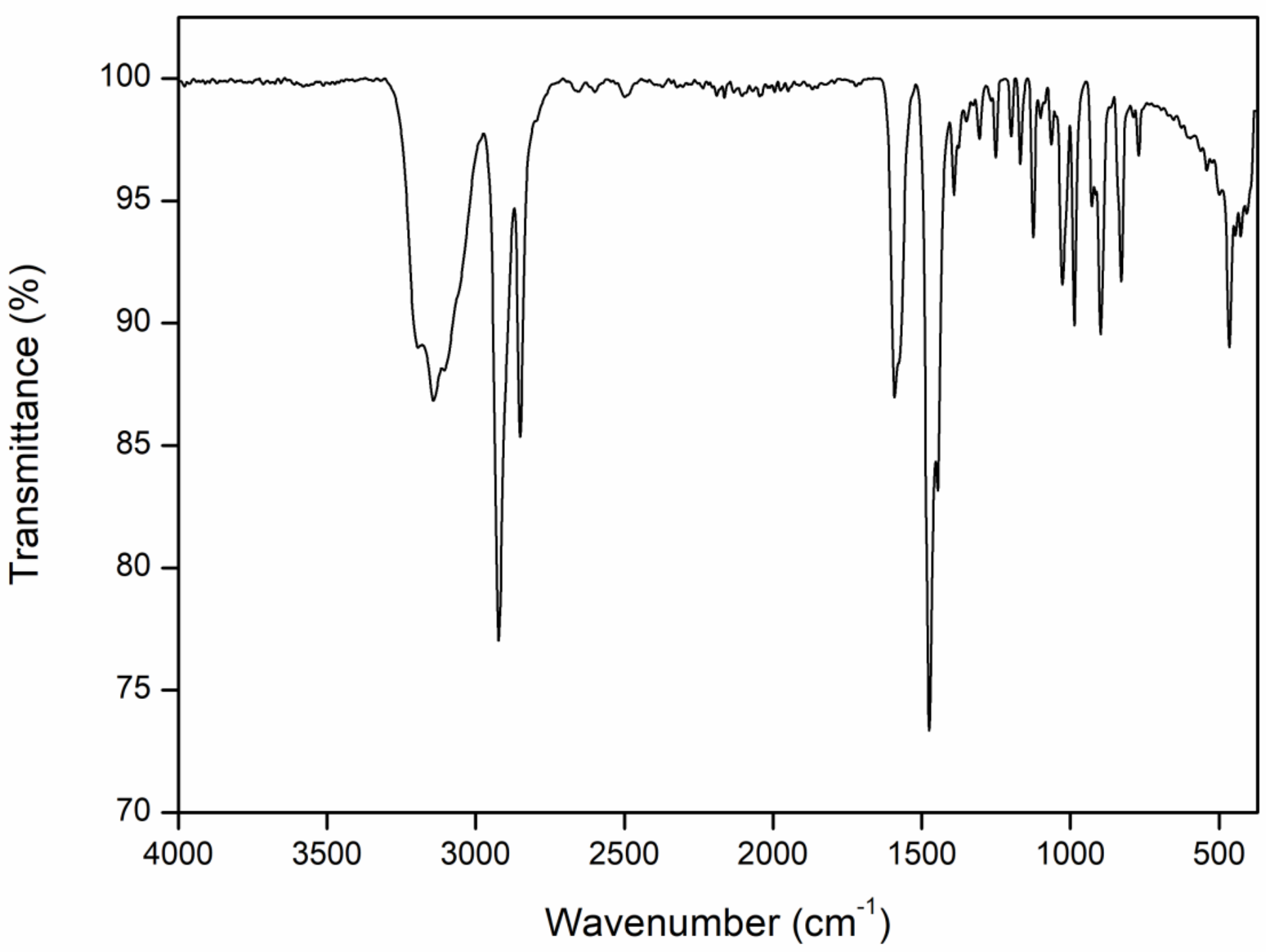

**Figure S30.** IR spectra of compound $[CMA]_4[Bi_2Cl_{10}]$ (**2**).

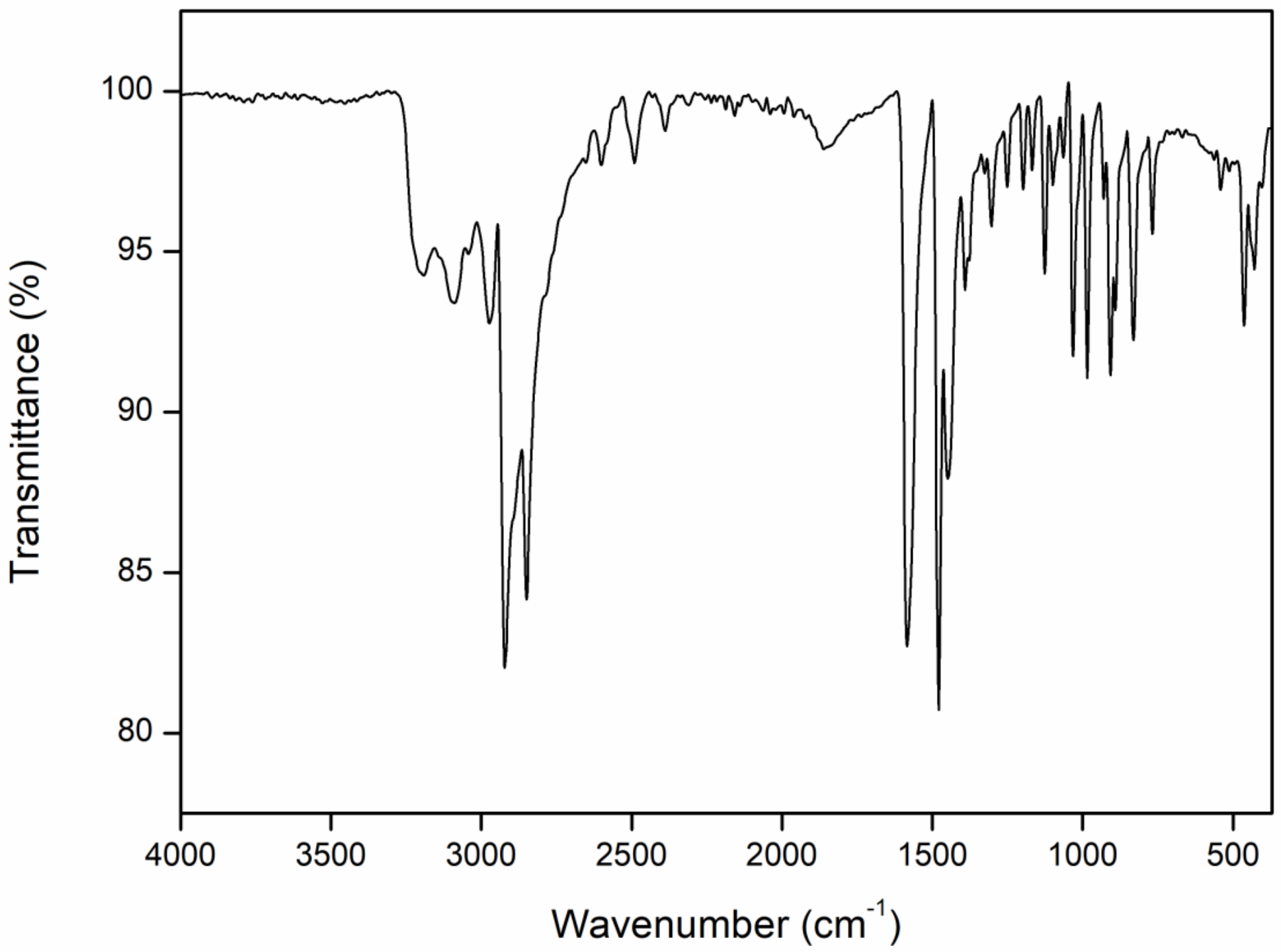

**Figure S31.** IR spectra of compound $[CMA]_3[SbBr_5]Br$ (**6**).

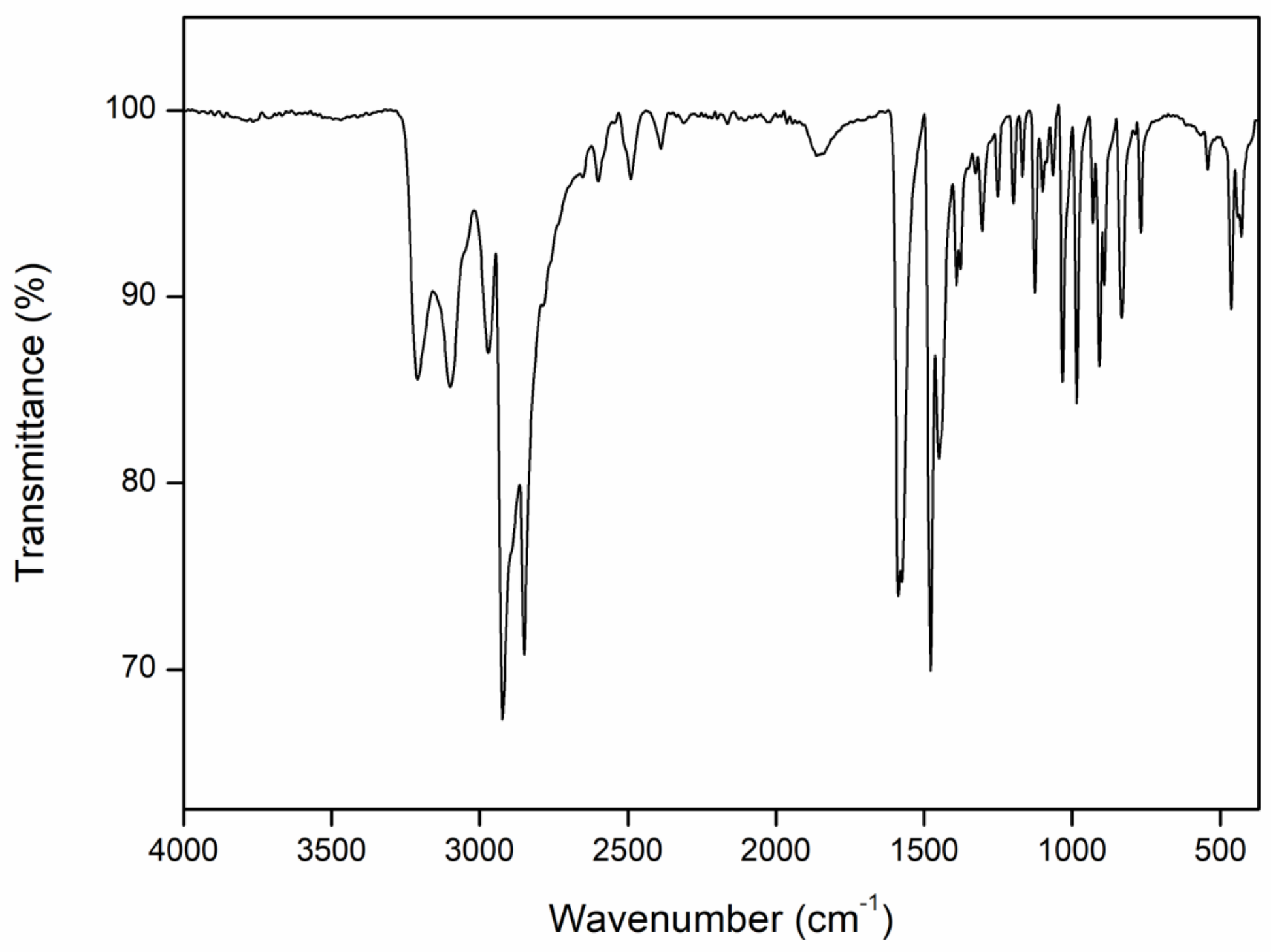

**Figure S32.** IR spectra of compound $[CMA]_3[BiBr_5]Br$ (**7**).

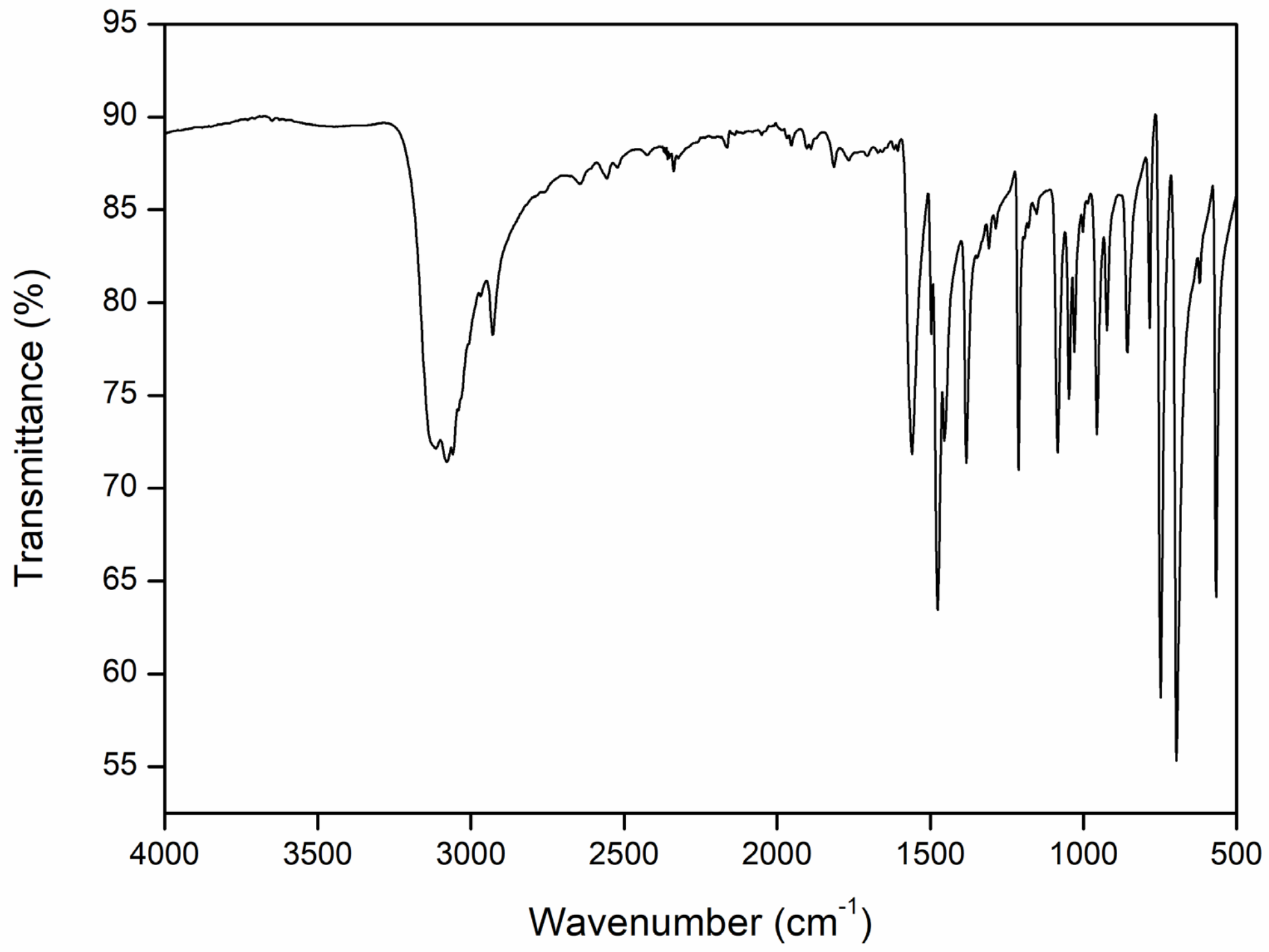

**Figure S33:** IR spectra of compound $[BZA]_6[Sb_2Cl_{10}]Cl_2$ (**3**).

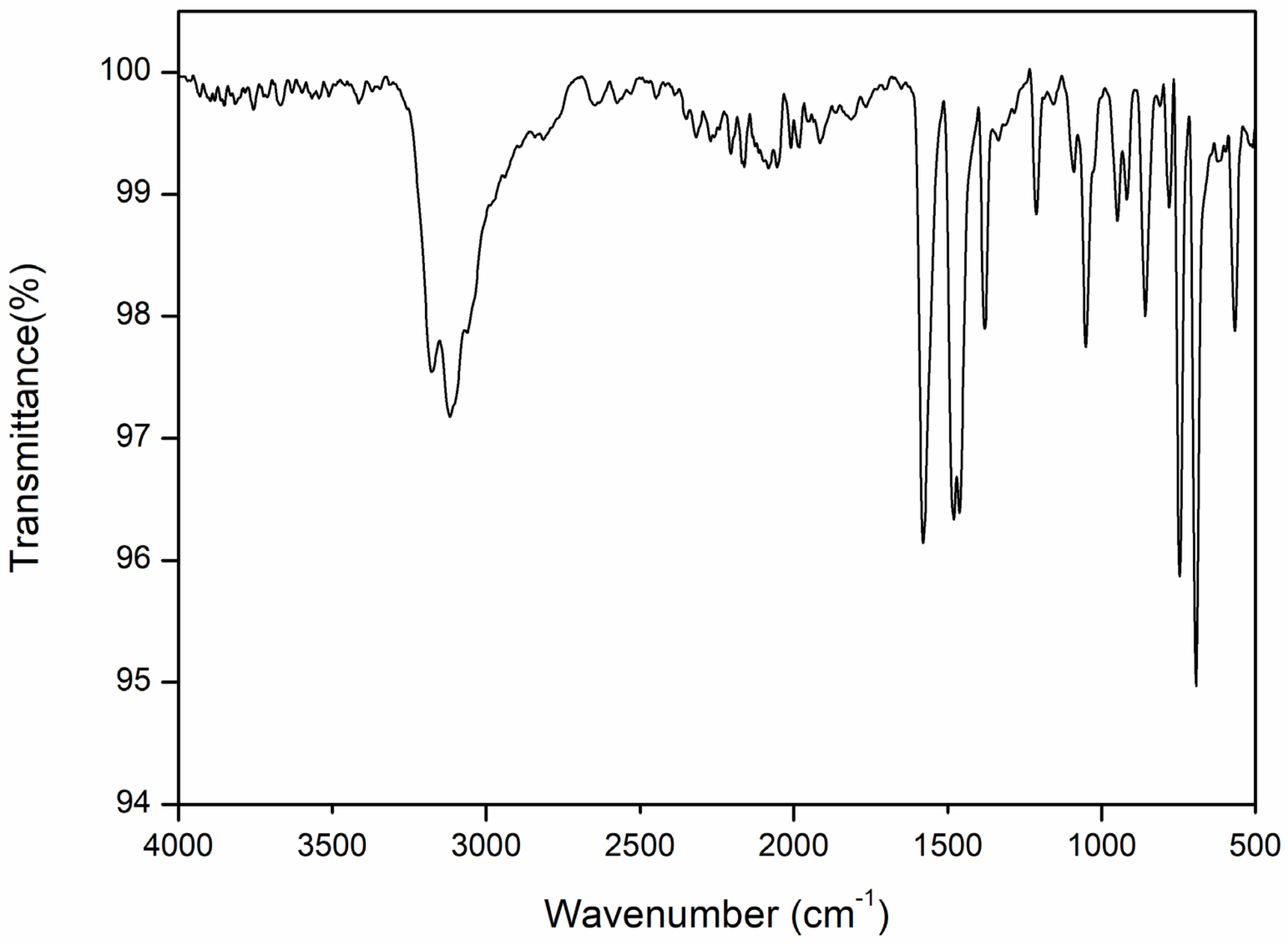

**Figure S34:** IR spectra of compound $[BZA]_4[Bi_2Cl_{10}]$ (**4**).

# 6 Thermal Analysis

Thermal behaviour was studied on a Netzsch STA 409 CD from 25 to 1000 °C with a heating rate of 10 °C $min^{-1}$ in a constant flow of 150 mL $min^{-1}$ $N_2$.

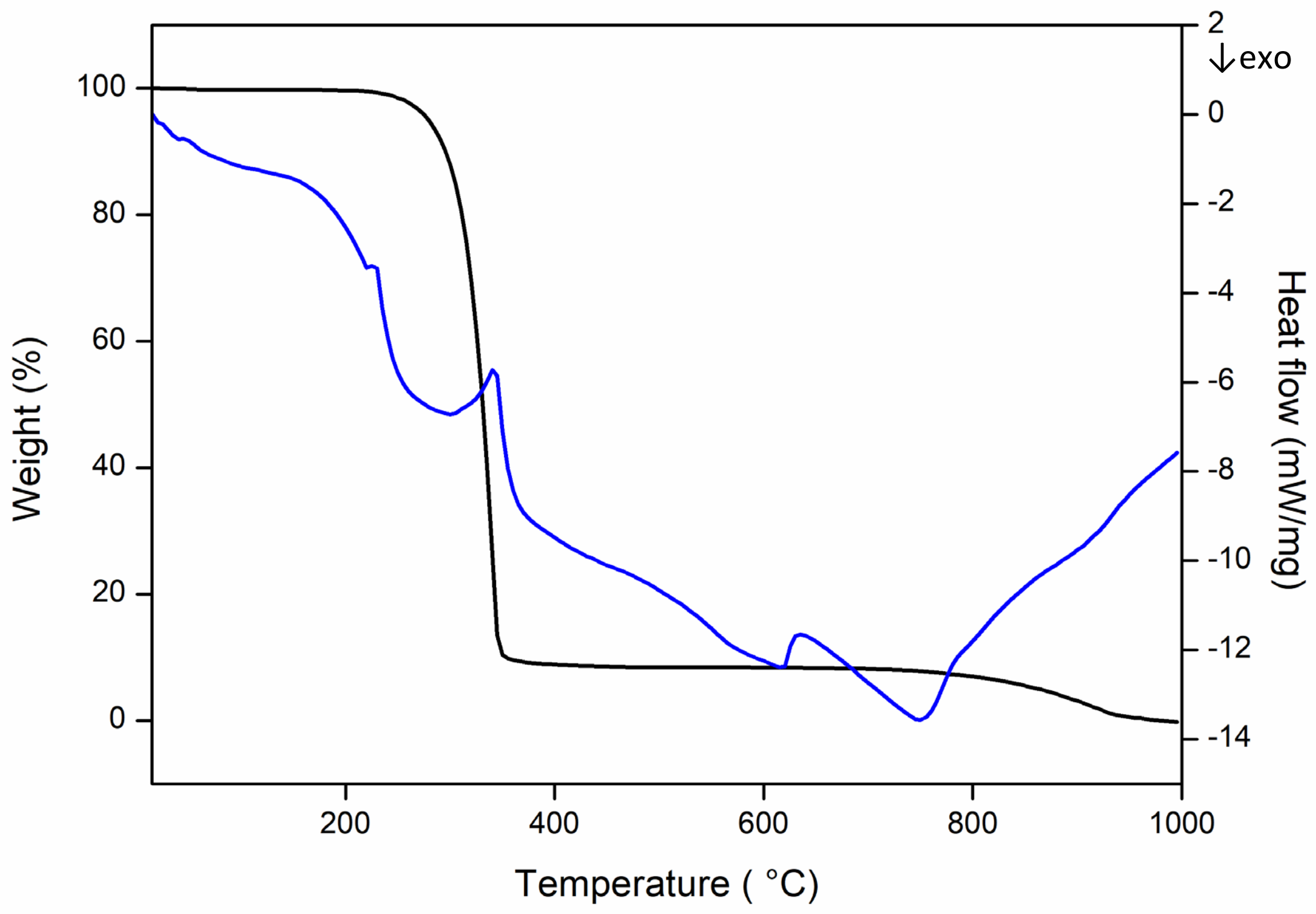

**Figure S35.** TGA and DSC of compound $[CMA]_4[Sb_2Cl_{10}]$ (**1**) with a mass loss of 99.05 % observed starting at 315.8 °C.

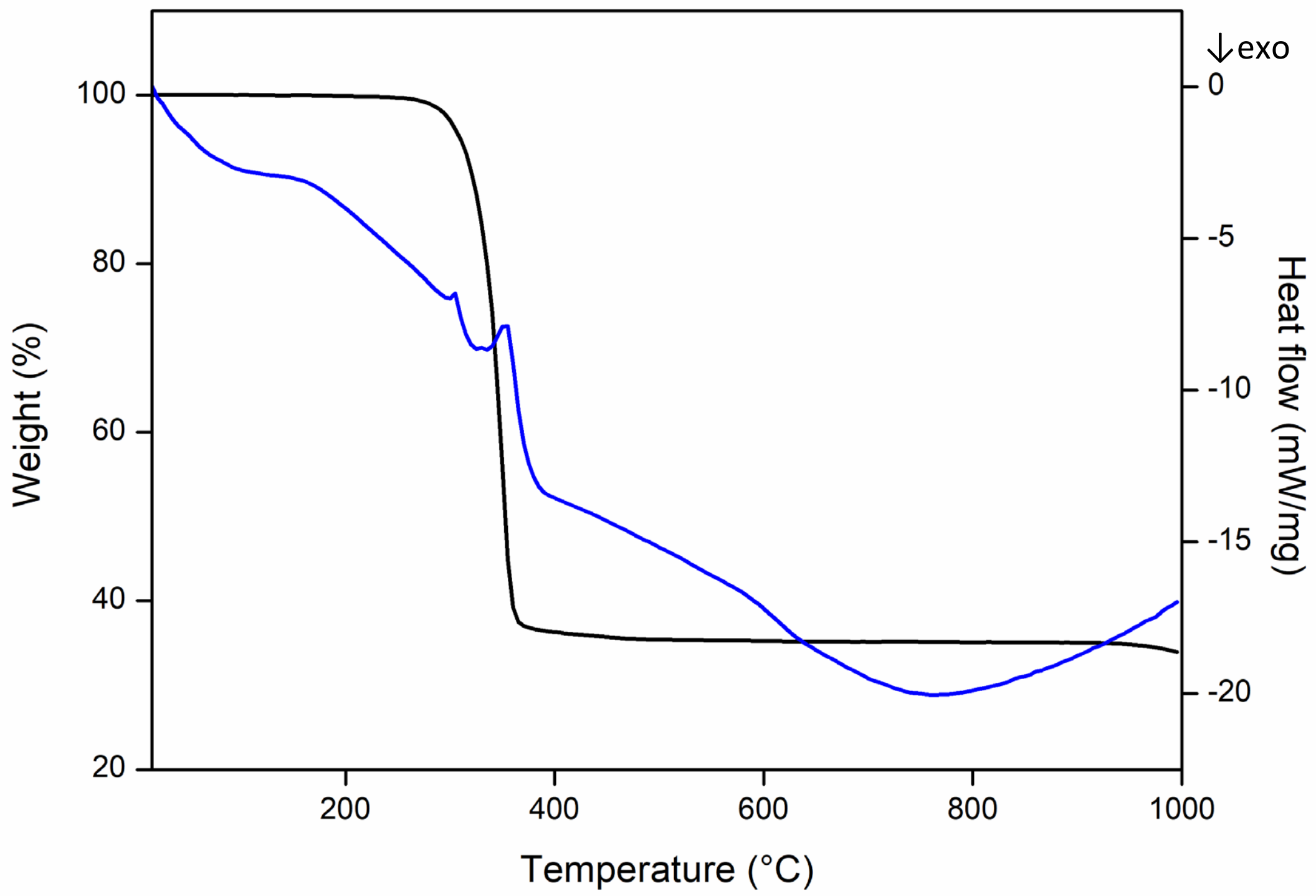

**Figure S36.** TGA and DSC of compound $[CMA]_4[Bi_2Cl_{10}]$ (**2**) with a mass loss of 64.5 % observed starting at 319.9 °C.

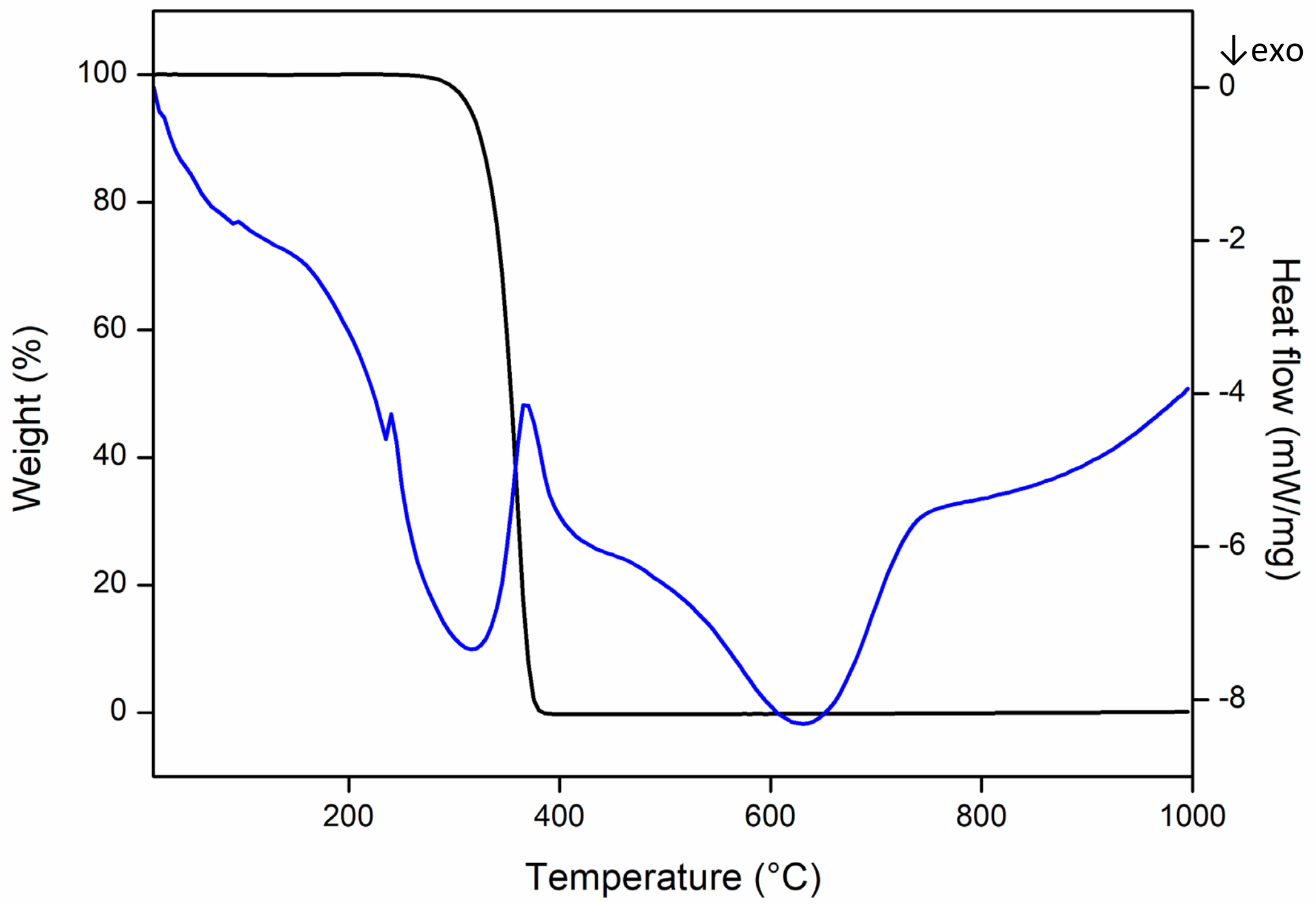

**Figure S37.** TGA and DSC of compound $[CMA]_3[SbBr_5]Br$ (**6**) with a mass loss of 99.94 % observed starting at 334.7 °C.

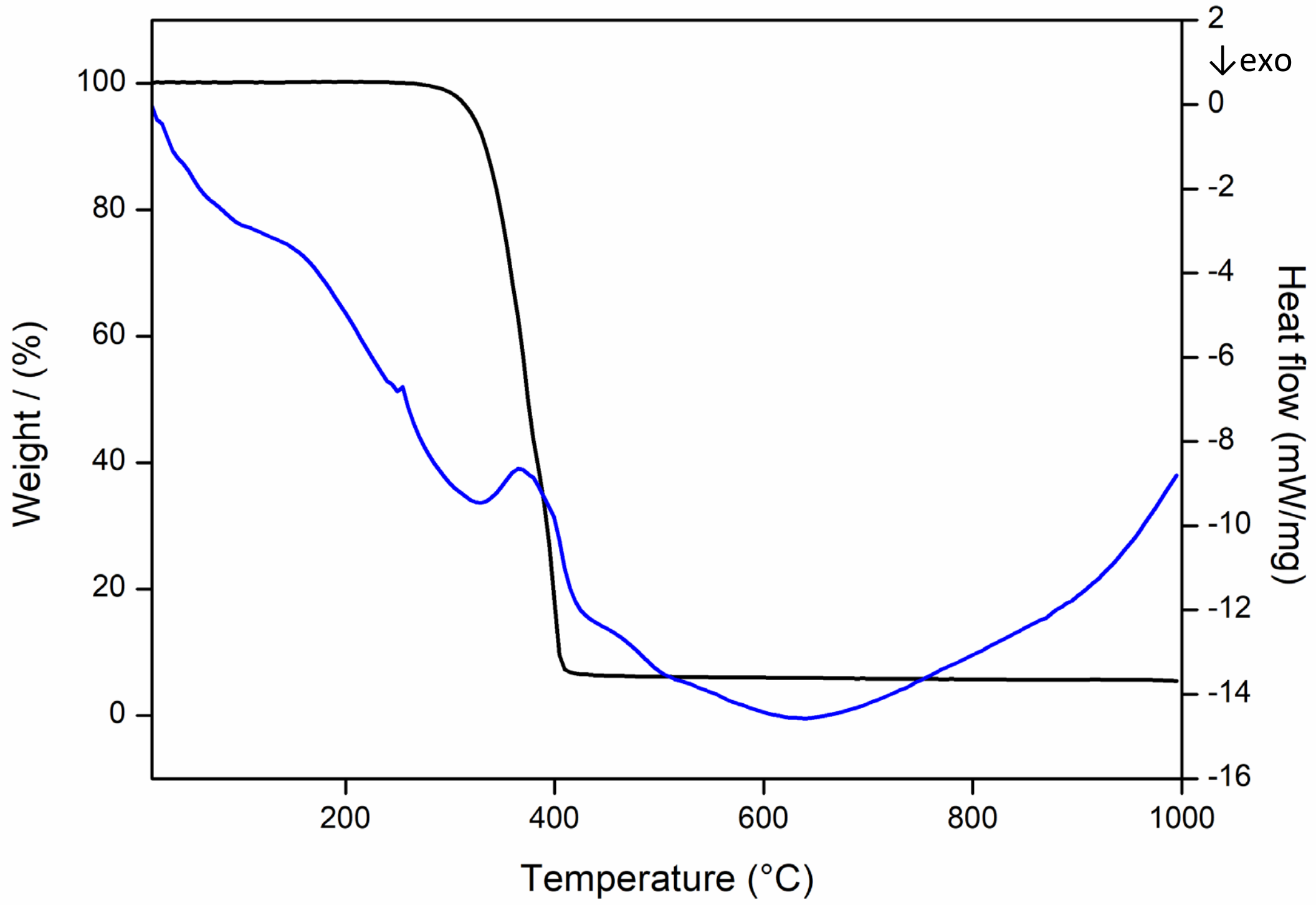

**Figure S38.** TGA and DSC of compound $[CMA]_3[BiBr_5]Br$ (**7**) with a mass loss of 93.98 % observed starting at 383.0 °C.

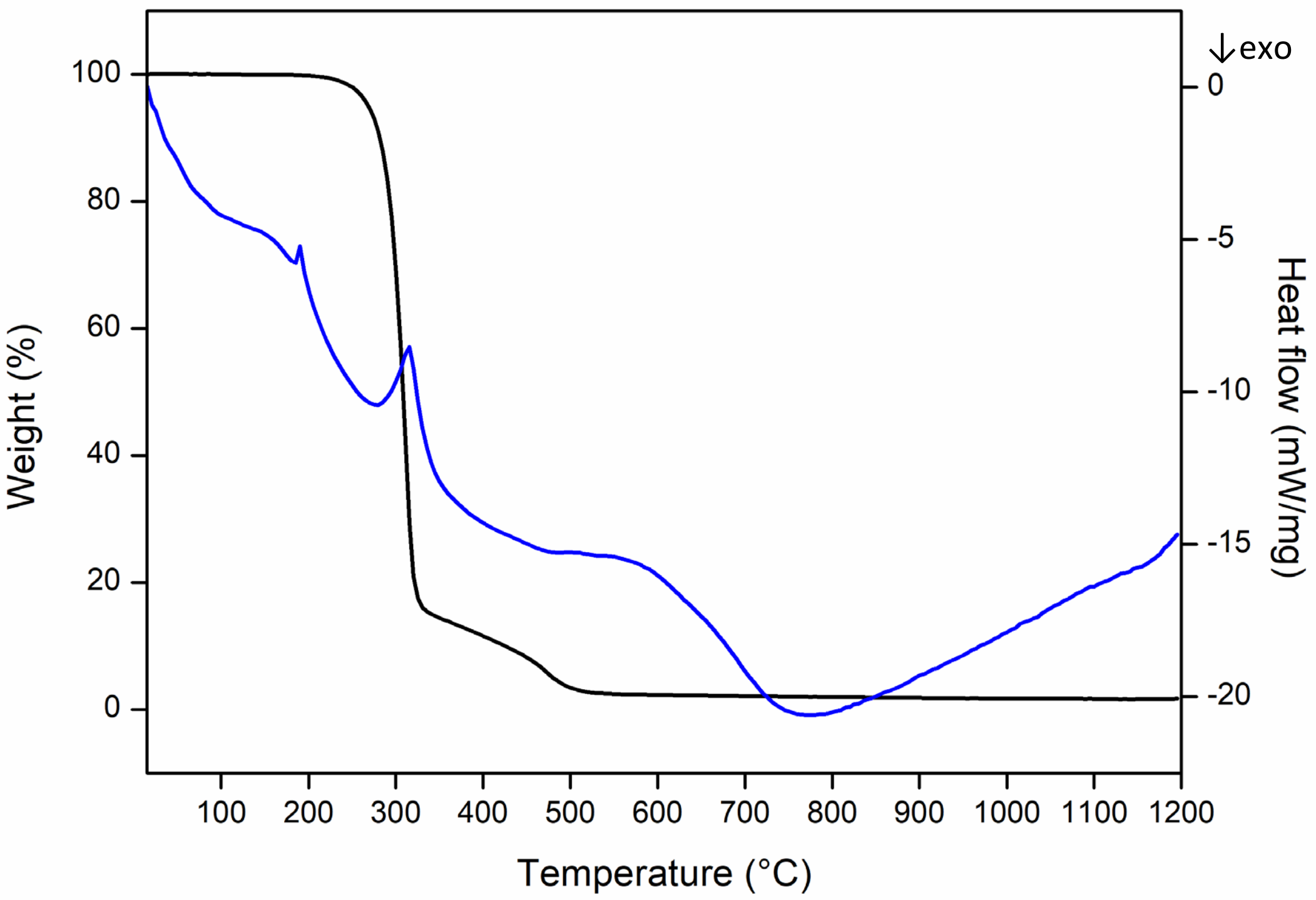

**Figure S39:** TGA and DSC of compound $[BZA]_6[Sb_2Cl_{10}]Cl_2$ (**3**) with a mass loss of 97.5 % observed starting at 188.1 °C.

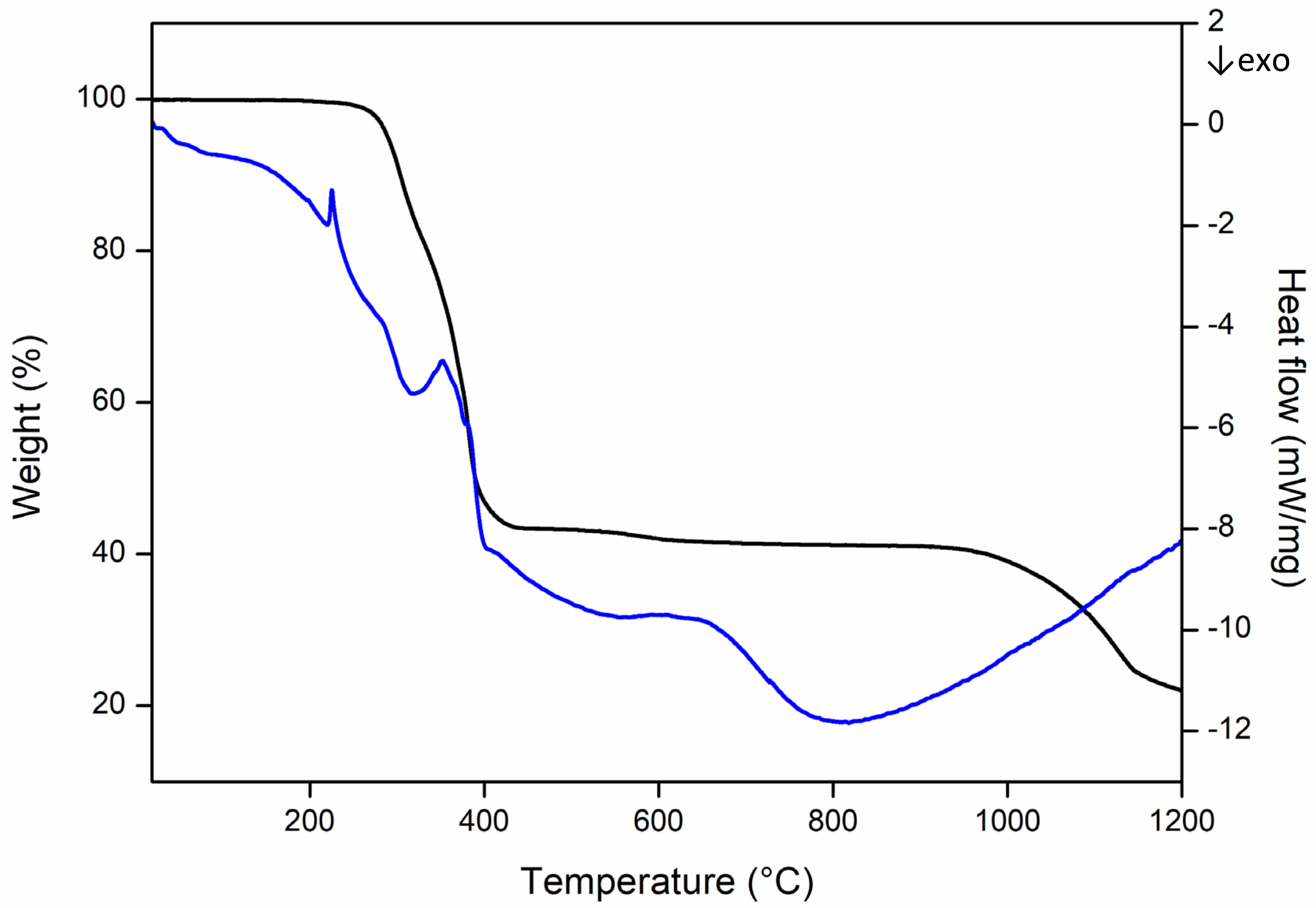

**Figure S40:** TGA and DSC of compound $[BZA]_4[Bi_2Cl_{10}]$ (**4**) with a mass loss of 76.6 % observed starting at 279.5 °C.

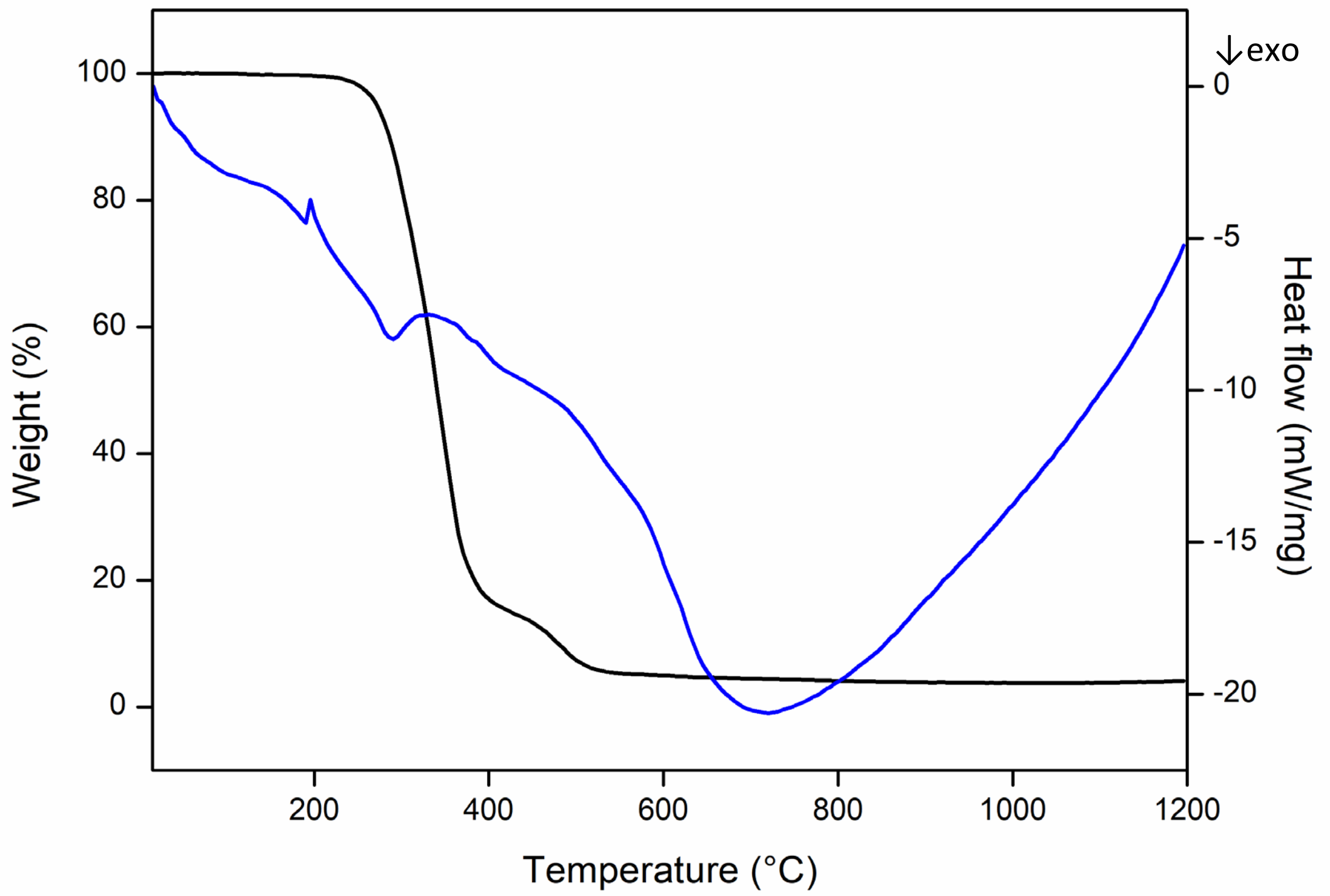

**Figure S41:** TGA and DSC of compound $[BZA]_2[SbBr_5]$ (**8**) with a mass loss of 94.6 % observed starting at 194.7 °C.

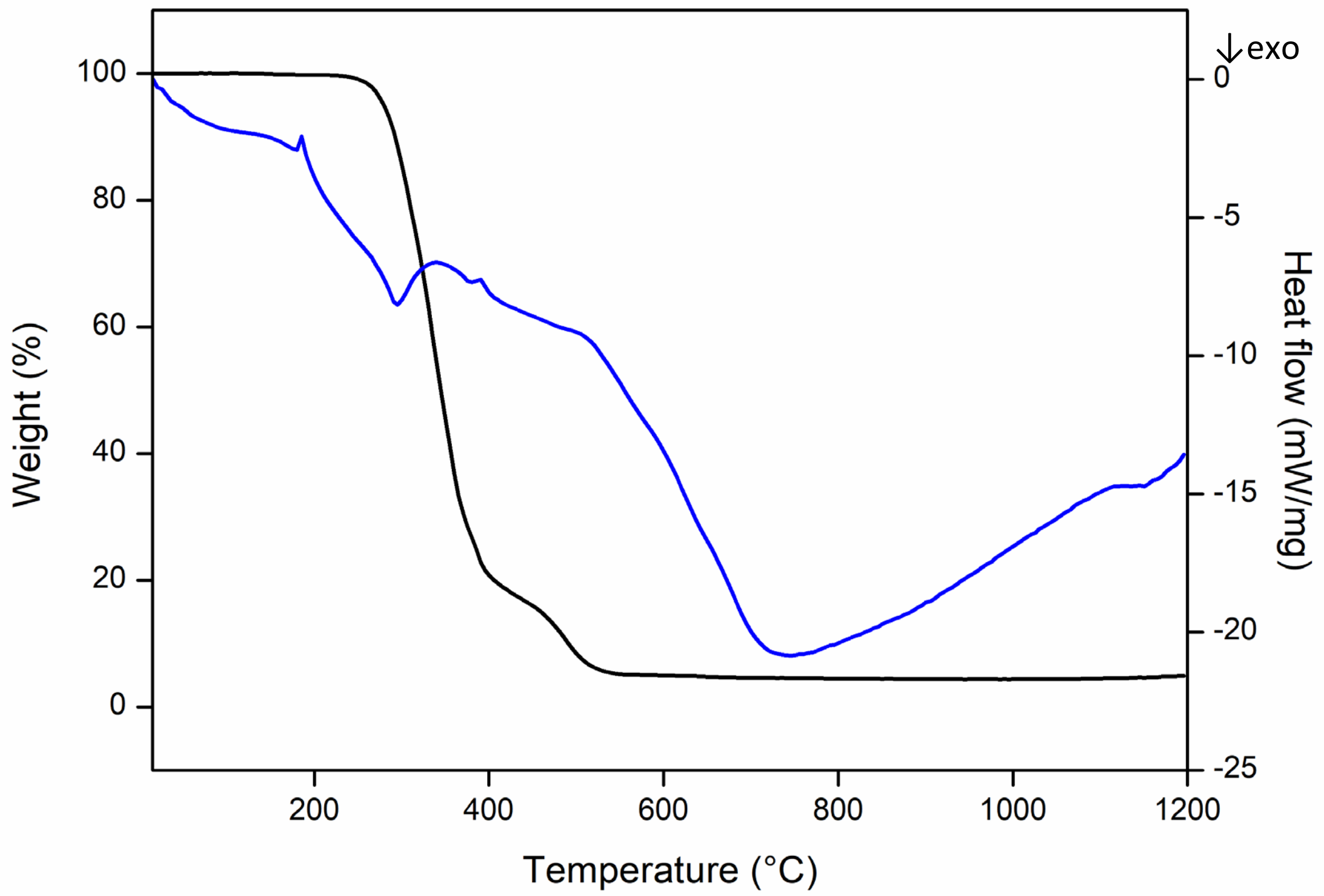

**Figure S42:** TGA and DSC of compound $[BZA]_3[SbBr_6]$ (**9**) with a mass loss of 94.9 % observed starting at 184.1 °C.

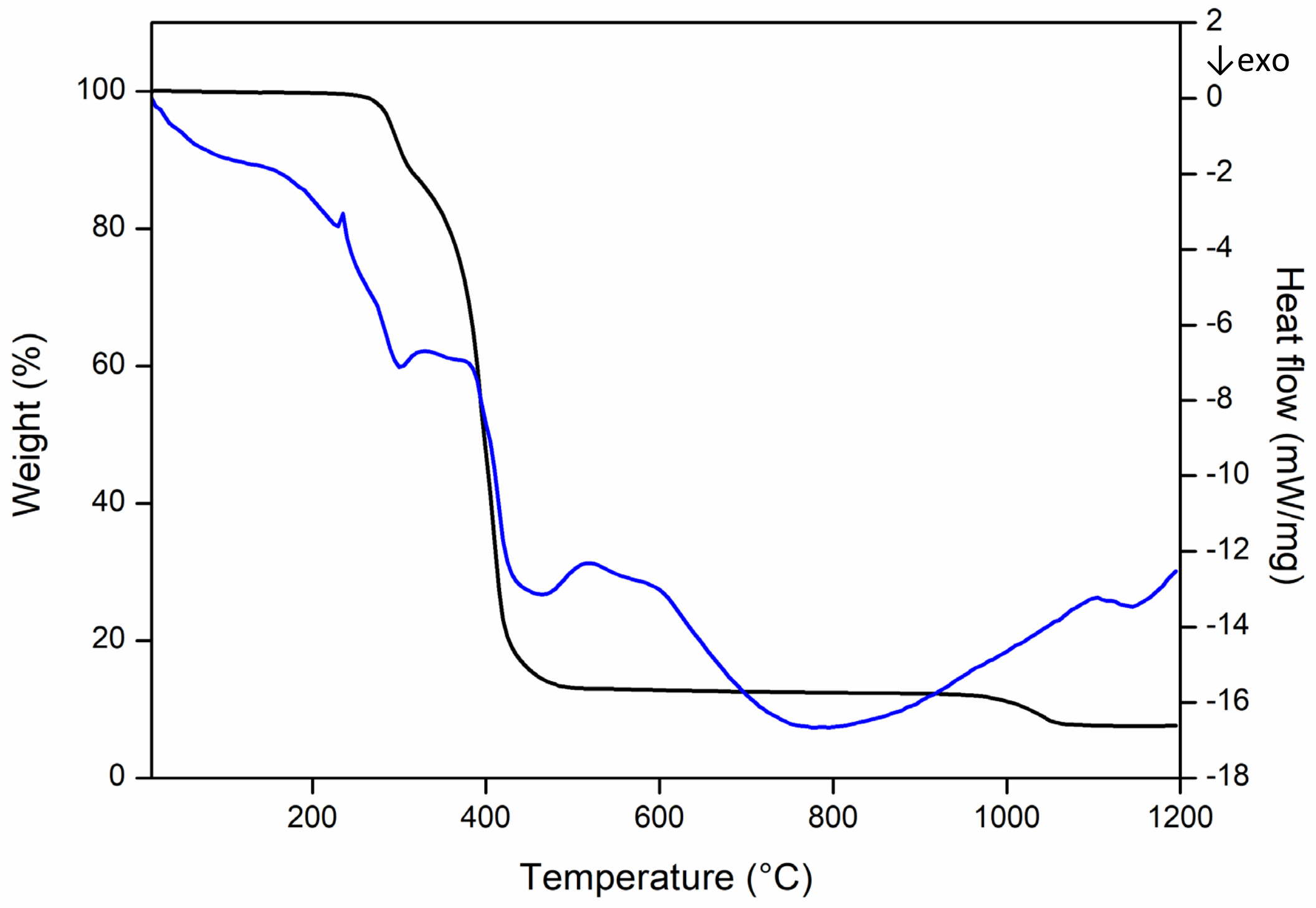

**Figure S43:** TGA and DSC of compound $[BZA]_3[BiBr_6]$ (**10**) with a mass loss of 91.4 % observed starting at 234.5 °C.

**Table S9:** Summary of the onsets of decomposition for all investigated compounds.

| **Compound** | $T_{Decomp.}$ **(°C)** |
|---|---|
| $[CMA]_4[Sb_2Cl_{10}]$ (**1**) | 315 |
| $[CMA]_4[Bi_2Cl_{10}]$ (**2**) | 320 |
| $[BZA]_6[Sb_2Cl_{10}]Cl_2$ (**3**) | 188 |
| $[BZA]_4[Bi_2Cl_{10}]$ (**4**) | 280 |
| $[BZA_3[BiCl_5]Cl$ (**5**)[1] | 257 |
| $[CMA]_3[SbBr_5]Br$ (**6**) | 335 |
| $[CMA]_3[BiBr_5]Br$ (**7**) | 383 |
| $[BZA]_2[SbBr_5]$ (**8**) | 195 |
| $[BZA]_3[SbBr_6]$ (**9**) | 184 |
| $[BZA]_3[BiBr_6]$ (**10**) | 235 |

# 7 Band gap analysis

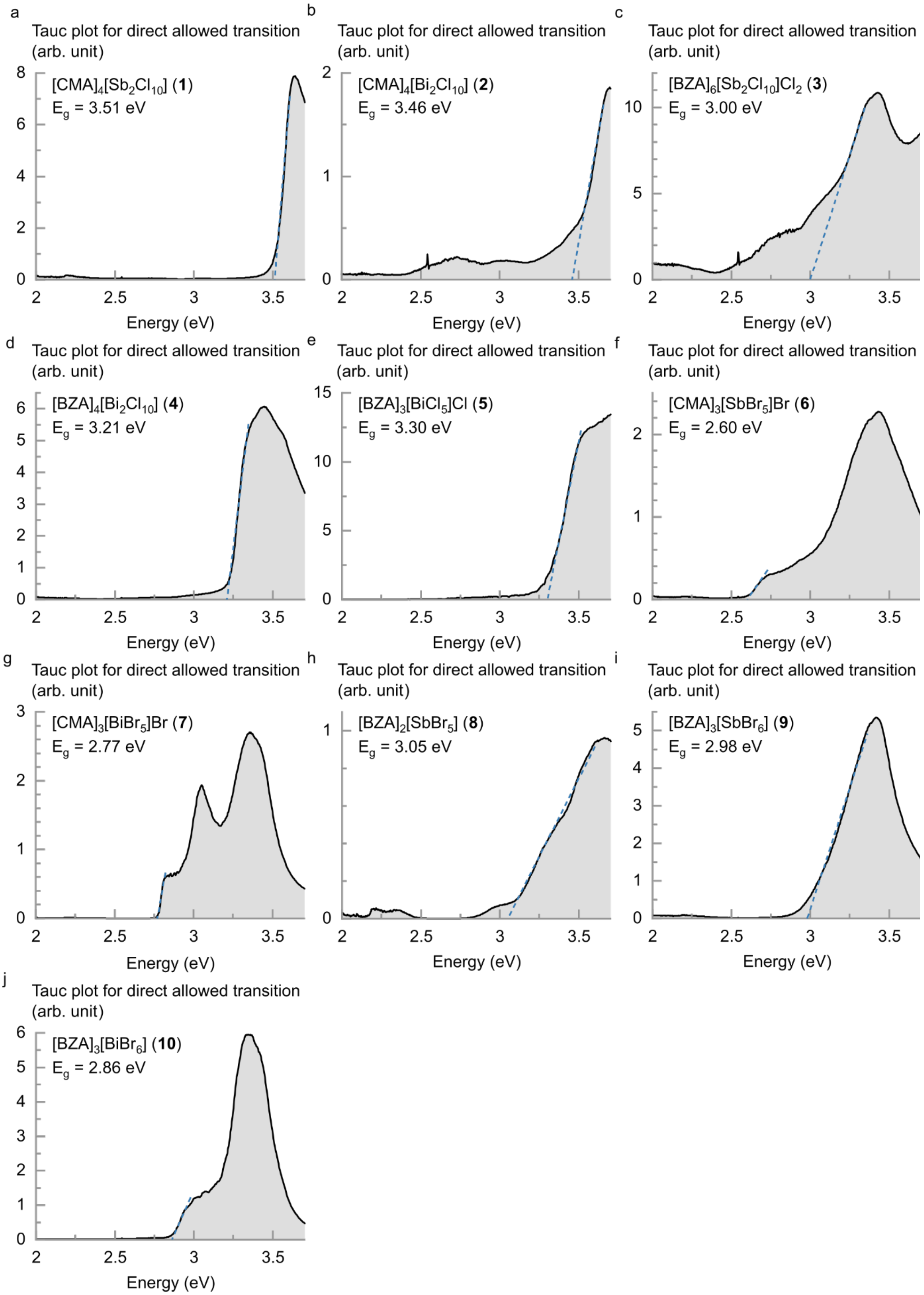

**Figure S44:** Tauc plots for a direct allowed transition for compounds **1**–**10** with the respective band-gap energy stated in each figure (a–j).

## 8 Additional PLE data

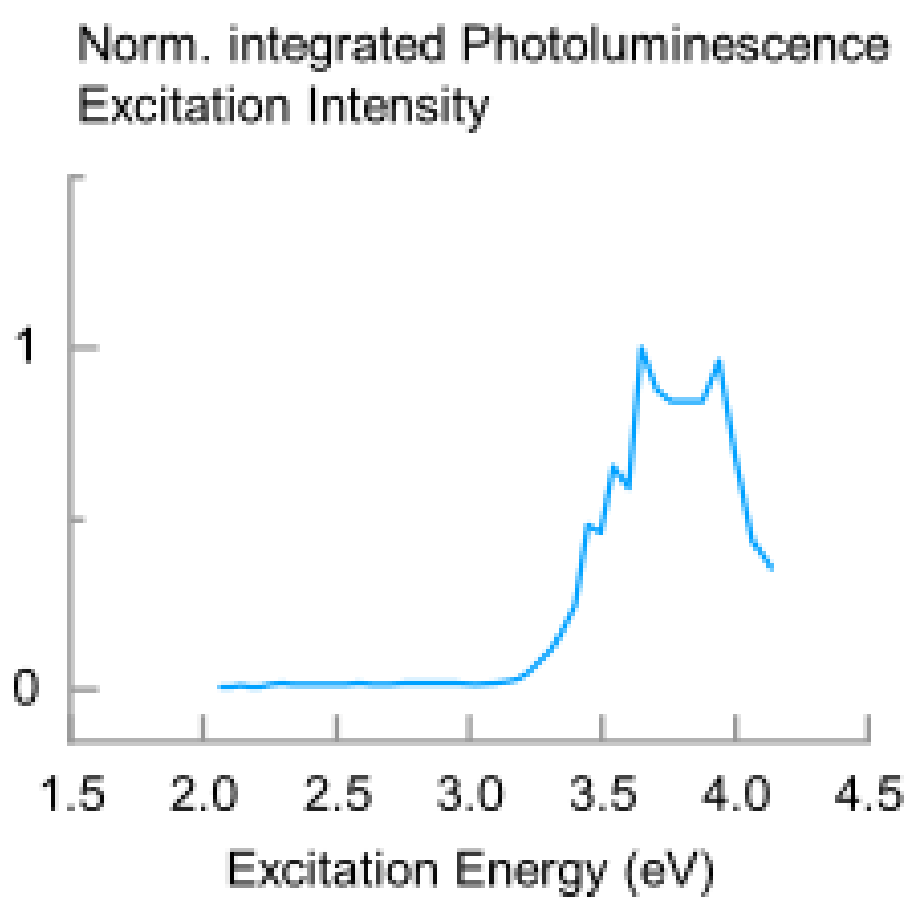

**Figure S45:** Photoluminescence excitation spectrum of $[BZA]_2[SbBr_5]$ (**8**).

# 9 Huang Rhys model

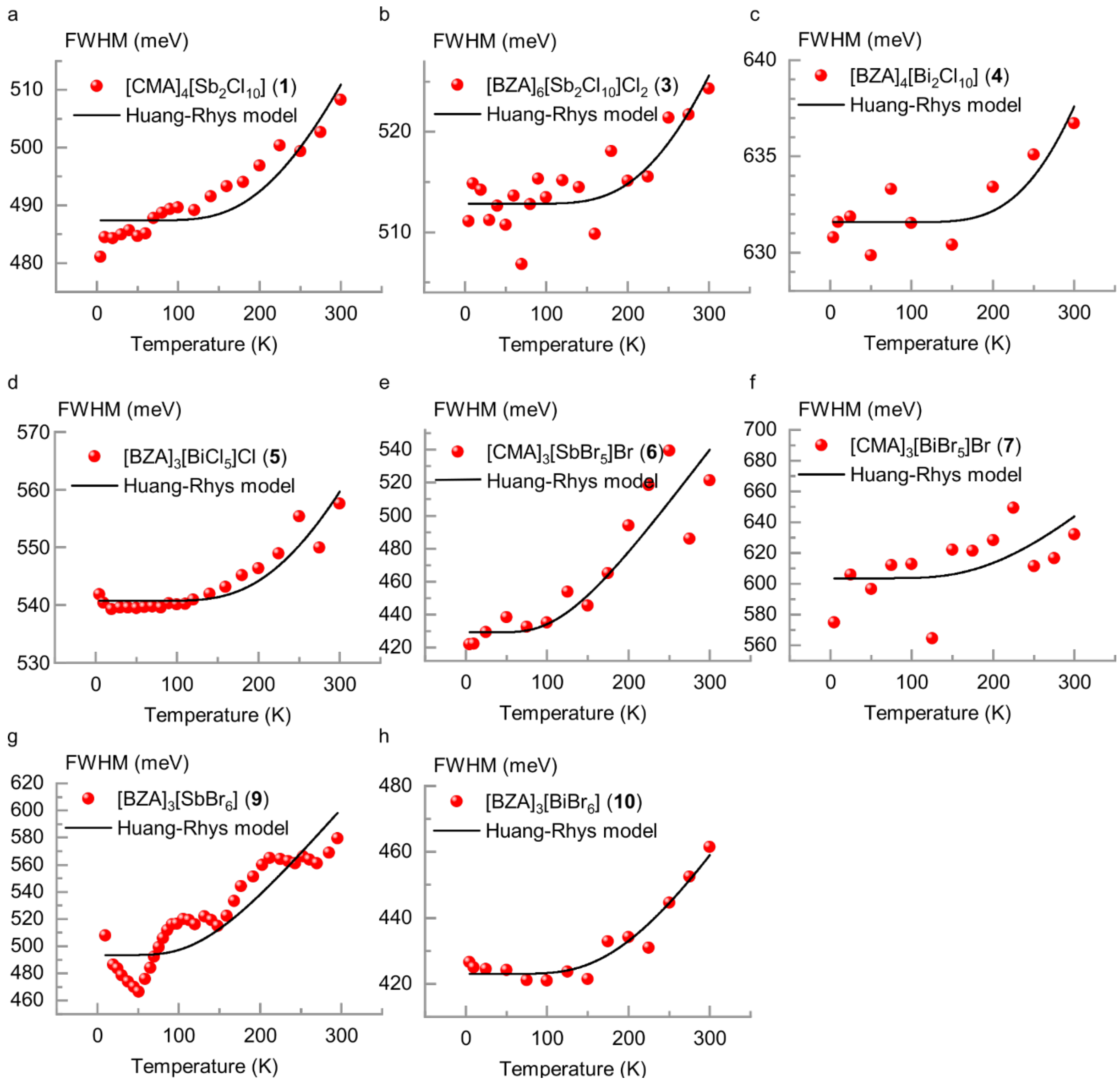

**Figure S46:** (a–h) Temperature-dependent FWHM (red dots) for compounds **1**–**10** fitted with a Huang-Rhys model (black line).

## 10 Stokes compensation

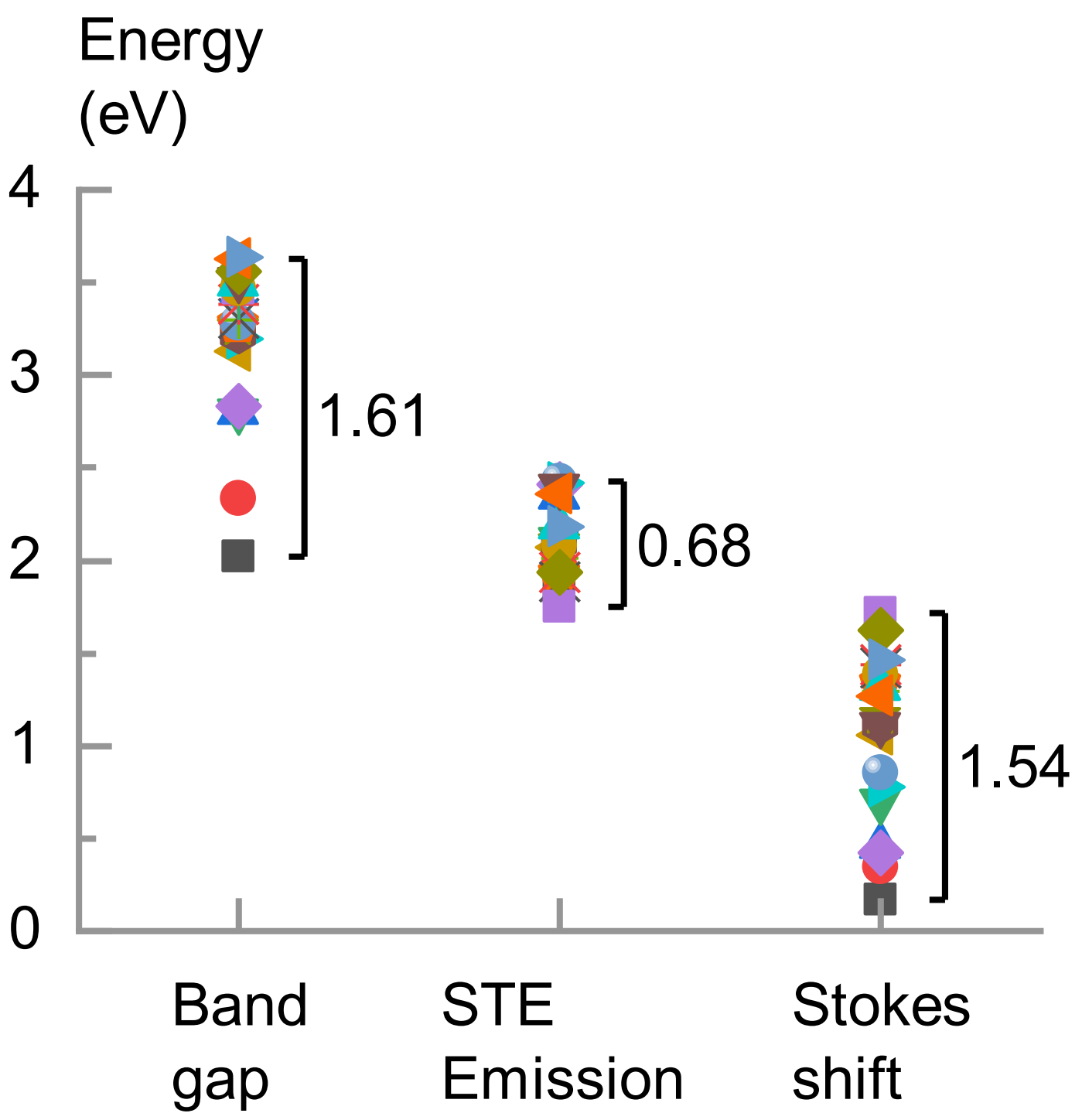

**Figure S47:** Band-gap energies, emission centers and Stokes shifts of literature-known hybrid Antimony and Bismuth bromides and chlorides with 0D and 1D octahedral metal coordination from Refs. [3,13–23] confirming the narrow range of emission energies in hybrid metal halides.

# 11 Optical Spectroscopy Setup and Measurement

µ-Absorption and µ-Photoluminescence (PL) measurements were conducted in a He liquid-flow cryostat (Oxford Instruments, Microstat HiRes) at 5 K with the sample in vacuum. For reflection spectroscopy radiation from a combined tungsten lamp source was used for reflection measurements. A silicon oxide substrate was used as a reference. To determine the optical band gaps the raw data was transformed from reflectance *R* to absorption according to the Kubelka-Munk function

$$F(R) = \frac{(1-R)^2}{2R}$$

and then plotted as a Tauc-plot, where $(F(R) \cdot h\nu)^{1/n}$ is plotted against radiation energy. For a direct band gap $n$ would be $\frac{1}{2}$, for in indirect band bap 2. [24,25] Since the transition in the region of interest was generally far more pronounced when choosing $n = 1/2$, we assume that all analyzed substances feature a direct band gap.

For cw PL measurements, excitation light from a He-Cd laser operating at a wavelength of 325 nm (3.82 eV) was focused on the single crystal compounds using a microscope objective lens (Nikon, CFI Plan Fluor ELWD) with a 40x magnification and NA=0.63. The PL signal was spectrally filtered through a reflection grating monochromator (Andor, Kymera 328i) and detected on a Si charge-coupled device (Andor, DU420A-BU2). For temperature-dependent measurements of the PL FWHM the temperature was set between 5 K and 300 K. Excitation power density was carefully chosen to avoid decomposition of the compounds. For PLE measurements, the samples were studied at ambient conditions and a 5 kHz regenerative amplifier laser system (Spectra Physics Solstice Ace) with a variable excitation wavelength was used. The wavelength was varied between 310 nm (4.0 eV) and 600 nm (2.1 eV) and a pulse shaper, consisting of a reflective grating, lens and spatial filters, limited the spectral width to 5 nm. For wavelengths 300-400 nm step sizes between measurements were 5 nm, for 400-500 nm 10 nm, and for 500-600 nm 20 nm wide. The laser beam was focused onto the sample and the PLE signal was spectrally filtered using a grating spectrometer (Andor, Kymera 193i-B2) and detected on a Si charge-coupled device (Andor, DU-401-BR-DD) and the excitation power was set to be 1.8 µW. At each excitation wavelength step, the measured PL signal was integrated in the range of 1.7 eV to 2.0 eV to yield the PLE intensity.

# 12 DFT Calculations

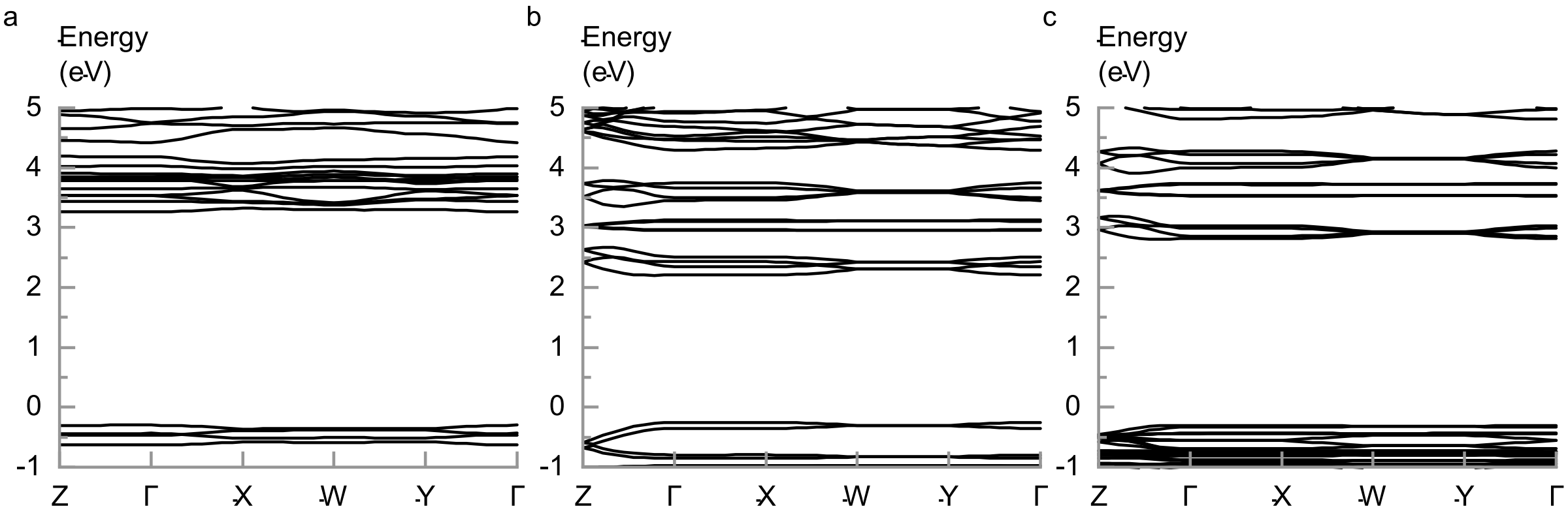

**Figure S48:** Band structures of a) compound **1**, b) compound **6** and c) compound **7**.

# 13 References

(1) Klement, P.; Dehnhardt, N.; Dong, C.-D.; Dobener, F.; Bayliff, S.; Winkler, J.; Hofmann, D. M.; Klar, P. J.; Schumacher, S.; Chatterjee, S.; Heine, J. Atomically Thin Sheets of Lead‐Free 1D Hybrid Perovskites Feature Tunable White‐Light Emission from Self‐Trapped Excitons. *Adv. Mater.* **2021**, *33* (23), 2100518. DOI: 10.1002/adma.202100518.

(2) Anyfantis, G. C.; Ganotopoulos, N.-M.; Savvidou, A.; Raptopoulou, C. P.; Psycharis, V.; Mousdis, G. A. Synthesis and characterization of new organic–inorganic hybrid compounds based on Sb, with a perovskite like structure. *Polyhedron* **2018**, *151*, 299–305. DOI: 10.1016/j.poly.2018.05.024.

(3) Da Chen; Dai, F.; Hao, S.; Zhou, G.; Liu, Q.; Wolverton, C.; Zhao, J.; Xia, Z. Crystal structure and luminescence properties of lead-free metal halides (C 6 H 5 CH 2 NH 3 ) 3 MBr 6 (M = Bi and Sb). *J. Mater. Chem. C* **2020**, *8* (22), 7322–7329. DOI: 10.1039/D0TC00562B.

(4) Sheldrick, G. M. SHELXT - integrated space-group and crystal-structure determination. *Acta Crystallogr., Sect. A: Found. Adv.* **2015**, *71* (Pt 1), 3–8. DOI: 10.1107/S2053273314026370.

(5) Sheldrick, G. M. Crystal structure refinement with SHELXL. *Acta Crystallogr., Sect. C: Struct. Chem.* **2015**, *71* (Pt 1), 3–8. DOI: 10.1107/S2053229614024218.

(6) Sheldrick, G. M. A short history of SHELX. *Acta Crystallogr., Sect. A: Found. Crystallogr.* **2008**, *64* (Pt 1), 112–122. DOI: 10.1107/S0108767307043930.

(7) Dolomanov, O. V.; Bourhis, L. J.; Gildea, R. J.; Howard, J. A. K.; Puschmann, H. OLEX2 : a complete structure solution, refinement and analysis program. *J. Appl. Crystallogr.* **2009**, *42* (2), 339–341. DOI: 10.1107/S0021889808042726.

(8) Brandenburg, K. *Diamond*; Crystal Impact GbR, 2005.

(9) Hu, X.; Zhu, Y.; Wang, J.; Zheng, G.; Yao, D.; Lin, B.; Tian, N.; Zhou, B.; Long, F. Stable organic-inorganic hybrid bismuth-halide: Exploration of crystal-structural, morphological, thermal, spectroscopic and optoelectronic properties. *J. Mol. Struct.* **2022**, *1264*, 133102. DOI: 10.1016/j.molstruc.2022.133102.

(10) Li, X.-N.; Li, P.-F.; Wang, Z.-X.; Shi, P.-P.; Tang, Y.-Y.; Ye, H.-Y. The structural phase transition in a hybrid layered perovskite: [C7H16N]2[SnI4]. *Polyhedron* **2017**, *129*, 92–96. DOI: 10.1016/j.poly.2017.03.025.

(11) Lv, X.-H.; Liao, W.-Q.; Wang, Z.-X.; Li, P.-F.; Mao, C.-Y.; Ye, H.-Y. Design and Prominent Dielectric Properties of a Layered Phase-Transition Crystal: (Cyclohexylmethylammonium) 2 CdCl 4. *Cryst. Growth Des.* **2016**, *16* (7), 3912–3916. DOI: 10.1021/acs.cgd.6b00480.

(12) Brittain, H. G. Vibrational Spectroscopic Studies of Cocrystals and Salts. 4. Cocrystal Products formed by Benzylamine, α-Methylbenzylamine, and their Chloride Salts. *Cryst. Growth Des.* **2011**, *11* (6), 2500–2509. DOI: 10.1021/cg2002628.

(13) Das, S. S.; Rana, R.; Srinivas, C.; Mishra, A.; Iyer, P. K.; Samal, S. L. Synthesis and Luminescence Properties of (TMS) 4 Bi 1– x Sb x Br 7 : A Series of 0D Hybrid Metal Halides Containing S-Based Organic Cations. *Cryst. Growth Des.* **2025**, *25* (4), 1053–1065. DOI: 10.1021/acs.cgd.4c01393.

(14) Deng, C.; Hao, S.; Liu, K.; Molokeev, M. S.; Wolverton, C.; Fan, L.; Zhou, G.; Da Chen; Zhao, J.; Liu, Q. Broadband light emitting zero-dimensional antimony and bismuth-based hybrid halides with diverse structures. *J. Mater. Chem. C* **2021**, *9* (44), 15942–15948. DOI: 10.1039/D1TC04198C.

(15) Wang, Z.; Dan, Q.; Zhao, R.-Y.; Xu, R.-D.; Liu, G.-N.; Li, C. Optical properties of two bismuth(III) halide hybrids with pyridyl sulfide derivative counter cations. *Inorg. Chem. Commun.* **2020**, *111*, 107632. DOI: 10.1016/j.inoche.2019.107632.

(16) Zhang, W.; Tao, K.; Ji, C.; Sun, Z.; Han, S.; Zhang, J.; Wu, Z.; Luo, J. (C6H13N)2BiI5: A One-Dimensional Lead-Free Perovskite-Derivative Photoconductive Light Absorber. *Inorg. Chem.* **2018**, *57* (8), 4239–4243. DOI: 10.1021/acs.inorgchem.8b00152.
(17) Biswas, A.; Bakthavatsalam, R.; Mali, B. P.; Bahadur, V.; Biswas, C.; Raavi, S. S. K.; Gonnade, R. G.; Kundu, J. The metal halide structure and the extent of distortion control the photo-physical properties of luminescent zero dimensional organic-antimony(iii ) halide hybrids. *J. Mater. Chem. C* **2021**, *9* (1), 348–358. DOI: 10.1039/D0TC03440A.
(18) Li, X.; Peng, C.; Xiao, Y.; Xue, D.; Luo, B.; Huang, X.-C. Guest-Induced Reversible Phase Transformation of Organic–Inorganic Phenylpiperazinium Antimony (III) Chlorides with Solvatochromic Photoluminescence. *J. Phys. Chem. C* **2021**, *125* (45), 25112–25118. DOI: 10.1021/acs.jpcc.1c07725.
(19) Lin, F.; Wang, H.; Liu, W.; Li, J. Zero-dimensional ionic antimony halide inorganic–organic hybrid with strong greenish yellow emission. *J. Mater. Chem. C* **2020**, *8* (22), 7300–7303. DOI: 10.1039/c9tc05868k.
(20) Morad, V.; Yakunin, S.; Kovalenko, M. V. Supramolecular Approach for Fine-Tuning of the Bright Luminescence from Zero-Dimensional Antimony(III) Halides. *ACS Mater. Lett.* **2020**, *2* (7), 845–852. DOI: 10.1021/acsmaterialslett.0c00174.
(21) Wang, Z.; Zhang, Z.; Tao, L.; Shen, N.; Hu, B.; Gong, L.; Li, J.; Chen, X.; Huang, X. Hybrid Chloroantimonates(III): Thermally Induced Triple-Mode Reversible Luminescent Switching and Laser-Printable Rewritable Luminescent Paper. *Angew. Chem. Int. Ed.* **2019**, *58* (29), 9974–9978. DOI: 10.1002/anie.201903945.
(22) Zhou, C.; Lin, H.; Tian, Y.; Yuan, Z.; Clark, R.; Chen, B.; van de Burgt, L. J.; Wang, J. C.; Zhou, Y.; Hanson, K.; Meisner, Q. J.; Neu, J.; Besara, T.; Siegrist, T.; Lambers, E.; Djurovich, P.; Ma, B. Luminescent zero-dimensional organic metal halide hybrids with near-unity quantum efficiency. *Chem. Sci.* **2018**, *9* (3), 586–593. DOI: 10.1039/c7sc04539e.
(23) Zhao, J.-Q.; Han, M.-F.; Zhao, X.-J.; Ma, Y.-Y.; Jing, C.-Q.; Pan, H.-M.; Li, D.-Y.; Yue, C.-Y.; Lei, X.-W. Structural Dimensionality Modulation toward Enhanced Photoluminescence Efficiencies of Hybrid Lead ‐ Free Antimony Halides. *Adv. Opt. Mater.* **2021**, *9* (19), 2100556. DOI: 10.1002/adom.202100556.
(24) Makuła, P.; Pacia, M.; Macyk, W. How To Correctly Determine the Band Gap Energy of Modified Semiconductor Photocatalysts Based on UV-Vis Spectra. *J. Phys. Chem. Lett.* **2018**, *9* (23), 6814–6817. DOI: 10.1021/acs.jpclett.8b02892.
(25) Michalow, K. A.; Logvinovich, D.; Weidenkaff, A.; Amberg, M.; Fortunato, G.; Heel, A.; Graule, T.; Rekas, M. Synthesis, characterization and electronic structure of nitrogen-doped TiO2 nanopowder. *Catal. Today* **2009**, *144* (1-2), 7–12. DOI: 10.1016/j.cattod.2008.12.015.